\documentclass[11pt]{article}
\pdfoutput=1 

\usepackage{calc}
\usepackage{rotating}
\usepackage[english]{babel}
\usepackage{graphicx}
\usepackage{subfig}
\usepackage{float}
\usepackage{amsmath}
\usepackage{amssymb}
\usepackage{amsthm}
\usepackage{latexsym}
\usepackage{dcolumn}
\usepackage{geometry}
\usepackage{color}
\usepackage{jheppub}

\newcommand{\newc}{\newcommand*}

\long\def\begincomment#1\endcomment{%
        \begingroup\sf\baselineskip12pt#1\endgroup}

\newc{\etal}{\textrm{et al.}} 
\newc{\eg}{\textrm{e.g.}} 
\newc{\ie}{\textrm{i.e.}}
\newc{\etc}{\textrm{etc}}
\newc\vs{\textrm{vs.}}
\newc{\cl}{\rm {CL}}

\newc{\ev}{\ensuremath{\,\mathrm{eV}}}
\newc{\kev}{\ensuremath{\,\mathrm{keV}}}
\newc{\mev}{\ensuremath{\,\mathrm{MeV}}}
\newc{\gev}{\ensuremath{\,\mathrm{GeV}}}
\newc{\tev}{\ensuremath{\,\mathrm{TeV}}}

\newc{\MeV}{\mev} 
\newc{\TeV}{\tev}
\newc{\invpb}{\ensuremath{/\text{pb}}}
\newc{\invfb}{\ensuremath{/\text{fb}}}

\newc\nb{\ensuremath{\,\mathrm{nb}}} \newc\pb{\ensuremath{\,\mathrm{pb}}} \newc\fb{\ensuremath{\,\mathrm{fb}}}

\newc\pc{\ensuremath{\,\mathrm{pc}}}
\newc\kpc{\ensuremath{\,\mathrm{kpc}}}
\newc\mpc{\ensuremath{\,\mathrm{Mpc}}}

\newc\ps{\ensuremath{\,\mathrm{ps}}} 

\newc\cmeter{\ensuremath{\,\mathrm{cm}}} 
\newc\meter{\ensuremath{\,\mathrm{m}}} 
\newc\kmeter{\ensuremath{\,\mathrm{km}}}

\newc\second{\ensuremath{\,\mathrm{s}}}
\newc\msecond{\ensuremath{\,\mathrm{ms}}}
\newc\nsecond{\ensuremath{\,\mathrm{ns}}}
\newc\psecond{\ensuremath{\,\mathrm{ps}}}

\newc{\chisqmin}{\ensuremath{\chi^2_{\mathrm{min}}}}
\newc{\Delchisq}{\ensuremath{\Delta\chi^2}}
\newc{\chisq}{\ensuremath{\chi^2}}

\newc{\like}{\ensuremath{\mathcal{L}}}

\newc\lsim{\ensuremath{\mathrel{\rlap{\lower4pt\hbox{\hskip1pt$\sim$}}\raise1pt\hbox{$<$}}}}
\newc\gsim{\ensuremath{\mathrel{\rlap{\lower4pt\hbox{\hskip1pt$\sim$}}\raise1pt\hbox{$>$}}}}

\newc{\VEV}[1]{\ensuremath{\langle #1 \rangle}}

\newc{\dl}{\ensuremath{\stackrel{\leftarrow}{D}}}
\newc{\dr}{\ensuremath{\stackrel{\rightarrow}{D}}}

\newc{\bcenter}{\begin{center}}    \newc{\ecenter}{\end{center}}
\newc{\bfl}{\begin{flushleft}}    \newc{\efl}{\end{flushleft}}
\newc{\bfr}{\begin{flushright}}    \newc{\efr}{\end{flushright}}

\newc{\bi}{\begin{itemize}}
\newc{\ei}{\end{itemize}}
\newc{\bed}{\begin{description}}
\newc{\eed}{\end{description}}
\newc{\ben}{\begin{enumerate}}
\newc{\een}{\end{enumerate}}

\newc{\be}{\begin{equation}}
\newc{\ee}{\end{equation}}
\newc{\bea}{\begin{eqnarray}}
\newc{\eea}{\end{eqnarray}}
\newc{\ra}{\rightarrow}

\newc{\alphas}{\ensuremath{\alpha_s}}
\newc{\alphatwo}{\ensuremath{\alpha_2}}
\newc{\alphaone}{\ensuremath{\alpha_1}}
\newc{\alphai}[1]{\ensuremath{\alpha_{#1}}}
\newc{\alphaem}{\ensuremath{\alpha_{\mathrm{em}}}}

\newc{\alphaeff}{\ensuremath{\alpha_{\mathrm{eff}}}}

\newc{\sineff}{\ensuremath{\sin \theta_{\mathrm{eff}}}}
\newc{\sinsqeff}{\ensuremath{\sin^2 \theta_{\mathrm{eff}}}}
\newc{\dalphahad}{\ensuremath{\Delta \alpha_{\mathrm{had}}}}

\newc{\yt}{\ensuremath{h_t}} \newc{\yb}{\ensuremath{h_b}} \newc{\ytau}{\ensuremath{h_{\tau}}}

\newc\mz{\ensuremath{M_Z}} 
\newc\mw{\ensuremath{m_W}}
\newc\mZ{\mz}        \newc\mW{\mw}

\newc\mhsm{\ensuremath{ m_{H_{\mathrm{SM}}}}}

\newc{\mtop}{\ensuremath{ m_t}}               \newc{\mtpole}{\ensuremath{ M_t}}
\newc{\mbottom}{\ensuremath{ m_b}} 
\newc{\mtau}{\ensuremath{ m_{\tau}}}
\newc{\mt}{\mtpole}
\newc{\mb}{\mbottom} 

\newc{\rtwogg}{\ensuremath{R_{h_2}(\gamma\gamma)}}
\newc{\rtwozz}{\ensuremath{R_{h_2}(ZZ)}}
\newc{\ronegg}{\ensuremath{R_{h_1}(\gamma\gamma)}}
\newc{\ronezz}{\ensuremath{R_{h_1}(ZZ)}}
\newc{\rsiggg}{\ensuremath{R_{h_\textrm{sig}}(\gamma\gamma)}}
\newc{\rsigzz}{\ensuremath{R_{h_\textrm{sig}}(ZZ)}}

\newc{\llbar}{\ensuremath{\ell\bar{\ell}}}
\newc{\tauptaum}{\ensuremath{ \tau^+\tau^-}}

\newc{\qqbar}{\ensuremath{ q\bar{q}}} \newc{\ppbar}{\ensuremath{ p\bar{p}}}
\newc{\bbbar}{\ensuremath{ b\bar{b}}} \newc{\ttbar}{\ensuremath{ t\bar{t}}}
\newc{\ffbar}{\ensuremath{ f\bar{f}}} \newc{\tautaubar}{\ensuremath{ \tau\bar{\tau}}}

\newc{\mchi}{\ensuremath{m_\neutone}}
\newc{\squark}{\ensuremath{\tilde{q}}}
\newc{\slepton}{\ensuremath{\tilde{l}}}
\newc{\gluino}{\ensuremath{\tilde{g}}} 
\newc{\mgluino}{\ensuremath{{m_{\gluino}}}}

\newc{\sthw}{\ensuremath{ \sin\theta_W}}              \newc{\cthw}{\ensuremath{\cos\theta_W}}
\newc{\tanthw}{\ensuremath{ \tan\theta_W}}              \newc{\cotthw}{\ensuremath{\cot\theta_W}}

\newc{\ssqthw}{\ensuremath{\sin^2 \theta_W}}

\newc{\msbar}{\ensuremath{\overline{MS}}} \newc{\drbar}{\ensuremath{\overline{DR}}}

\newc{\mtmtsmmsbar}{\ensuremath{ m_t(m_t)^{\msbar}_{{\mathrm{SM}}}}}
\newc{\mtmtsmdrbar}{\ensuremath{ m_t(m_t)^{\drbar}_{{\mathrm{SM}}}}}
\newc{\mtmtmssmdrbar}{\ensuremath{ m_t(m_t)^{\drbar}_{{\mathrm{SUSY}}}}}

\newc{\mbmbmsbar}{\ensuremath{ m_b(m_b)^{\msbar} }}

\newc{\mbmbsmmsbar}{\ensuremath{ m_b(m_b)^{\msbar}_{{\mathrm{SM}}}}}
\newc{\mbmzsmmsbar}{\ensuremath{ m_b(\mz)^{\msbar}_{{\mathrm{SM}}}}}
\newc{\mbmzsmdrbar}{\ensuremath{ m_b(\mz)^{\drbar}_{{\mathrm{SM}}}}}
\newc{\mbmzmssmdrbar}{\ensuremath{ m_b(\mz)^{\drbar}_{{\mathrm{SUSY}}}}}

\newc{\mtaumzsmmsbar}{\ensuremath{ m_{\tau}(\mz)^{\msbar}_{{\mathrm{SM}}}}}
\newc{\mtaumzsmdrbar}{\ensuremath{ m_{\tau}(\mz)^{\drbar}_{{\mathrm{SM}}}}}
\newc{\mtaumzmssmdrbar}{\ensuremath{ m_{\tau}(\mz)^{\drbar}_{{\mathrm{SUSY}}}}}

\newc{\alphasmzms}{\ensuremath{\alpha_s(M_Z)^{\overline{MS}}}}
\newc{\alphaimzms}[1]{\ensuremath{\alpha_{#1}(M_Z)^{\overline{MS}}}}

\newc{\alphaemmz}{\ensuremath{\alpha_{\mathrm{em}}(M_Z)^{\overline{MS}}}}

\newc{\mzero}{\ensuremath{{m_0}}}
\newc{\mhalf}{\ensuremath{ m_{1/2}}}
\newc{\tanb}{\ensuremath{\tan\beta}}
\newc{\azero}{\ensuremath{ A_0}}
\newc{\signmu}{\ensuremath{\rm{sgn}\,\mu}}
\newc{\atau}{\ensuremath{{A_{\tau}}}}

\newc{\mueff}{\ensuremath{\mu_{\rm{eff}}}}

\newc{\lam}{\ensuremath{{\lambda}}}
\newc{\kap}{\ensuremath{{\kappa}}}

\newc{\alam}{\ensuremath{{A_{\lambda}}}}
\newc{\akap}{\ensuremath{{A_{\kappa}}}}

 \newc{\hs}{\ensuremath{ H_s}}      
\newc{\mhs}{\ensuremath{ m_{H_s}}} 

\newc{\mgut}{\ensuremath{ M_{\rm GUT}}}
\newc{\mplanck}{\ensuremath{ M_{\rm P}}}      \newc{\mpl}{\ensuremath{ M_{\rm Pl}}}
\newc{\msusy}{\ensuremath{ M_{\rm SUSY}}}      \newc{\ms}{\ensuremath{ M_{\rm S}}}

 \newc{\hu}{\ensuremath{ H_u}}       \newc{\hd}{\ensuremath{ H_d}}
 \newc{\mhu}{\ensuremath{ m_{H_u}}}       \newc{\mhd}{\ensuremath{ m_{H_d}}}
 \newc{\mhuew}{\ensuremath{ m^{\ast}_{H_u}}}       \newc{\mhdew}{\ensuremath{ m^{\ast}_{H_d}}}
 \newc{\mhuewsq}{\ensuremath{ m^{\ast\, 2}_{H_u}}}       \newc{\mhdewsq}{\ensuremath{ m^{\ast\, 2}_{H_d}}}
 \newc{\mhl}{\ensuremath{m_\hl}} 
 \newc{\mhone}{\ensuremath{m_{h_1}}} 
 \newc{\mhtwo}{\ensuremath{m_{h_2}}} 
 \newc{\mglu}{\ensuremath{m_{\tilde g}}} 
 \newc{\mul}{\ensuremath{m_{\tilde{u}_L}}} 
 \newc{\mtone}{\ensuremath{m_{\tilde{t}_1}}} 
 \newc{\ma}{\ensuremath{m_A}} 
 \newc{\maone}{\ensuremath{m_{a_1}}} 
 \newc{\matwo}{\ensuremath{m_{a_2}}}
 \newc{\hone}{\ensuremath{h_1}}
 \newc{\htwo}{\ensuremath{h_2}}
 \newc{\aone}{\ensuremath{a_1}}
 \newc{\atwo}{\ensuremath{a_2}}

\newc{\sigsip}{\ensuremath{\sigma^{\rm SI}_{p}}}	\newc{\sigsin}{\ensuremath{\sigma^{\rm SI}_{n}}}
\newc{\sigsdp}{\ensuremath{\sigma^{\rm SD}_{p}}}	\newc{\sigsdn}{\ensuremath{\sigma^{\rm SD}_{n}}}
\newc{\sigsi}{\ensuremath{\sigma^{\rm SI}}}	\newc{\sigsd}{\ensuremath{\sigma^{\rm SD}}}

\newc{\abund}{\ensuremath{ \Omega h^2}}
\newc{\omegadm}{\ensuremath{ \Omega_{{\rm DM}}}}     \newc{\abunddm}{\ensuremath{ \Omega_{{\rm DM}} h^2}} 
\newc{\omegam}{\ensuremath{ \Omega_{{\rm m}}}}       \newc{\abundm}{\ensuremath{ \Omega_{{\rm m}} h^2}}
\newc{\omegab}{\ensuremath{ \Omega_{{\rm b}}}}	\newc{\abundb}{\ensuremath{ \Omega_{{\rm b}} h^2}}
\newc{\omegatot}{\ensuremath{ \Omega_{{\rm TOT}}}}
\newc{\omegacdm}{\ensuremath{ \Omega_{{\rm CDM}}}}   \newc{\abundcdm}{\ensuremath{ \Omega_{{\rm CDM}} h^2}}
\newc{\omegalambda}{\ensuremath{ \Omega_{\Lambda}}} \newc{\abundlambda}{\ensuremath{ \Omega_{\Lambda} h^2}}
\newc{\omegarad}{\ensuremath{ \Omega_{{\rm rad}}}}  \newc{\abundrad}{\ensuremath{ \Omega_{{\rm rad}} h^2}}

\newc{\rhocrit}{\ensuremath{ \rho_{\rm crit}}}
\newc{\rhochi}{\ensuremath{ \rho_{\chi}}}

\newc{\abunchi}{\ensuremath{\Omega_\chi h^2}}
\newc{\abundlsp}{\ensuremath{\Omega_{\rm LSP}h^2}}

\newc{\amu}{\ensuremath{ a_{\mu}}}        \newc{\amususy}{\ensuremath{ a_{\mu}^{\mathrm{SUSY}}}}
\newc{\amuexpt}{\ensuremath{ a_{\mu}^{\mathrm{expt}}}}        \newc{\amusm}{\ensuremath{ a_{\mu}^{\mathrm{SM}}}}
\newc\deltaamu{\ensuremath{\Delta a_{\mu}}} \newc{\deltaamususy}{\ensuremath{\delta a_{\mu}^{\mathrm{SUSY}}}}
\newc\gmtwo{\ensuremath{ (g-2)_{\mu}}} 
\newc{\deltagmtwomususy}{\ensuremath{\delta\left(g-2\right)_{\mu}^{\mathrm{SUSY}}}}
\newc{\deltagmtwomu}{\ensuremath{\delta\left(g-2\right)_{\mu}}}

\newc\BR{\ensuremath{\rm BR}}

\newc\bsgamma{\ensuremath{ b\rightarrow s \gamma }}
\newc\bxsgamma{\ensuremath{\overline{B}\rightarrow X_{s}\gamma}}

\newc\brbsgamma{\ensuremath{\BR\left(\bsgamma\right)}}
\newc\brbxsgamma{\ensuremath{\BR\left(\bxsgamma\right)}}

\newc\bsmumu{\ensuremath{B_s\to\mu^+\mu^-}}
\newc\brbsmumu{\ensuremath{\BR\left(B_s\to\mu^+\mu^-\right)}}

\newc\bdmmumu{\ensuremath{\overline{B}_d\to\mu^+\mu^-}}

\newc\bbbarmix{\ensuremath{\overline{B}_s\mbox{-}B_s}}      
\newc\delmbs{\ensuremath{\Delta M_{B_s}}}

\newc{\butaunu}{\ensuremath{B_u \rightarrow \tau \nu}}
\newc{\brbutaunu}{\ensuremath{\BR\left(B_u \rightarrow \tau \nu\right)}}

\newcommand*{\neutone}{\ensuremath{\chi}}

\newcommand*{\micromegas}{MicrOMEGAs}

\newcommand*{\superiso}{\text{SuperIso}}

\newcommand*{\nmssmtools}{\text{NMSSMTools}}

\let\oldcite\cite
\renewcommand*{\cite}{~\oldcite}

\newcommand*{\hl}{\ensuremath{h}}

\title{\bf A light NMSSM pseudoscalar Higgs boson at the LHC redux}
\author[a]{N-E. Bomark,}
\author[b]{S. Moretti,}
\author[c]{S. Munir,}
\author[a,1]{and L. Roszkowski\note{On leave of absence from the University of
    Sheffield, U.K.}}

\affiliation[a]{National Centre for Nuclear Research,
  Ho{\. z}a 69, 00-681 Warsaw, Poland}
\affiliation[b]{School of Physics \& Astronomy,
University of Southampton, Southampton SO17 1BJ, UK}
\affiliation[c]{Department of Physics and Astronomy,
Uppsala University, Box 516, SE-751 20 Uppsala, Sweden}

\emailAdd{Nbomark@fuw.edu.pl}
\emailAdd{S.Moretti@soton.ac.uk}
\emailAdd{Shoaib.Munir@physics.uu.se}
\emailAdd{L.Roszkowski@sheffield.ac.uk}

\abstract{The Next-to-Minimal Supersymmetric Standard Model (NMSSM) contains a
singlet-like pseudoscalar Higgs boson in addition to the
doublet-like pseudoscalar of the Minimal Supersymmetric Standard Model. This new pseudoscalar can
    have a very low mass without violating the LEP exclusion
    constraints and it can potentially provide a hallmark signature of
non-minimal supersymmetry at the LHC. In this
analysis we revisit the light pseudoscalar in the NMSSM with partial
universality at some high unification scale. We delineate the regions of the model's
parameter space that are consistent with the up-to-date theoretical
and experimental constraints, from both Higgs boson searches and elsewhere
(most notably $b$-physics), and examine to what extent they can be
probed by the LHC. To this end we review the most important production channels of such
a Higgs state and assess the scope of its observation at the
forthcoming Run-2 of the LHC. We conclude that the
$b\bar{b}$-associated production of the pseudoscalar,
which has been emphasised in previous studies, does not carry
much promise anymore, given the measured mass of the Higgs boson at the LHC.
However, the decays of one of the heavier scalar Higgs bosons of
the NMSSM can potentially lead to the discovery of its light pseudoscalar.
Especially promising are the decays of one or both of the two lightest scalar states
into a pseudoscalar pair and of the heaviest scalar into
a pseudoscalar and a $Z$ boson. Since the latter channel has not been
explored in detail in the literature so far, we provide details of some
benchmark points which can be probed for establishing its signature.}

\begin{document}
\maketitle
\flushbottom

\section{\label{intro}Introduction}

The Next-to-Minimal Supersymmetric Standard Model
(NMSSM)\cite{Fayet:1974pd,Ellis:1988er,Durand:1988rg,Drees:1988fc,Miller:2003ay}
 contains a singlet Higgs field in addition to the two doublet fields
 of the Minimal Supersymmetric Standard Model (MSSM).
This results in two additional neutral mass eigenstates, one scalar
and one pseudoscalar, in the Higgs sector, on top of the three MSSM-like ones.
Naturally, these new Higgs bosons are singlet-dominated, implying that
their couplings to the
fermions and gauge bosons of the Standard Model (SM) are typically
much smaller than those of the doublet-dominated Higgs bosons.
Their masses are thus generally very weakly
constrained by the Higgs boson data from the Large Electron Positron (LEP) 
collider as well as the Large Hadron Collider (LHC), and can be
as low as a few \gev. Evidently, the observation of any of these potentially light states,
in addition to the SM-like Higgs boson discovered already\cite{Aad:2012tfa,Chatrchyan:2012ufa}, will provide a clear
indication of not just physics beyond the SM but also of a non-minimal
nature of supersymmetry (SUSY).

In particular, the lightest NMSSM pseudoscalar, $A_1$,
is a crucial probe of new physics as it can be tested at the already
established searches for a
supersymmetric pseudoscalar Higgs boson performed by many
experimental groups\cite{Love:2008aa,Tung:2008gd,
Lees:2012iw,Lees:2012te,Lees:2013vuj,Peruzzi:2014ksa}.
These searches are sensitive only to very light pseudoscalars Higgs
bosons which could result from the decays of heavy mesons.
The results from the CLEO
experiment\cite{Love:2008aa} have in fact been used in the past to directly constrain the
mass of the NMSSM $A_1$\cite{Domingo:2008rr}.
Heavier, $\gtrsim 5$\gev, pseudoscalars have also been probed in
the possible decay of a heavy SM-like scalar Higgs boson at
LEP2\cite{Schael:2010aw}. More recently, the LHC has searched
for a light pseudoscalar decaying in the $\mu^+\mu^-$ channel and
produced either singly in $pp$ collisions\cite{Chatrchyan:2012am}
or in pairs from the decays of a non-SM-like Higgs boson\cite{Chatrchyan:2012cg}.

In the context of the NMSSM, the production of $A_1$ via decays of
other heavy Higgs bosons has been the subject of a number of
studies\cite{Ellwanger:2003jt,Ellwanger:2005uu,Moretti:2006hq,Forshaw:2007ra,Belyaev:2008gj,Almarashi:2011te},
aiming at establishing a ``no-lose'' theorem for the discovery of the
NMSSM Higgs bosons (see\cite{Ellwanger:2013ova} for a recent review).
Other possible production processes have also been
investigated in the literature. In\cite{Cheung:2008rh} the production of a light $A_1$ in
association with a light neutralino via the decay of heavier
neutralinos was discussed.
$b\bar{b}$-associated production of the NMSSM $A_1$ at the
LHC followed by decays in the $\tau^+\tau^-$ and the $\mu^+\mu^-$
channels was studied in\cite{Almarashi:2010jm,Almarashi:2011hj}, respectively.
Assuming the same production process, it was established
in\cite{Almarashi:2011bf} that an $A_1$, with
$20\gev \lesssim m_{A_1} \lesssim 80$\gev, could be observed with
very large luminosity at the LHC in the $b\bar{b}$ decay channel.
This is made possible by an enhanced coupling of
$A_1$ to $b\bar{b}$ pairs for certain configurations of the model
parameters\cite{Almarashi:2012ri}.
$b\bar{b}$-associated production also affords the possibility to search for the
 $H_{\rm SM} \rightarrow A_1Z$ decay mode\cite{Almarashi:2011qq},
where $H_{\rm SM}$ is our notation for
the generic SM-like Higgs boson of the model. We should point out
here that in the NMSSM both $H_1$
and $H_2$, the lightest and next-to-lightest CP-even Higgs bosons,
respectively, can alternatively play the role of $H_{\rm
 SM}$\cite{Ellwanger:2011aa,King:2012is,Ellwanger:2012ke,Gherghetta:2012gb,Cao:2012fz}.

All the above mentioned analyses were, however, performed prior to the discovery of
the Higgs boson at the LHC\cite{Chatrchyan:2012ufa,Aad:2012tfa}.
Thenceforth, in\cite{Kim:2012az} the $A_1\rightarrow\gamma\gamma$
decay channel was studied for a light $A_1$.
In\cite{Munir:2013wka} it was noted that in the NMSSM the $A_1$ could in fact be
degenerate in mass with the SM-like Higgs boson. It could thus cause an
enhancement in the Higgs boson signal rates near 125\,\gev\ in the $\gamma\gamma$,
$b\bar{b}$ and $\tau^+\tau^-$ channels simultaneously provided, again,
that it is produced in association with a $b\bar{b}$ pair.
In\cite{Cerdeno:2013cz} the $H_{\rm SM} \rightarrow A_1A_1
\rightarrow 4\ell$ (with $\ell$ denoting $e^\pm$ and $\mu^\pm$) process at the LHC has been
studied in detail. The production of $A_1$ via neutralino decays has also been recently
revisited in\cite{Stal:2011cz,Das:2012rr,Cerdeno:2013qta}.
Finally, NMSSM benchmark proposals
capturing much of this phenomenology also exist\cite{Djouadi:2008uw,King:2012tr,Curtin:2013fra,King:2014xwa}.
In this article we analyse in detail some of the production processes
that yield a sizeable cross section and could potentially lead to
the detection of a light NMSSM $A_1$ at the LHC with $\sqrt{s} =14$\tev.
We perform parameter scans of the NMSSM with partial
universality at the Grand Unification Theory (GUT) scale to find regions where a light,
$\lesssim 150$\gev, $A_1$ can be obtained. In these scans we require
the mass of $H_{\rm SM}$ to lie around 125\gev\ and its signal
rates in the $\gamma\gamma$ and $ZZ$ channels to be
consistent with the SM expectations. We study in detail the
two possibilities, ${H_{\rm SM}} = H_1$ and
 ${H_{\rm SM}} = H_2$, as two separate cases.
Moreover, we divide the production mode of the $A_1$ into two main
categories: 1) direct production in association with a $b\bar{b}$
pair, which we refer to as the $bbA_1$ mode in the following, and 2) via
decay of a heavy scalar Higgs boson of the model.

In the second category mentioned above, we include the two main
production channels of the pseudoscalar, namely
$A_1A_1$ and $A_1Z$. Furthermore, the decaying heavier Higgs boson can
be any of the three neutral scalars, $H_1$, $H_2$ and $H_3$.
$A_1$'s thus produced decay into either $b\bar b$ or
$\tau^+\tau^-$ pairs. The former decay channel is always the
dominant one as the ratio of the Branching Ratios (BRs) for these channels
is given approximately by the ratio of the $b$ and $\tau$ masses,
but the latter can be equally important due to a relatively small $\tau^+\tau^-$ background.
In case of the $A_1Z$ decay channel, we only consider the leptonic
($e^+e^-$ and $\mu^+\mu^-$) decays of the $Z$ boson.
To study the prospects for the discovery of an $A_1$ at the LHC in all
these production and decay channels, we employ
hadron level Monte Carlo (MC) simulations. We perform a detailed
signal-to-background analysis for each process of interest, employing
a jet substructure method for detecting the
$b$ quarks originating from an $A_1$ decay.

The article is organised as follows. In section~\ref{model}, we will
briefly discuss the model under consideration and some important
properties of the singlet-like pseudoscalar in it. In
section~\ref{method} we will explain our methodology for NMSSM
parameter space scans as well as for our signal-to-background analyses
of the $A_1$ production and decay processes considered. In section~\ref{results} we
will discuss our results in detail. Finally, we will present our conclusions in section~\ref{concl}.

\section{\label{model} The light pseudoscalar {\boldmath $A_1$} in the NMSSM}

The scale-invariant superpotential of the NMSSM (see,
e.g.,\cite{Ellwanger:2009dp,Maniatis:2009re} for reviews) is defined in terms of
the two NMSSM Higgs doublet superfields $\widehat{H}_u$ and $\widehat{H}_d$ along
with an additional Higgs singlet superfield $\widehat{S}$ as
\begin{equation}
\label{eq:superpot}
W_{\rm NMSSM}\ =\ {\rm MSSM\;Yukawa\;terms} \: +\:
\lambda \widehat{S} \widehat{H}_u \widehat{H}_d \: + \:
\frac{\kappa}{3}\ \widehat{S}^3\,,
\end{equation}
where \lam\ and \kap\ are dimensionless Yukawa couplings. Upon spontaneous
symmetry breaking, the superfield $\widehat{S}$ develops a vacuum
expectation value (VEV), $s\equiv\langle \widehat{S}\rangle$,
generating an effective $\mu$-term, $\mueff=\lam s$. The soft
SUSY-breaking terms in the scalar Higgs sector are then given by
\begin{eqnarray}
V_{\rm{soft}}=m_{\hu}^2|\hu|^2+m_{\hd}^2|\hd|^2+m_{S}^2|S|^2+\left(\lam\alam
  S \hu \hd+\frac{1}{3}\kap\akap
  S^3+\textrm{h.c.}\right)\,.\label{eq:soft}
\end{eqnarray}

Due to the presence of the additional singlet field, the NMSSM contains several new
parameters besides the 150 or so parameters of the MSSM. However,
assuming the sfermion
mass matrices and the scalar trilinear coupling matrices to be diagonal
reduces the parameter space of the model considerably. One can further
impose universality conditions on the dimensionful parameters at the
GUT scale, leading to the so-called
Constrained NMSSM (CNMSSM). Thus all the scalar soft SUSY-breaking
masses in the superpotential are unified into a generic mass parameter \mzero, the
gaugino masses into \mhalf, and all the trilinear couplings, including
\alam$^*$ and \akap$^*$ (defined at the GUT scale), into \azero.
Then, given that the correct $Z$ boson mass, $m_Z$, is
known, \mzero, \mhalf, \azero,
the coupling \lam, taken as an input at the SUSY-breaking scale, \msusy, and the sign
of \mueff\ constitute the only free parameters of the CNMSSM.

As noted in\cite{Kowalska:2012gs}, the fully constrained NMSSM
struggles to achieve the correct mass for the assumed SM-like Higgs
boson, particularly in the presence of other important experimental
constraints. Furthermore, the parameters governing the
mass of the singlet-like pseudoscalar Higgs boson, which will be discussed below,
are not input parameters themselves, but are calculated at the electroweak (EW)
scale starting from the four GUT-scale parameters. In order to avoid these issues
the unification conditions noted above need to be relaxed.
In a partially unconstrained version of the model
the soft masses of the Higgs fields, $m_{H_u}$, $m_{H_d}$ and $m_S$,
are disunified from \mzero\ and taken as free
parameters at the GUT scale. Through the minimisation
conditions of the Higgs potential these three soft masses can then be
traded at the EW scale for the parameters \kap, \mueff\ and
\tanb. Similarly, the soft trilinear coupling
parameters \alam$^*$ and \akap$^*$, though still input at the GUT
scale, are disunified from \azero. The model is thus defined in terms of the following nine continuous input parameters:
\begin{center}
\mzero, \mhalf, \azero, \tanb, \lam, \kap, \mueff, \alam$^*$, \akap$^*$,
\end{center}
where $\tanb \equiv v_u/v_d$, with $v_u$ being the VEV of the $u$-type
Higgs doublet and $v_d$ that of the $d$-type one. This version of the
model serves as a good approximation of the most general
EW-scale NMSSM as far as the phenomenology of the Higgs sector is concerned.
In this way, one can minimise the number of free parameters in a physically
motivated way instead of imposing any ad-hoc conditions, as would be
needed for the general NMSSM. We, therefore, adopt this model to
analyse the phenomenology of the light pseudoscalar here. We refer to
it as the CNMSSM-NUHM, where NUHM stands for
non-universal (soft) Higgs masses, in the following.

\subsection{\label{a1mass} Mass of {$A_1$}}

The presence of an extra singlet Higgs field in the NMSSM results in a total of five
neutral Higgs mass eigenstates, scalars $H_{1,2,3}$ and pseudoscalars
$A_{1,2}$, and a charged pair $H^\pm$, after rotating away the Goldstone bosons.
The tree-level mass of $A_1$ can be given by the approximate
expression
\begin{equation}
\label{eq:ma1}
m_{A_1}^2 \simeq \lam(\alam+4\kap s)\frac{v^2 \sin 2\beta}{2s}-3\kappa
sA_\kappa - \frac{M^4_{P,12}}{M^2_{P,11}}
\end{equation}
where $v \equiv \sqrt{v_u^2 + v_d^2} \simeq 174$\gev\ and all the
parameters are defined at \msusy. $M^2_{P,12}$ and $M^2_{P,11}$ in the
above equation correspond to the off-diagonal and the
doublet-like diagonal elements, respectively, of the symmetric $2\times
2$ pseudoscalar mass
matrix. Note that since $m_{A_1}^2$ is proportional to $\lam\sin 2\beta$,
the effect of an increase in \lam\ with fixed
$\tan\beta$ is analogous to that of a decrease in $\tan\beta$
with fixed \lam. Thus, in the following, whenever the dependence of
$m^2_{A_1}$ on \lam\ is analysed, the inverse dependence on $\tan\beta$ is implicit.
 Assuming negligible singlet-doublet mixing, one can
 ignore the third term
on the right hand side of eq.~(\ref{eq:ma1}) and rewrite it as

\begin{equation}
\label{eq:ma2}
m_{A_1}^2 \simeq \frac{\alam}{2s}  v^2 \lam\sin 2\beta + \kap (2  v^2 \lam
\sin 2\beta - 3s\akap)\,.
\end{equation}

Then, for given signs and absolute values of \alam\ and \akap, the
mass of $A_1$ depends on the sizes of \lam\ and \kap\ (which are both
taken to be positive here along with \mueff, and hence $s$). This
leads to four possible scenarios, as explained below. \\

\noindent {\bf $\akap<0$}.  The second term on the right hand side
  of eq.~(\ref{eq:ma2}) is positive. In this case one has the following.
\begin{itemize}
\item $\alam >0$ leads to a positive first term also. $m^2_{A_1}$ increases with increasing \lam\ and/or \kap.
\item $\alam <0$ gives a negative first term. Increasing \lam\
increases the size of this term as well as of the positive
second term. Thus, in order to avoid an overall negative $m_{A_1}^2$,
\kap\ ought to be large enough so that the
second term dominates over the negative first term.
$m_{A_1}^2$ then increases further with increasing \kap, while  the size of
\lam\ is much less significant.
\end{itemize}

\noindent {\bf $\akap>0$}. In this case one has the following.
\begin{itemize}
\item $|2\lam\ v^2 \sin 2\beta| > |3s\akap|$ implies a positive second
 term. The dependence of $m_{A_1}^2$ on \alam, \lam\ and \kap\ is similar to the
 $\akap<0$ case above.

\item $|2\lam\ v^2 \sin 2\beta| < |3s\akap|$ results in a negative second term.
For $\alam > 0$ the first term is positive and can dominate over the
second term for small enough \kap. Decreasing \kap\ thus increases
$m_A^2$, almost irrespectively of the size of \lam.
$\alam < 0$ is not allowed as it results in $m_A^2 < 0$ owing to
a negative first term also.
\end{itemize}

\subsection{\label{detect} Production of {$A_1$}}

At the LHC, $A_1$ can either be produced directly in the
conventional Higgs boson production modes or, alternatively, through decays of the
other Higgs bosons of the NMSSM. We study in detail both of these
possibilities. The $A_1$ thus produced can then decay via a long list
of available channels, out of which only the final states with $b$ and $\tau$ pairs are of numerical
relevance. \\

\noindent {\bf Direct production}. ~As noted earlier, the $gg \rightarrow bbA_1$ channel is the preferred
direct mode of producing $A_1$ at the LHC, owing to the
possibility of a considerably enhanced $b\bar{b}A_1$ coupling compared to
the $ggA_1$ effective coupling in the NMSSM. We refer the reader
to\cite{Munir:2013wka} for details of the parameter
configurations that can yield such an enhancement. Here we shall discuss
the detectability of a light, $\lesssim 150$\gev, $A_1$ produced in this mode
at the LHC. \\

\noindent {\bf Indirect production}. ~We refer to the processes covered by this production category generically
as $H'' \rightarrow A_1 A_1/Z$, and consider only the gluon fusion
(GF) channel for the production of the parent Higgs boson, $H''$, at the
LHC. There are two further distinct
possibilities regarding $H''$ for a given model point. It can either be
$H_{\rm SM}$, in which case its mass measurement from the LHC serves as an additional kinematical
handle. Alternatively, the parent Higgs boson can be one
of the other two CP-even states, which we denote collectively by
$H'$. This kind of processes are particularly important for
$m_{A_1} > m_{H_{\rm SM}}/2$, although removing
the condition on the final states to have a combined invariant mass
close to 125\gev\ reduces the experimental sensitivity by  a factor of $2$ to $3$.

For $m_{A_1} < m_{H''} - m_Z$ both $H''\to A_1Z$ and $H''\to A_1A_1$
processes can be accessible. For the $A_1Z$ channel we only take the $Z\rightarrow \ell^+\ell^-$
decay into account, where $\ell^+\ell^-$  ($2\ell$) stands for $\mu^+\mu^-$ and
$e^+e^-$ combined (due to the ease of their detection we do not
separate these two final states). This is
because the experimental sensitivity for these states is much better
than that for the $\tau^+\tau^-$ and $q\bar q$ pairs.
We ignore the $H''\rightarrow A_1Z^*$ decay, i.e., with the $Z$ boson
off mass shell. The reason is that the width of the
$H''\rightarrow A_1Z$ decay process is already very
small when it is allowed kinematically and becomes almost negligible
for an off-shell $Z$. Thus, for
$m_{A_1} > m_{H''} - m_Z$ only the $H''\to A_1A$ decay channel can be
exploited (as long as $m_{A_1} < m_{H''}/2$). For this channel we consider
the $b\bar b b\bar b$ (4$b$), $b\bar b\tau^+\tau^-$ (2$b$2$\tau$) and
$\tau^+\tau^- \tau^+\tau^-$ (4$\tau$) final state combinations.

\section{\label{method} Methodology}

We performed several scans of the NMSSM parameter space to search for
regions yielding $m_{A_1} \lesssim 150$\gev, using the nested sampling package
MultiNest-v2.18\cite{Feroz:2008xx}. We used the public package
\nmssmtools -v4.2.1\cite{NMSSMTools} for computing the SUSY
mass spectrum and BRs of the Higgs bosons for each model point.
In our scans, we also required either $H_1$ or $H_2$ to have a mass
near 125\gev\ and SM-like signal rates.\footnote{This is a very loose
  requirement imposed to keep the scans away from excluded regions 
of the parameter space. The exact experimental constraints are imposed 
later and are much more restrictive.} The signal rate, $R_X$, for a given decay
channel $X$, is defined as 
 \begin{equation}
\label{eq:rggh}
R_X \equiv  \frac{\sigma(gg\rightarrow H_i)\times {\rm BR}(H_i\rightarrow
  X)}{\sigma(gg\rightarrow h_{\rm SM})\times {\rm BR}(h_{\rm SM} \rightarrow X)}\,,
\end{equation}
where ${h_{\rm SM}}$ denotes the SM Higgs boson with a mass equal to that
of $H_i$, the NMSSM Higgs boson under consideration.\footnote{$h_{\rm
    SM}$ should not be confused with $H_{\rm SM}$,
which is the assumed SM-like Higgs boson in the NMSSM.} Since GF is
by far the dominant Higgs boson production mode at the LHC, 
$R_X$ serves as a good approximation for the inclusive theoretical
counterpart of the experimentally measured signal strength, $\mu_X $, defined  as
\begin{equation}
\mu_X =  \frac{\sigma(pp\rightarrow H_i \rightarrow
  X)}{\sigma(pp\rightarrow h_{\rm SM} \rightarrow X)}\,.
\end{equation}

The program \nmssmtools\ provides the values of $R_X$ for the dominant
decay channels of each NMSSM Higgs boson as an output for a given
model point. It is calculated in terms of the reduced couplings of a 
NMSSM Higgs boson to various particle pairs, i.e., couplings
normalised to the corresponding ones of a SM Higgs boson.
However, since the $WW$ and $ZZ$ decays of each $H_i$ in 
the NMSSM depend on the same $VVH_i$ reduced coupling, \nmssmtools\ 
computes a unique value of the signal rate, $R_{VV}$, for both these channels. 
But note that, at the LHC, the $\gamma\gamma$ and $ZZ$ 
decay channels remain the only ones so far where the observed
significance of the signal exceeds the expected one, with
the latest measurements of the signal strengths in these channels being
\begin{equation}
\label{eq:CMS-mu}\mu_{\gamma \gamma} = 1.13 \pm 0.24\,,~~~\mu_{ZZ} = 1.0 \pm 0.29\,,
\end{equation}
according to the CMS analyses\cite{CMS-PAS-HIG-14-009}, and
\begin{equation}
\label{eq:ATLAS-mu}
\mu_{\gamma \gamma} = 1.57^{+0.33}_{-0.28}\,, ~~~
\mu_{ZZ} = 1.44^{+0.40}_{-0.35}\,,
\end{equation}
in the ATLAS analyses\cite{ATLAS-CONF-2014-009}. 
We therefore identify the calculated $R_{VV}$ with the
signal rate for the $ZZ$ channel and ignore the experimental result
for the $WW$ channel. 

In our final analysis of the $A_1$ production and decay channels at
the LHC, for $H_{SM}$ to be consistent with the ATLAS Higgs boson
data, its $R_{\gamma\gamma/ZZ}$ is required to lie within the range given in
eqs.~(\ref{eq:ATLAS-mu}). Similarly, for consistency with the CMS 
data, $R_{\gamma\gamma/ZZ}$
for $H_{\rm SM}$ should satisfy eqs.~(\ref{eq:CMS-mu}). Note that
these two requirements are imposed separately, since the
measurements from the two experiments are not mutually very consistent
 in all cases and can also be expected to fluctuate somewhat at
the next LHC run. It is for this reason that among the `good points' 
from our scans we will 
retain also those that do not comply with these robust requirements. 
All these good points are, however, required to be consistent with the
LEP and LHC exclusion limits, applicable on the other, non-SM-like, 
Higgs bosons of the model and tested using 
HiggsBounds-v4.1.3\cite{Bechtle:2008jh,Bechtle:2011sb,Bechtle:2013gu,Bechtle:2013wla}.

The output files of \nmssmtools\ contain a SLHA Block as input for
the HiggsBounds package, which is essentially composed of the squared 
reduced couplings of all $H_i$'s. Here an ambiguity arises due to the
fact that HiggsBounds uses these reduced couplings for calculating the
cross section ratios (similarly to eq.~(\ref{eq:rggh}) and its
equivalents for other Higgs boson production modes also) for imposing the exclusion
limits on the $H_i$'s. The $ggH_i$ reduced couplings are assumed to take care of also
the EW corrections\cite{Djouadi:1994ge,Degrassi:2004mx,Actis:2008ug},
since HiggsBounds effectively obtains the approximate partonic
cross section $\sigma(gg\rightarrow H_i)$ by multiplying $ggH_i$ with the SM 
counterpart, $\sigma(gg\rightarrow h_{\rm SM})$, evaluated
including the EW corrections\cite{Dittmaier:2011ti}. However, these 
corrections are not available in the NMSSM and hence are not
implemented in \nmssmtools, which only includes the next-to-leading 
order (NLO) QCD contributions to the $ggH_i$ reduced coupling. 
Since the EW corrections reach only up to 8\% with respect to the
leading order (LO) cross section and the non-SM-like Higgs bosons 
generally have very poor signal rates, we ignore this slight ambiguity in the
implementation of the collider exclusion limits on them.

During our scans, while the experimental constraints specific to the
Higgs sector implemented in \nmssmtools\ were retained,
all the other phenomenological constraints were ignored. Instead, the
$b$-physics observables were calculated explicitly for each point using the
dedicated package \superiso -v3.3\cite{superiso}, for improved theoretical
precision. The good points to be shown in our results
are also consistent with the following $b$-physics
constraints, based on\cite{Beringer:1900zz}:
\begin{itemize}
\item $\brbsmumu = (3.2\pm1.35\pm 0.32) \times 10^{-9}$, where the last
quantity implies a 10\% theoretical error in the numerical evaluation,
\item $\brbutaunu = (1.66\pm 0.66 \pm 0.38) \times 10^{-4}$,
\item $\brbxsgamma = (3.43\pm 0.22 \pm0.21) \times
10^{-4}$.
\end{itemize}
In addition, these points also satisfy the Dark Matter relic density constraint,
 $\Omega_\chi h^2 < 0.131$, assuming a $+10$\% theoretical error around the
central value of 0.119 measured by the PLANCK
telescope\cite{Ade:2013zuv}. For calculating the theoretical model prediction for
$\Omega_\chi h^2$ we used the public package \micromegas
-v2.4.5\cite{micromegas}. Finally, the program \nmssmtools\ by default
discards a SUSY point if the perturbativity constraint is violated or if the global
minimum is not a physical one. 

\subsection{Event analysis}

For each process of interest we carry out a dedicated
signal-to-background analysis based on MC event
generation for proton-proton collisions at 14\,TeV centre-of-mass
energy. The backgrounds for all the processes are computed
with MadGraph5\_aMC$@$NLO\cite{Alwall:2014hca} using its default factorisation
and renormalisation scale and the CTEQ6L1\cite{cteq6} library for parton
distribution functions. The signal cross sections for the
GF and $bbh_{\rm SM}$ production modes are computed using
SusHi-v1.1.1\cite{Harlander:2012pb} for the SM Higgs
boson. These cross sections are then rescaled using the reduced $bbA_1$ and $ggH_i$
couplings in the NMSSM and multiplied by the relevant BRs of $H_i$, all of
which are obtained from \nmssmtools. 

These GF cross section obtained from SusHi contains the QCD contributions up to
the next-to-next-to-leading order
(NNLO)\cite{Djouadi:1991tka,Dawson:1990zj,
Spira:1995rr,Harlander:2002wh,Anastasiou:2002yz,
Ravindran:2003um,Marzani:2008az,Harlander:2009mq,
Pak:2009dg}. However, we make sure to turn off the EW
corrections since these are not included in the $ggH_i$ reduced
couplings obtained from \nmssmtools, which is evaluated taking only
the NLO QCD effects into account, as pointed out earlier. 
In the case of the $bbh_{\rm SM}$ production, note that SusHi
calculates only the inclusive $b\bar b \to h_{\rm SM}$ cross section at
the NNLO in QCD\cite{Maltoni:2003pn,Harlander:2003ai} and not the
fully exclusive $gg\to b\bar b h_{\rm SM}$ process. However, these
two cross sections have been found to show reasonable numerical
agreement when including higher order
corrections\cite{Dittmaier:2003ej,Liu:2012qu}. To our advantage,
the $b\bar b \to h_{\rm SM}$ cross section can be conveniently rescaled
 with the $bbA_1$ reduced coupling obtained from \nmssmtools. 
Such a rescaling would not be straightforward in the case of the 
$gg\to b\bar b h_{\rm SM}$ process at NLO, due to diagrams with top loops.\footnote{Note also that
 \nmssmtools\ calculates the $H'' \rightarrow A_1A_1$ decay widths at
 the LO. The higher order corrections to the triple Higgs couplings
 have been calculated only recently
  in\cite{Nhung:2013lpa}. In our analysis we ignore such corrections,
  since their impact on our overall conclusions is expected to be insignificant.}

Both the signal and the background for each process are hadronised and
fragmented using Pythia 8.180\cite{Sjostrand:2007gs} interfaced with
 FastJet-v3.0.6\cite{Cacciari:2011ma} for jet clustering and jet substructure analysis.
The parton-level acceptance cuts used in the event generation in MadGraph are
\begin{itemize}
\item $|\eta|<$ 2.5 for all final state objects,
\item $p_T>15$\gev\ far all final state objects,
\item $\Delta R \equiv \sqrt{(\Delta\eta)^2+(\Delta\phi)^2}>0.2$ for all $b$-quark pairs,
\item $\Delta R >0.4$ for all other pairs of final state objects,
\end{itemize}
where $p_T$, $\eta$, $\phi$ are the transverse momentum,
pseudorapidity and azimuthal angle, respectively. The looser
cut on $\Delta R$ for $b$-quark pairs is to allow the use of the jet
substructure method in the later analyses.
In table~\ref{tab:Acceptance} we show the background cross sections
after the acceptance cuts have been applied at the parton level.
Since the signal events are generated directly in Pythia, the above
acceptance cuts are implemented not at the parton level but
at the hadron level after jet clustering. For this reason we do not
provide any exact efficiency values for the signal processes.
However, in general, from around 0.05\%\ for $b\bar b Z$  and
10\% for $b\bar b b\bar b$, to almost 100\%\ of the signal
events meet the acceptance requirements at the parton level,
depending on the masses involved.

\begin{table}[tbp]
\begin{center}
\begin{tabular}{|l|c|}\hline
Channel  &  Background cross section   \\\hline
$b\bar b b\bar b$  &  3400 pb    \\
$b\bar b \tau^+\tau^-$  &  3.1 pb    \\
$\tau^+\tau^-\tau^+\tau^-$   &  5.4 fb   \\
$b\bar b Z$    &   126 pb     \\
$\tau^+\tau^- Z$  &  0.46 pb    \\\hline
\end{tabular}
\caption{Parton-level cross sections for the various backgrounds after
  applying acceptance cuts, as given by MadGraph.}\label{tab:Acceptance}
\end{center}
\end{table}%

We do not perform any proper $b$- or $\tau$-tagging but use the
MC truth to identify jets that stem from $B$-mesons and
$\tau$-leptons, respectively. The single tagging efficiency,
50\%\ for both $b$- and $\tau$-jets, is then taken into
account by a scaling of the cross sections, assuming no knowledge of
the charge-sign of the jet. Note also that, although
we do not include any detector effects or smearing, the
hadronisation and consequent jet clustering have a similar effect of
broadening the studied objects. We therefore expect the inclusion of
smearing due to the detector resolution to only marginally change our results.

In case of the $\tau$-jets in the final state, we use only their
visible parts to represent the $\tau$'s. However, we note here that the
use of more sophisticated tools for this purpose (see,
e.g.,\cite{Elagin:2010aw,Gripaios:2012th,Bianchini:2014vza}), may help
improve the $\tau$-reconstruction efficiency further. When there are only two
$\tau$-jets in the final state, one could employ, e.g., the collinear
approximation\cite{Ellis:1987xu} to improve the
mass resolution. However, in our case the $\tau$'s tend to be too
back-to-back for this to work properly.

To identify the $A_1$ bosons decaying to $b\bar b$ pairs over the QCD background,
we employ the jet substructure method
of\cite{Butterworth:2008iy}, used recently in\cite{deLima:2014dta} to
demonstrate the detectability of pair-produced $h_{\rm SM}$ in the
$4b$ final state. In this method, we first cluster all final
state visible particles using the Cambridge-Aachen (CA)
algorithm\cite{Cambridge,Aachen} with $R=1.2$. For
each resulting jet, $j$, that has $p_T > 30$\gev\ and an invariant
mass $>12$\gev, we go back in the
clustering sequence until we find  two subjets, $j_1$ and $j_2$, with
relatively similar respective invariant masses,
$m_{j_1}$ and $m_{j_2}$, and transverse momenta, $p_{Tj_1}$ and
$p_{Tj_2}$. Furthermore, both $m_{j_1}$ and $m_{j_2}$ should be
significantly lower than the invariant mass, $m_j$, of the jet
they would subsequently merge into (i.e., the declustered jet $j$).
Technically this means that the two subjets are required to satisfy
the conditions
\begin{equation}
m_{j_{1,2}}/m_j <0.67~~~ {\rm and} ~~~
  \frac{\min(p_{Tj_1}^2,p_{Tj_2}^2)}{m_j^2}\Delta R^2(j_1,j_2) > 0.09\,.
\end{equation}
%
These two subjets are then taken to be potentially coming from an
$A_1$ and their constituents are reclustered with the CA
algorithm with $R=\max(\min(\Delta R(j_1,j_2)/2,0.3),0.2)$. If the two
hardest of the jets thus obtained are $b$-tagged and the three hardest jets
together have an invariant mass  $>12$ GeV, the combination of these
three subjets is considered a fat jet resulting from the
$A_1 \rightarrow b\bar b$ decay.
The constituents of the fat jet are then removed from the event and
the remaining particles are reclustered using the
antikT\cite{antikT} algorithm with $R=0.4$ in order to find single
$b$- and other jets.

We now have three possible signatures for a decaying $A_1$: one fat
jet, two single $b$-jets and two $\tau$-jets.
For the $H'' \rightarrow A_1A_1$ decay, one then has six possible
combinations to look for. In all cases
we require that the difference between the invariant masses of the two $A_1$
candidates is less than 10\gev. For the special case of $H''=H_{\rm
  SM}$ we additionally require that the combined invariant mass of the
two $A_1$ candidates should be $125\pm 20$ GeV.
For the $H''\rightarrow A_1 Z (\rightarrow  A_1 \ell^+\ell^-)$ process, we
have three possible final states which correspond to three distinct
signatures of the $A_1$. In the special case of $H'' =H_{\rm SM}$ we
require the combined invariant mass of the $Z$ and
the $A_1$ candidate to be $125\pm10$ GeV. The higher precision
required for this process is due to the greater accuracy in the lepton momentum measurements.

From MadGraph 5 we obtain the parton-level events and cross sections for
all possible background processes, including $pp\to b\bar b b\bar b$,
$pp\to b\bar b\tau^+\tau^-$, $pp\to \tau^+\tau^-\tau^+\tau^-$,
$pp\to Z b\bar b$ and $pp\to Z\tau^+\tau^-$. These background events
 are also subjected to the analyses explained above and compared with
 the signal samples. We then calculate the expected cross sections for
 the signal processes which
yield $S/\sqrt{B}>5$ for three benchmark accumulated luminosities,
$\mathcal{L}=30$/fb,\,300/fb and 3000/fb,
 assumed for the LHC. The experimental sensitivity thus obtained for a
 given luminosity is divided by 0.9 ($\simeq {\rm BR}(A_1\to b\bar
 b)$)  for each $b\bar b$ pair and by 0.1 ($\simeq {\rm BR} (A_1\to
 \tau^+\tau^-)$) for each $\tau^+\tau^-$ pair in the
final state.\footnote{In principle, these BRs may vary slightly from one
  point to another in the model parameter space, but the assumed average
  values serve as very good approximations of the true
  values. Note also that for the $b\bar b \tau^+\tau^-$ final state there
  is an additional factor of 2 coming from the combinatorics of the event.} This way the final expected sensitivities can be compared
directly with the calculated values of $\sigma(gg\to
H'')\times {\rm BR}(H''\to A_1A_1)$
and $\sigma(gg\to H'')\times {\rm BR}(H''\to A_1Z)$ for each
model point. Henceforth we will refer to such $\sigma \times BR$ for a given
process as its total cross section.

\begin{figure}[tbp]
\centering
\subfloat[]{%
\label{fig:-a}%
\includegraphics*[width=7.3cm]{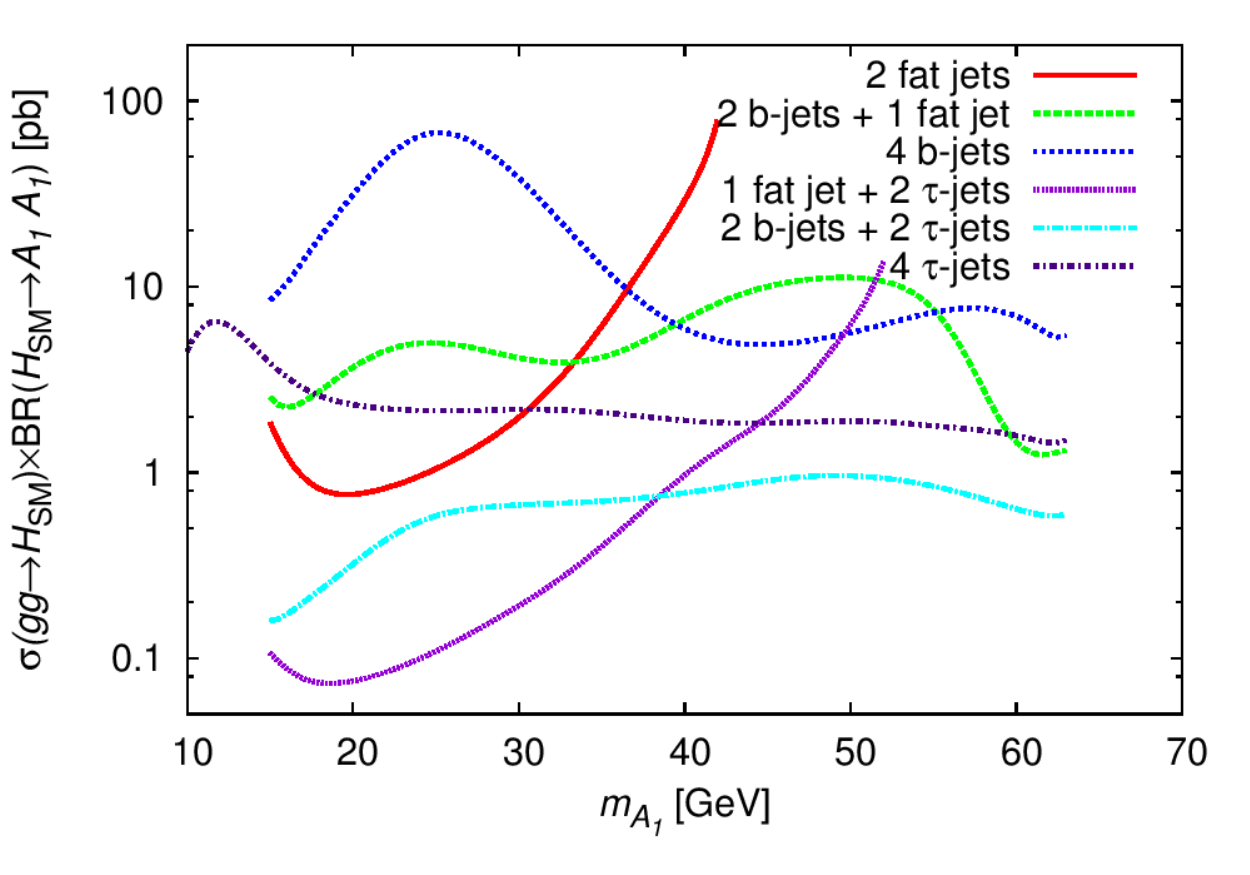}
}%
\hspace{0.5cm}%
\subfloat[]{%
\label{fig:-b}%
\includegraphics*[width=7.3cm]{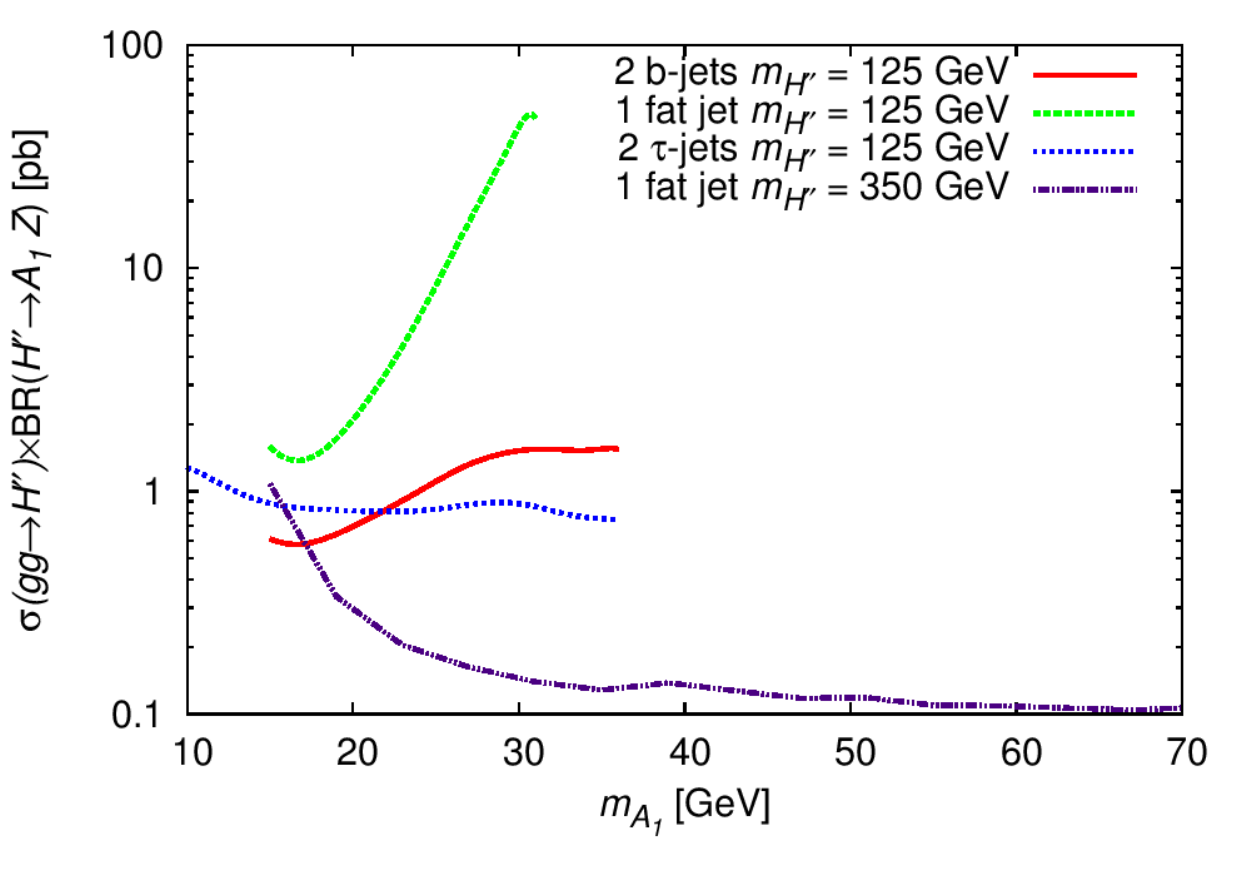}
}%
\caption{Expected experimental sensitivities as functions of $m_{A_1}$,
  in various possible final state combinations for (a)
the $gg \to H_{\rm SM}\rightarrow A_1A_1$ process and (b) the $gg \to H''\rightarrow A_1
Z$ process.}
\label{fig:Sens}
\end{figure}

The sensitivities obtained in the various final states are shown
in figure~\ref{fig:Sens}(a) for $H_{\rm SM}$ ($m_{H_{\rm SM}}= 125\gev$) for 3000/fb
assumed integrated luminosity. In figure~\ref{fig:Sens}(b), in addition
to the sensitivities
with $H'' = H_{\rm SM}$, we also show the sensitivity
for the fat jet analysis with $m_{H'}=350$\gev\ to illustrate the
importance of the $H''$ mass for the $ZA_1$ channel. Note that in case of
$H''=H_{\rm SM}$, the limited phase space makes the $b$-jets very soft,
rendering the fat jet analysis rather insensitive to this process. Increasing the
parent Higgs boson mass results in a much larger phase space which
allows one to reach much higher sensitivities through the fat jet analysis,
despite losing the benefit of constraining the combined invariant
mass to be near 125\gev.

In general, we see in figure~\ref{fig:Sens}(b) that the fat jet analysis
can be very effective when the $A_1$ is much lighter than the $H''$,
but gets worse as $m_{A_1}$ increases and, in fact, soon becomes
relatively useless (the corresponding curves are thus cut off at
the mass above which the analysis becomes ineffective).
This is due to the fact that the fat jet analysis assumes boosted $b$-quark pairs.
We do not perform any $b$-quark pair analyses below
$m_{A_1}< 15$\gev\ because the limited phase space means
Pythia struggles with the hadronisation of the $b$ quarks. One can
also see (especially in the curve with $m_{H'}=350$\gev) that, if
the ${A_1}$ mass becomes too small compared to the $H''$ mass, the
sensitivity diminishes due to the $b$-jets becoming too collinear
to be separable even with jet substructure methods.
In the upper end, the cut-offs (for sensitivity curves other than those
relying on the fat jet analysis) are determined by the
kinematical upper limit for the given channel,
i.e., $m_{A_1}\approx 62.5$\gev\ for $H_{\rm SM}\to A_1A_1$
and $m_{A_1}\approx 35$\gev\ for $H_{\rm SM}\to A_1Z$.

In order to keep the figures readable, in the following sections we
will only show the curves corresponding to the analyses with the
highest sensitivities for a given
channel. To get an impression of the sensitivity obtainable via
other analyses one may compare with figure~\ref{fig:Sens}. In that case
one should bear two things in mind. First, not constraining the combined invariant mass of
the final states to be close to 125\gev\ (when $H''\neq H_{\rm SM}$)
reduces the sensitivity by a factor of $2$ to $3$, as stated earlier, although it leaves the relative
sensitivities more or less intact. And second, increasing $m_{H''}$ means
that the fat jet analysis works better, especially at higher $m_{A_1}$.

Finally, we do not include triggers in our
analyses. Especially for the $4b$ final state, this is a complicated
issue that we leave for the experimentalists to address. Here we just
point out that the triggers used in\cite{CMS-PAS-HIG-14-013} require
too high a $p_T$ of the $b$-jets to be applicable here. For the final
states including $\tau$'s, one could trigger on the missing transverse
energy, which could be one more argument in their favour.

\section{\label{results} Results and discussion}

In this section we present the numerical results of our analysis.
As noted in the Introduction, in the NMSSM $H_1$
and $H_2$ can both have masses around $125$\gev\ and
SM-like properties, and can thus alternatively play the role of the
$H_{\rm SM}$. An SM-like $H_1$ with mass around 125\gev\ can be obtained over wide
regions of the CNMSSM-NUHM parameter space, defined in section~\ref{model}.
However, the additional requirement of $m_{A_1} \lesssim 150$\gev\
somewhat constrains these regions. An SM-like $H_2$, in contrast,
requires much more specific parameter combinations. In the
following we discuss the LHC phenomenology of a light $A_1$ in the two
cases, with $H_{\rm SM} = H_1$ and with $H_{\rm SM} = H_2$, separately.
In principle, in the NMSSM the signal peak
observed at the LHC near 125\gev\ can also be interpreted as a
superposition of both $H_1$ and $H_2$ that are nearly degenerate in 
mass\cite{Gunion:2012gc,Gunion:2012he}. In such a case the signal rates
due to these two Higgs bosons should be combined for testing against the
LHC Higgs boson data, depending on the assumed
experimental mass resolution. However, in our scans 
such points occurred very rarely, covering a very insignificant portion of the parameter
space. We, therefore, ignore such a possibility in our analysis,
assuming the signal rates to be due only to one Higgs boson, the
$H_{\rm SM}$ in a given case.

\subsection{SM-like $H_1$}

\begin{figure}[tbp]
\centering
\subfloat[]{%
\label{fig:-a}%
\includegraphics*[width=7.7cm]{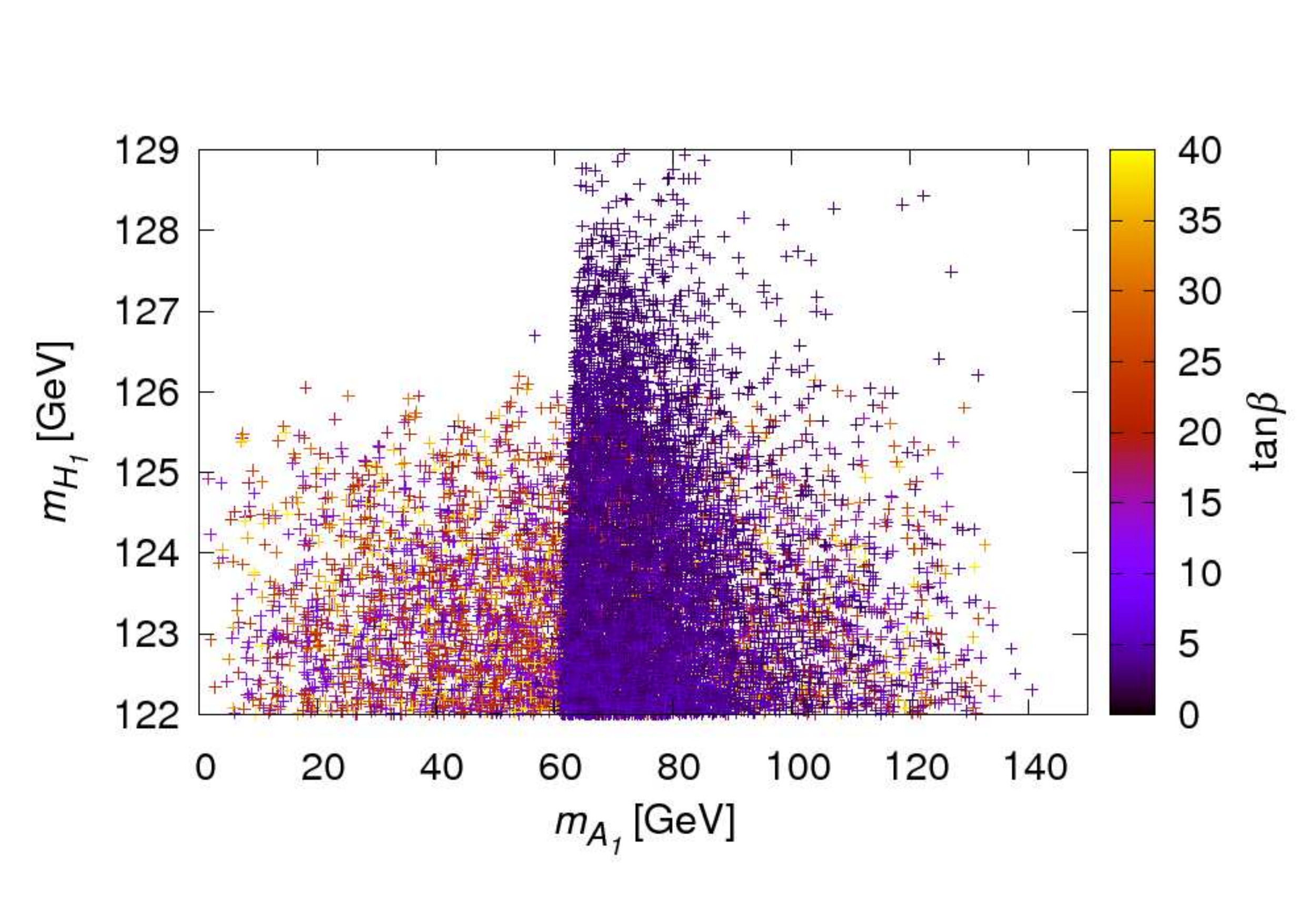}
}%
\subfloat[]{%
\label{fig:-b}%
\includegraphics*[width=7.7cm]{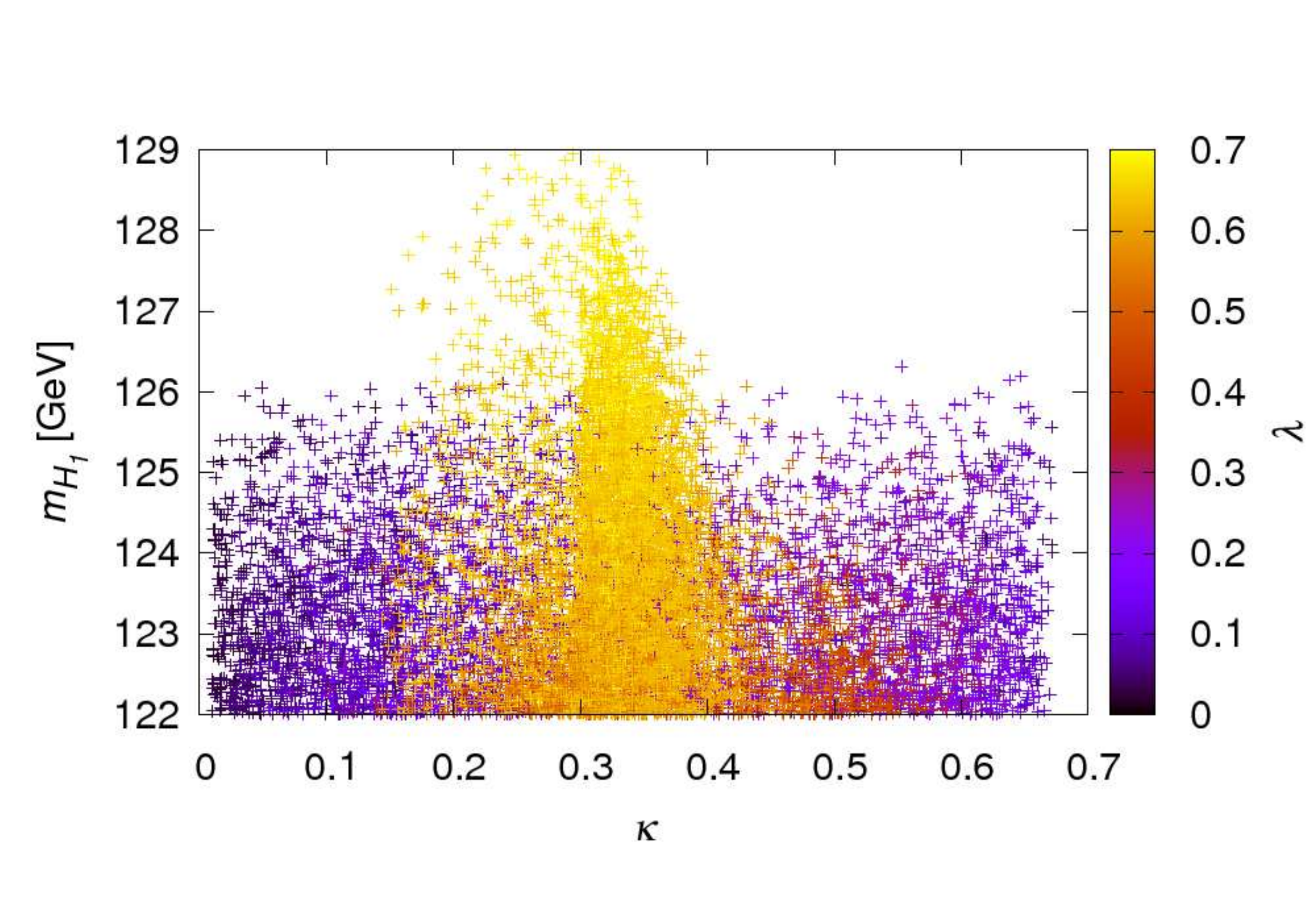}
}%
\caption[]{Case with $H_{\rm SM} = H_1$: (a) Mass of $H_1$ vs. that of
  $A_1$, with the heat map
  showing the distribution of \tanb; (b) $m_{H_1}$ as a
  function of the parameter \kap, with the heat map showing the distribution of the coupling \lam.}
\label{fig:h1params1}
\end{figure}

In figure~\ref{fig:h1params1}(a) we show the
distribution of the mass of $H_1$ against that of $A_1$ for the points
obtained in our scans assuming  $H_{\rm SM} = H_1$. The ranges of the
parameters scanned, are given in table~\ref{tab:params}.
Note that in this and in the following figures we
will allow the SM-like Higgs mass to be in the range
$122\gev \leq m_{H_{\rm SM}} \leq 129\gev$.
This is to take into account the experimental
as well possibly large theoretical uncertainties
in the model prediction of $m_{H_{\rm SM}}$,
given the Higgs boson mass measurement of 125\gev\ at the LHC.
The heat map in the figure corresponds to the parameter \tanb. One can
see a particularly dense population of points for $\tanb \sim
1-6$ in the figure, with the mass of $H_1$ reaching
comparatively larger values than elsewhere. However, $m_{A_1}$ for
such points almost never falls below $\sim 60$\gev.
In figure~\ref{fig:h1params1}(b) we show $m_{H_1}$ as a function of the
coupling \kap, with the heat map corresponding to the coupling
\lam. Again there is a clear strip of points with $\lam \gtrsim
0.6$ (and $\kap \sim 0.15 - 0.5$) for which
$m_{H_1}$ can be as high as 129\gev. These points are the ones lying
also in the small \tanb\ strip in figure~\ref{fig:h1params1}(a). The
rest of the points, corresponding to smaller \lam\ and larger \tanb,
can barely yield $m_{H_1}$ in excess of 126\gev. The reason for the behaviour of
$m_{H_1}$ observed in these figures is explained in the following.

\begin{figure}[tbp]
\centering
\subfloat[]{%
\label{fig:-a}%
\includegraphics*[width=7.7cm]{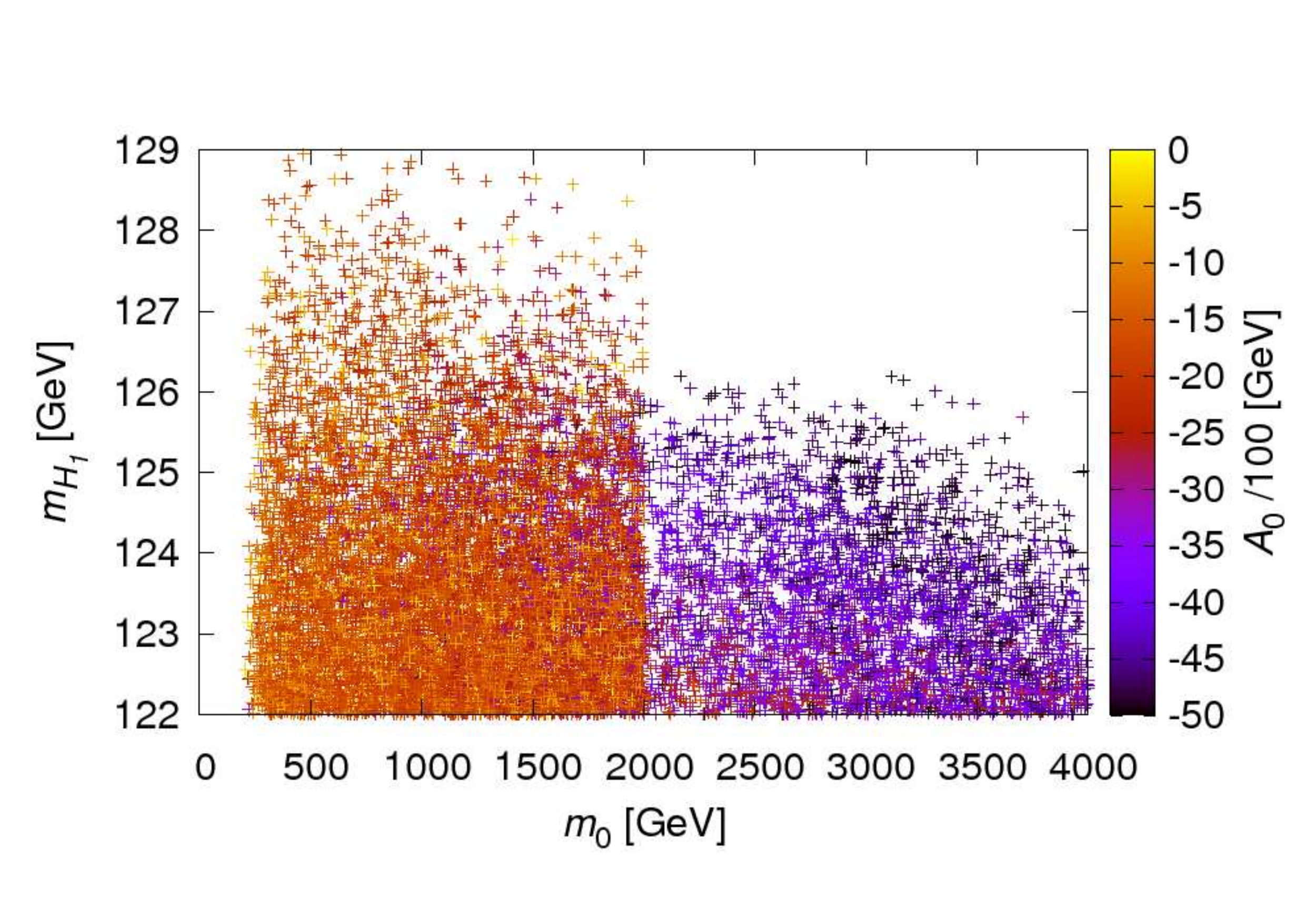}
}%
\subfloat[]{%
\label{fig:-b}%
\includegraphics*[width=7.7cm]{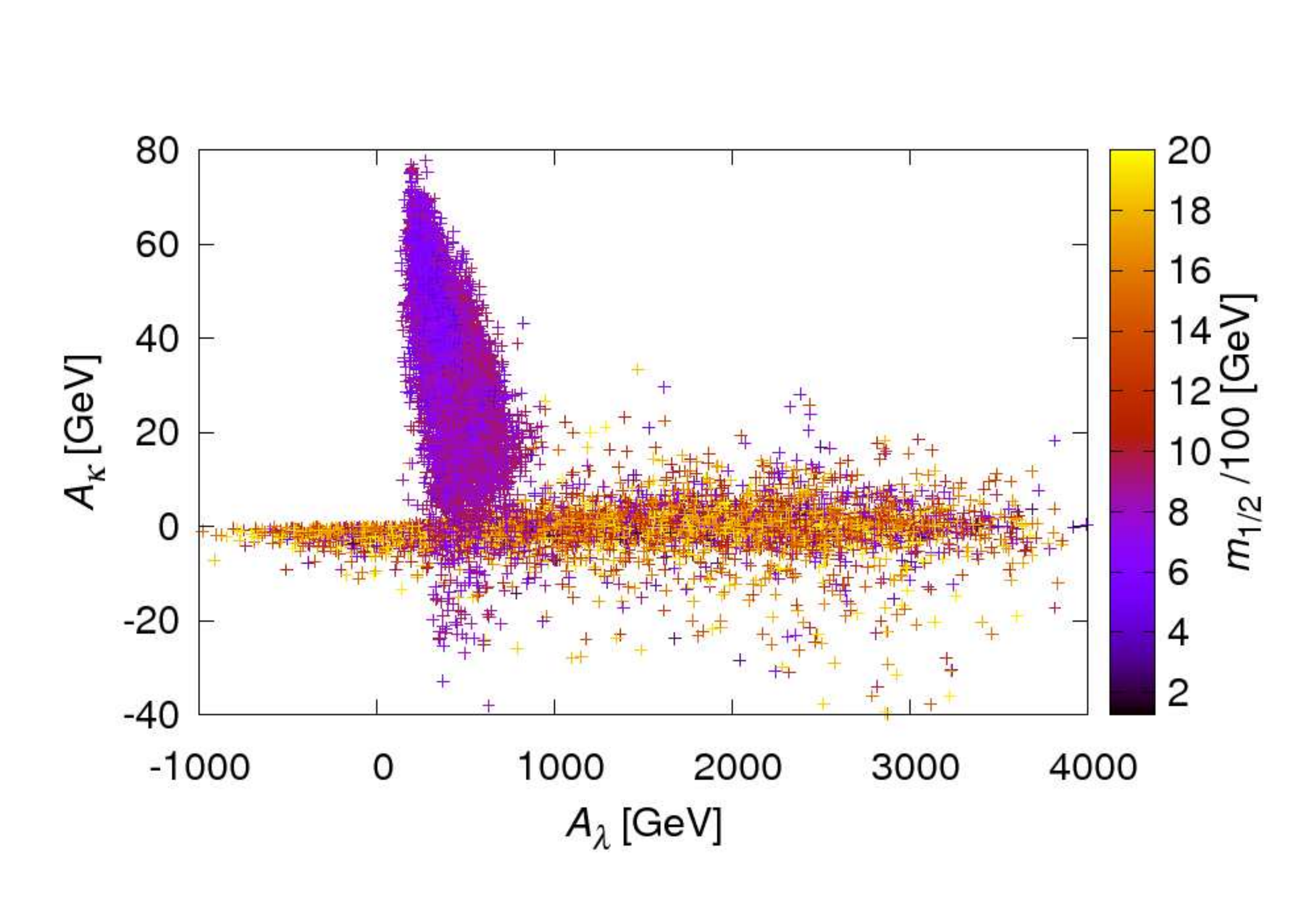}
}%
\caption[]{Case with $H_{\rm SM} = H_1$: (a) Mass of $H_1$ as a
  function of the parameter \mzero, with the heat map showing the
  distribution of the soft trilinear coupling \azero; (b) the parameter \alam\
  vs. the parameter \akap\, with the heat map showing the
  distribution of the soft mass \mhalf.}
\label{fig:h1params2}
\end{figure}

The tree-level mass of $H_{\rm SM}$ in the NMSSM is given by\cite{Ellwanger:2009dp}
\begin{equation}
\label{eq:mhlim}
m_{H_{\rm SM}}^2 \simeq m_Z^2\cos^2 2\beta
 + \lam^2 v^2 \sin^2 2\beta -
  \frac{\lam^2 v^2}{\kap^2}\left[\lam -\sin2\beta \left(\kap +
  \frac{\alam} {2s}\right)\right]^2 \,.
\end{equation}
For small \lam\ and large \tanb\ the negative third term on the right hand side of the
above equation can dominate over the positive second term, leading to
a reduction in the tree-level $m_{H_{\rm SM}}^2$. The mass
of $H_{\rm SM}$ can then reach values as
high as 125\gev\ or so only through large radiative corrections
from the stop sector, thus requiring the so-called maximal mixing
scenario and thereby invoking fine-tuning concerns. Alternatively
the correct mass of $H_{\rm SM}$, particularly when it is the $H_1$, can be obtained in a more
natural way through large \lam\ and small \tanb, implying a
reduced dependence on the radiative corrections. This enhances
the tree-level contribution to $m_{H_{\rm SM}}^2$ from the
positive second term and nullifies that from the negative third term.
We shall refer to this parameter configuration as the `naturalness
limit' in the following.\footnote{In our original scan with wide
parameter ranges, very few points belonging in the naturalness limit were obtained.
We therefore performed a dedicated scan of the
reduced parameter ranges corresponding to the naturalness limit, also
given in table\,\ref{tab:params}, and
merged the points from the two scans. This results in a relatively
high density of such points in strips with sharp edges seen in the figures.}
In figure~\ref{fig:h1params2}(a) we see that, for the points in
the strip corresponding to the naturalness limit, larger $m_{H_1}$ is
obtained without requiring either \mzero\, shown on the horizontal axis,
or \azero, shown by the heat map, to be too large. For points outside this
strip, the desired mass of $H_1$ can only be achieved with large \mzero\ or, in
particular, very large $-|\azero|$.

\begin{table}[tbp]
\begin{center}
\begin{tabular}{|c|c|c|}
\hline
 Parameter & Extended range & Reduced range  \\
\hline
\mzero\,(GeV) & 200 -- 4000 & 200 -- 2000 \\
\mhalf\,(GeV)  & 100 -- 2000 & 100 -- 1000 \\
\azero\,(GeV)  &  $-5000$ -- 0 & $-3000$ -- 0 \\
\mueff\,(GeV) & 100 -- 2000 & 100 -- 200 \\
\tanb & 1 -- 40 & 1 -- 6 \\
\lam  & 0.01 -- 0.7 & 0.4 -- 0.7  \\
\kap  & 0.01 -- 0.7 & 0.01 -- 0.7 \\
\alam$^*$\,(GeV) & $-2000$ -- 2000 & $-500$ -- 500 \\
\akap$^*$\,(GeV) & $-2000$ -- 2000 & $-500$ -- 500 \\
\hline
\end{tabular}
\caption{Ranges of the CNMSSM-NUHM input parameters scanned for
  obtaining a $\sim
  125$\gev\ $H_{\rm SM}$ along with a light $A_1$.}
\label{tab:params}
\end{center}
\end{table}

In figure~\ref{fig:h1params2}(b) we show the distributions of the
remaining three parameters,\footnote{Unlike \azero,
  \alam\ and \akap\ shown in the figures are the
 ones calculated at \msusy\ by \nmssmtools\ from \alam$^*$ and
\akap$^*$, respectively, input at the GUT scale.} \alam, \akap\
and \mhalf, on the horizontal and vertical axes and by the heat map, respectively.
Again one sees a dense strip of points with relatively smaller values of
\mhalf\, which corresponds to the naturalness limit. These points are also
restricted to comparatively smaller values of $+|\alam|$ but extend to a wider range
of \akap\ than the points outside the strip. The smallness and
positivity of \alam\ is warranted for further enhancing the
tree-level $m_{H_{\rm SM}}^2$ by reducing the size of the term
in the square brackets in eq.~(\ref{eq:mhlim}). The sign and size
of \akap\ is then mainly defined by the condition of smallness of
$m_{A_1}$. According to eq.~(\ref{eq:ma2}), for $\alam,\akap >
0$, increasing \akap\ reduces $m^2_{A_1}$ as long as $|2\lam\ v^2 \sin
2\beta| < |3s\akap|$. Moreover, since \lam\ and \tanb\ ought to be
large in order to maximise $m_{H_1}$, $m_{A_1}$ can only be reduced
further by reducing \kap.

However, going back to the figure~\ref{fig:h1params1}(a), it is evident that
$A_1$ never has a mass below $m_{H_1}/2$ in the naturalness limit, as
$m_{A_1}$ cuts off sharply at that value.
The reason for this is that the tree-level $H_iA_1A_1$ couplings are
proportional to $\lambda^2$ (see eq.~(A.17)
in\cite{Ellwanger:2009dp}). Thus, for very large \lam, required to
enhance the mass of $H_1$,\footnote{Note
  that \lam\ is bounded from above by the perturbativity
condition $\lam^2 + \kap^2 \lesssim 0.5$\cite{Miller:2003ay}.} when $m_{A_1}<m_{H_1}/2$,
the $H_1\to A_1A_1$ decay can be highly dominant over the other $H_1$ decay
channels. In fact the BR($H_1\to A_1A_1$) reaches unity for
such points, resulting in highly suppressed BRs for all other channels.
Consequently the signal rates of $H_1$ in the $\gamma\gamma$ and $ZZ$ channels
drop to unacceptably low values and the corresponding points are
rejected during our scans. This can in fact be used to constrain the decay of $H_{\rm SM}$
to lighter scalars or pseudoscalars\cite{Cao:2013gba}.

\subsubsection{$bbA_1$ production}

\begin{figure}[tbp]
\begin{center}
\includegraphics[width=7.3cm]{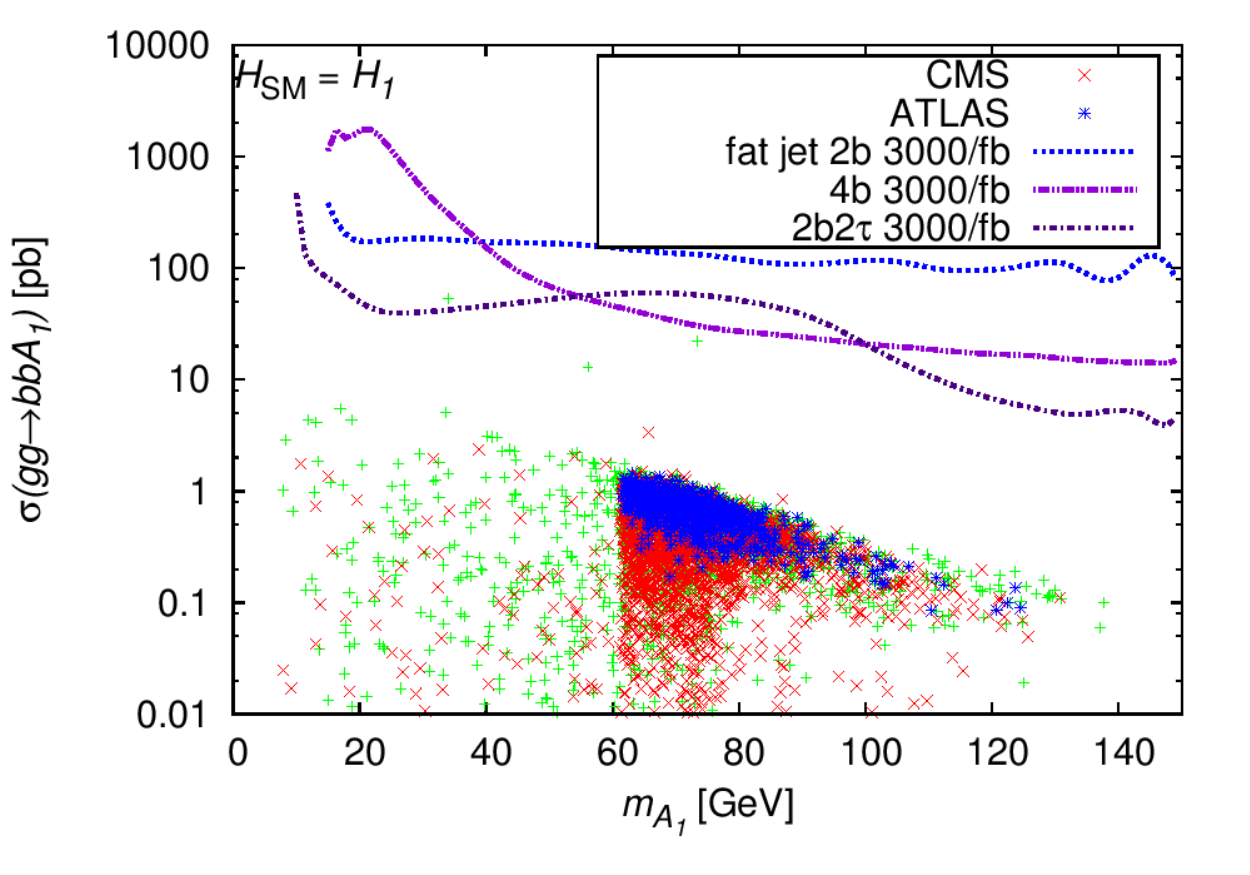}
\caption{$b\bar b A_1$ production cross section as a function of
  $m_{A_1}$ in the case with $H_{\rm SM} = H_1$. The red
and the blue points satisfy the CMS and the ATLAS constraints on
$R_{\gamma\gamma/ZZ}$, respectively. In green are the
unflitered points satisfying neither of these constraints. Also shown
are the sensitivity curves for various final state combinations. See text for details.}
\label{fig:bba1h1}
\end{center}
\end{figure}

We first analyse the $bbA_1$ production
process for this case. In figure~\ref{fig:bba1h1} we show the total cross section,
$\sigma(gg\rightarrow bbA_1)$, as a
function of $m_{A_1}$ for all the good points from our scan. The
red and blue points in the figure are the ones for which the calculated $R_X$
lies within the range of $\mu_X$ measured by CMS
and ATLAS, respectively. The
green points are then the `unfiltered' ones for which neither of these two
constraints are satisfied. We shall retain this color convention for all the
figures showing cross sections henceforth. One notices in the figure
that none of the points with $m_{A_1}$ below $\sim 60$\gev\ satisfies
the ATLAS constraints on $R_X$, while some of these comply with the CMS
ones. This is due to the fact that all these points belong
outside the naturalness limit, and hence possess very (MS)SM like
characteristics. According to eq.~(\ref{eq:ATLAS-mu}), the ATLAS
measurements of $\mu_{\gamma\gamma}$ and $\mu_{ZZ}$
are substantially higher than 1. Such high rates in the $\gamma\gamma$ and
$ZZ$ channels can only be obtained in the naturalness limit of the
NMSSM, as noted in\cite{Ellwanger:2011aa}. Hence one sees a large number of
points corresponding to the naturalness limit consistent with the ATLAS constraint in
the figure. Note also that the ATLAS collaboration has recently updated
its measurement of $\mu_{\gamma\gamma}$\cite{Aad:2014eha}, which is
now comparatively closer to the SM prediction. However, no updates on
$\mu_{ZZ}$ for the same data set have been released. This implies that even if
we use the newly released $\mu_{\gamma\gamma}$ value, $R_{ZZ}$ for a
given point still ought to be large to satisfy the older $\mu_{ZZ}$ range
from ATLAS.

Also shown in figure~\ref{fig:bba1h1} are the sensitivity
curves corresponding to the $2b2\tau$ and $4b$ final states for an
expected integrated luminosity of 3000/fb at the LHC.
We see that none of the good points from the scan have a cross section
large enough to be discoverable in any of the considered final state
combinations. This may seem to be in contrast with the results of earlier
studies\cite{Almarashi:2010jm,Almarashi:2011hj}. However, a closer
inspection of these previous results
reveals that for all the model points yielding a potentially detectable $A_1$,
$m_{H_1}$ is always below 120\gev. To further verify this observation,
we performed a test scan of the NMSSM parameter space with an earlier
version (2.3.1) of the \nmssmtools\
package and checked the surviving points with version 4.2.1
as well as with HiggsBounds. We noted that all the points with a large $bbA_1$
cross section obtained in the test scan were indeed ruled out by the latest
data from the Higgs searches at LEP and LHC. In addition to that, the
improvements in the calculation of the Higgs boson masses and reduced couplings
in the newer version may also be responsible for a more accurate
theoretical prediction of the cross section for this process.


\subsubsection{Production via $H_{\rm SM} \rightarrow A_1A_1/Z$}

Among the indirect production channels, we first analyse the
$H_1 \rightarrow A_1 A_1$ and $H_1 \rightarrow A_1Z$ processes.
As noted above in detail, for the points in the
naturalness limit, $m_{A_1}$ never falls below
$m_{H_1}/2$, making both these processes kinematically irrelevant.
This decay is then only possible for points obtained outside
this limit, i.e., for small \lam\ and large \tanb.

\begin{figure}[tbp]
\centering
\subfloat[]{%
\label{fig:-a}%
\includegraphics*[width=7.3cm]{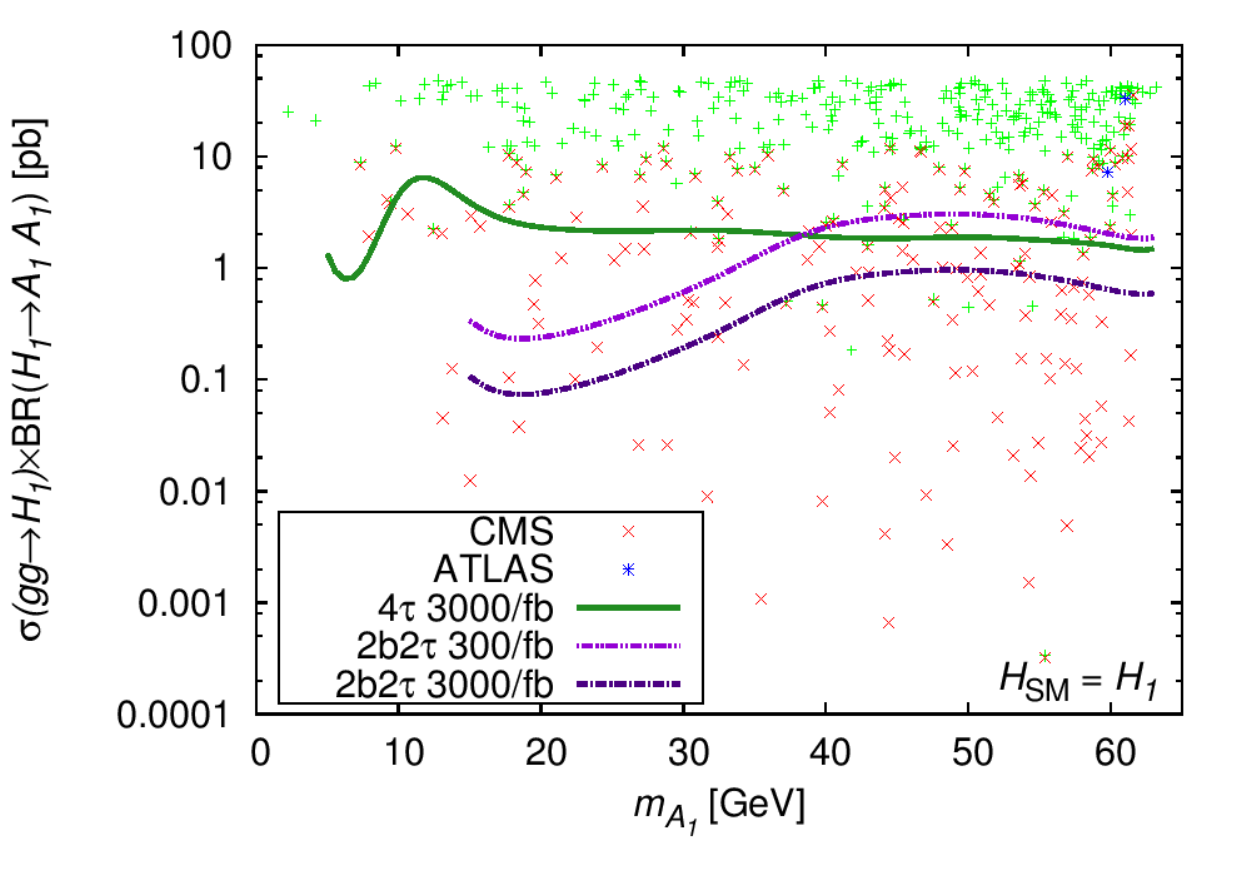}
}%
\hspace{0.5cm}%
\subfloat[]{%
\label{fig:-b}%
\includegraphics*[width=7.3cm]{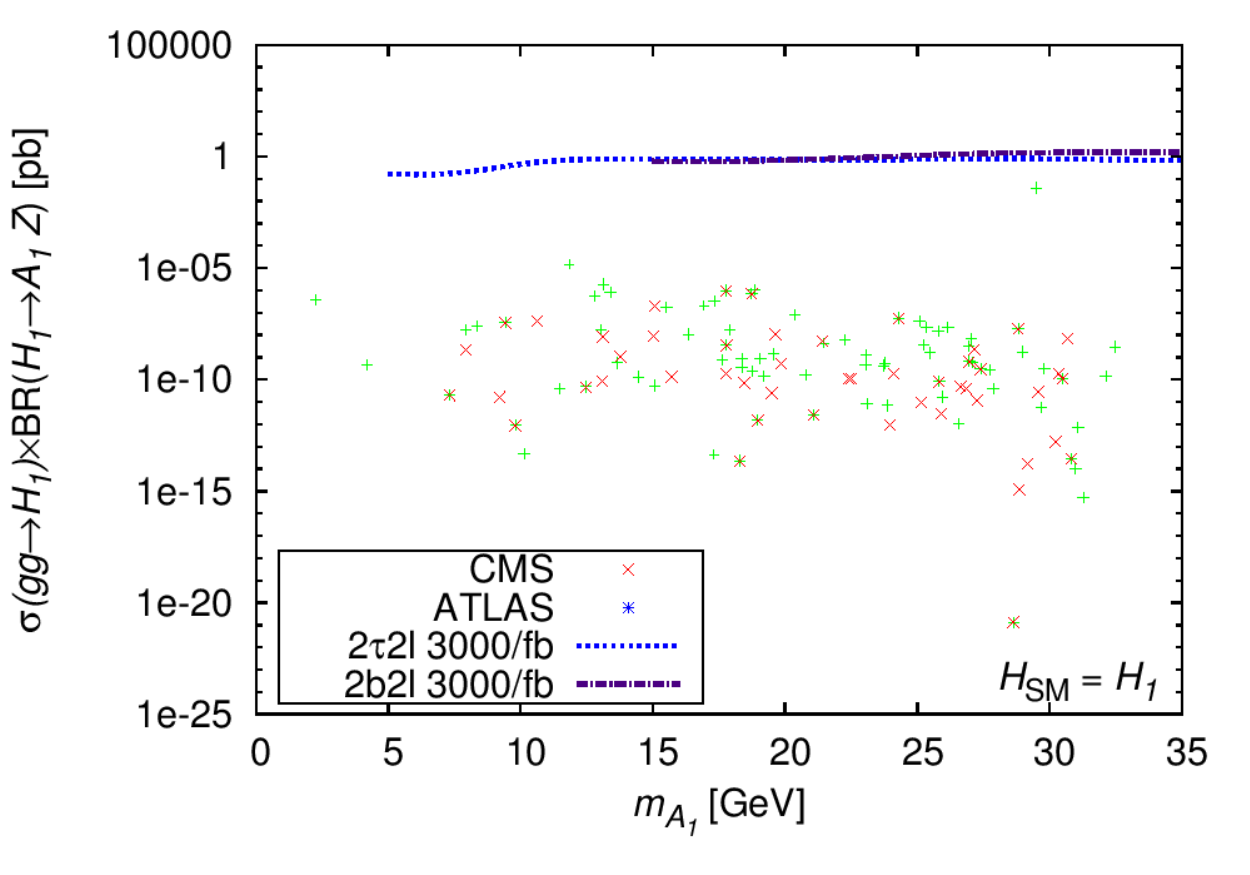}
}
\caption{Total cross sections in the case with $H_{\rm SM} =
 H_1$ for (a) the $gg \to H_{\rm SM}\to  A_1A_1$ process and (b)
 the $gg \to H_{\rm SM}\to A_1Z$ process. The color convention
for the points is the same as in figure
 \ref{fig:bba1h1}. See text for details about the sensitivity curves.}
\label{fig:h1a1zForH1}
\end{figure}

In figure~\ref{fig:h1a1zForH1}(a), we show the total cross section for
the $gg \rightarrow H_1 \rightarrow A_1 A_1$ channel. Also shown in
the figure are the curves corresponding to the sensitivity reaches in
the $2b2\tau$ as well as the $4\tau$
final states. For the $2b2\tau$ curve, we have used a
combination of the fat jet plus two $\tau$-jets and the two single $b$-jets
plus two $\tau$-jets analyses. The fat jet analysis has been used for low masses
while the two single $b$-jets analysis has been used when it becomes
more sensitive. The curve corresponding to the $4\tau$ final state extends
to lower $m_{A_1}$. This is because, for $m_{A_1}<10$ GeV, the $A_1\to b\bar b$
decay becomes kinematically disallowed, resulting in the $4\tau$ being the only applicable channel and causing
BR$(A_1\to \tau^+\tau^-)$ to rise from near 0.1 to around 0.9, thereby
causing a sharp dip in the $4\tau$ sensitivity curve. We note in the figure that while
a lot of unfiltered points should be accessible at the LHC in the
$2b2\tau$ final state at $\mathcal{L}=300$/fb, very few points consistent with the CMS constraint
on $R_X$ lie above the corresponding curve. None of the points
consistent with the corresponding ATLAS constraints appears in this figure,
for reasons noted earlier.

In figure~\ref{fig:h1a1zForH1}(b) we see that the
$A_1Z$ channel will not be accessible at the LHC in any final state
combination even for the maximum integrated luminosity assumed.
All the points where $m_{A_1}$ is low enough for this decay channel to
be relevant lie much below the sensitivity curves shown, which
correspond to the two single $b$-jets plus
$\ell^+\ell^-$ and the two $\tau$-jets plus $\ell^+\ell^-$ final states.
This is due to the fact that the BR$(H_1\to A_1Z)$ is extremely small for such
points. The reason for the smallness of this BR can be understood
by examining the $H_iA_jZ$ couplings, which are proportional to the factors
$S_{i1}P_{j1}-S_{i2}P_{j2}$ (see eq.~(A.17)
in\cite{Ellwanger:2009dp}). Here $S_{i1}$ and $S_{i2}$ are the
terms corresponding to the $H_{dR}$ and $H_{uR}$
weak eigenstates in the scalar mixing matrix, and $P_{j1}$ and $P_{j2}$ are the
terms corresponding to $H_{dI}$ and $H_{uI}$ weak eigenstates in the
pseudoscalar mixing matrix, with the indices $R$ and $I$ referring to
the real and imaginary parts, respectively, of the complex Higgs fields.
Owing to the structure of the scalar and the pseudoscalar mass matrices,
$P_{j1}/P_{j2}=\tanb$ for the singlet-like $A_j$, and for a very
SM-like $H_i=H_{\rm SM}$, $S_{i2}/S_{i1}\approx\tanb$. This causes the factor
$S_{i1}P_{j1}-S_{i2}P_{j2}$ to approach zero, resulting in an almost
vanishing $H_1A_1Z$ coupling.

\subsubsection{Production via $H' \rightarrow A_1 A_1/Z$}

\begin{figure}[tbp]
\centering
\subfloat[]{%
\label{fig:-a}%
\includegraphics*[width=7.3cm]{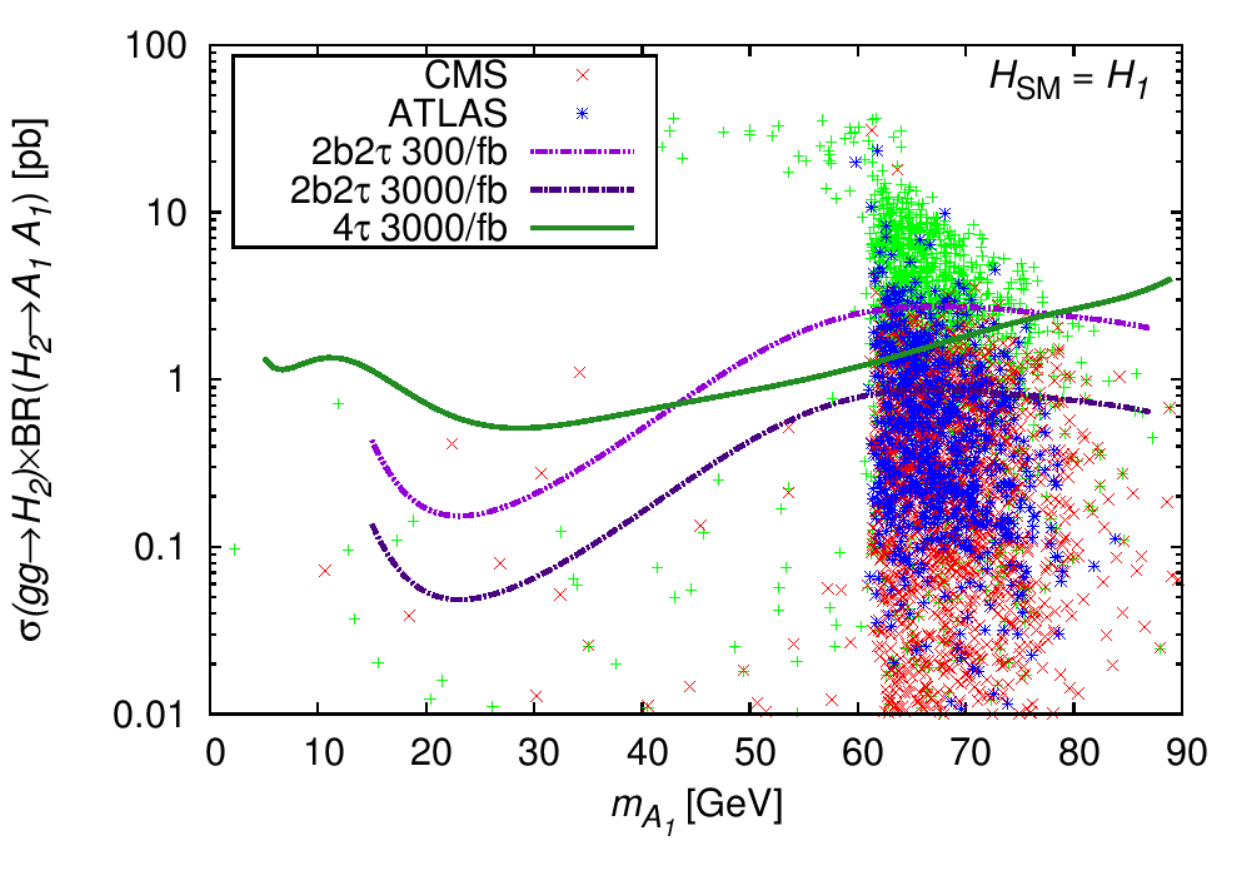}
}%
\hspace{0.5cm}%
\subfloat[]{%
\label{fig:-b}%
\includegraphics*[width=7.3cm]{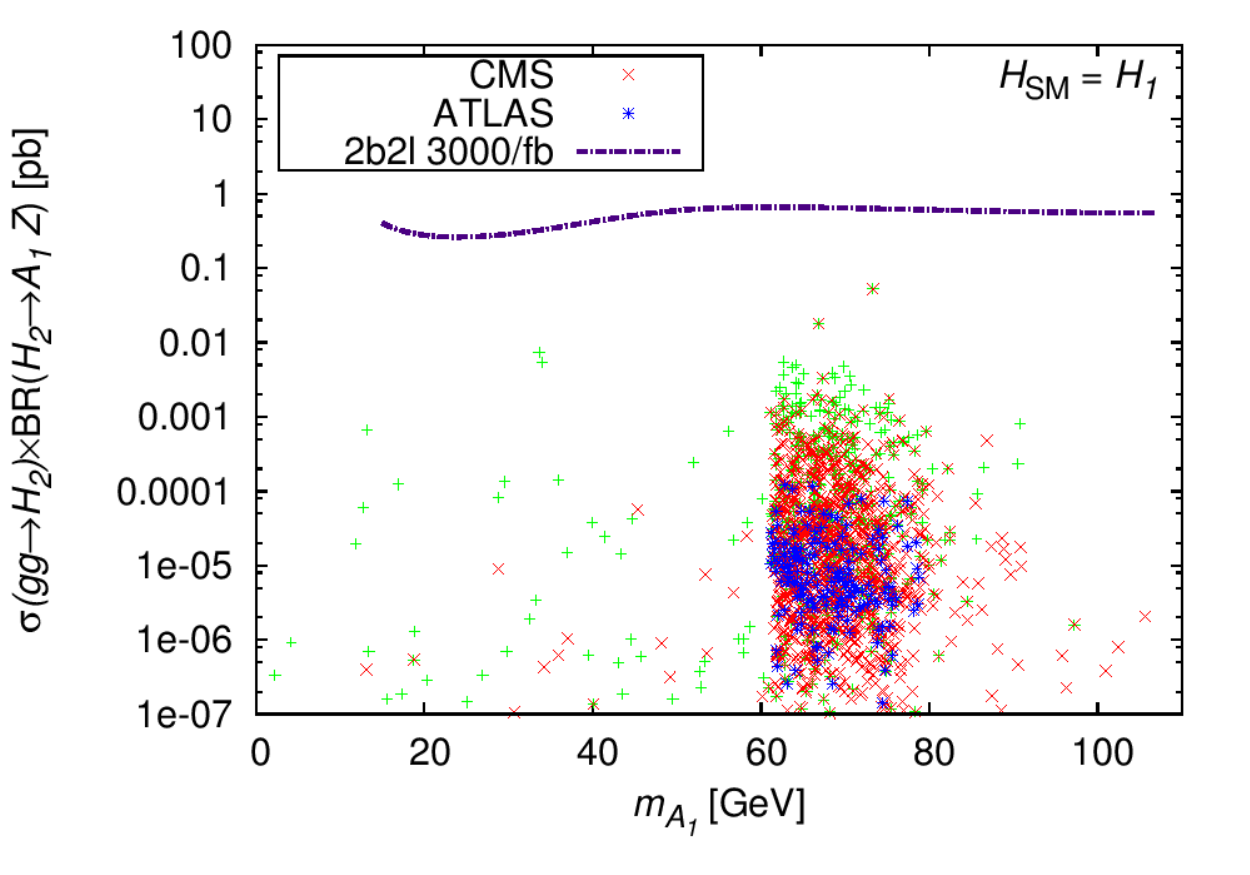}
}
\caption{Total cross sections in the case with $H_{\rm SM} =
 H_1$ for (a) the $gg \to H_2 \to  A_1A_1$ process and (b)
 the $gg \to H_2 \to A_1Z$ process. The color convention
for the points is the same as in figure
 \ref{fig:bba1h1}. See text for details about the sensitivity curves.}
\label{fig:h2a1zForH1}
\end{figure}

We next consider the decays of the other non-SM-like Higgs bosons in the
$A_1A_1$ and $A_1Z$ channels.
In figure~\ref{fig:h2a1zForH1}(a) and (b) we see that the situation for the $H_2
\to A_1A_1/Z$ decays is very similar to that for the $H_1 \to A_1A_1/Z$
decays, respectively. The sensitivity curves shown correspond to the same final state
combinations as those in figure~\ref{fig:h1a1zForH1}(a) and (b). But here the
difference is that the combined invariant mass of the final state
particles is not required to be $\sim 125$\gev. The $H_2$ mass used in
the lines for the $H_2\to A_1A_1$ decay is 175\gev, while for the
$H_2\to A_1 Z$ channel we use 200\gev\ in order to cover the whole
populated region of panel (b). The actual $H_2$ mass for the
points shown ranges from around 130\gev\ to around 300\gev, with a
large population below 150\gev, so for some points the sensitivities shown
might be somewhat overestimated.

We note that, although the prospects for the $H_2 \to A_1A_1$ channel are not great,
they are slightly better than those for the $H_1  \to A_1A_1$ mode
seen above. For the former, a significant number of points satisfying also the ATLAS
constraints on $R_X$ may be probed in the $b\bar b\tau^+\tau^-$ final
state combination, contrary to what was observed for the latter.
The $H_2\to A_1 Z$ channel, on the other hand, will not be accessible
at all in any final state combination even at 3000/fb integrated
luminosity. This is not surprising given the
singlet-like natures of both $A_1$ and $H_2$.

\begin{figure}[tbp]
\centering
\subfloat[]{%
\label{fig:-a}%
\includegraphics*[width=7.3cm]{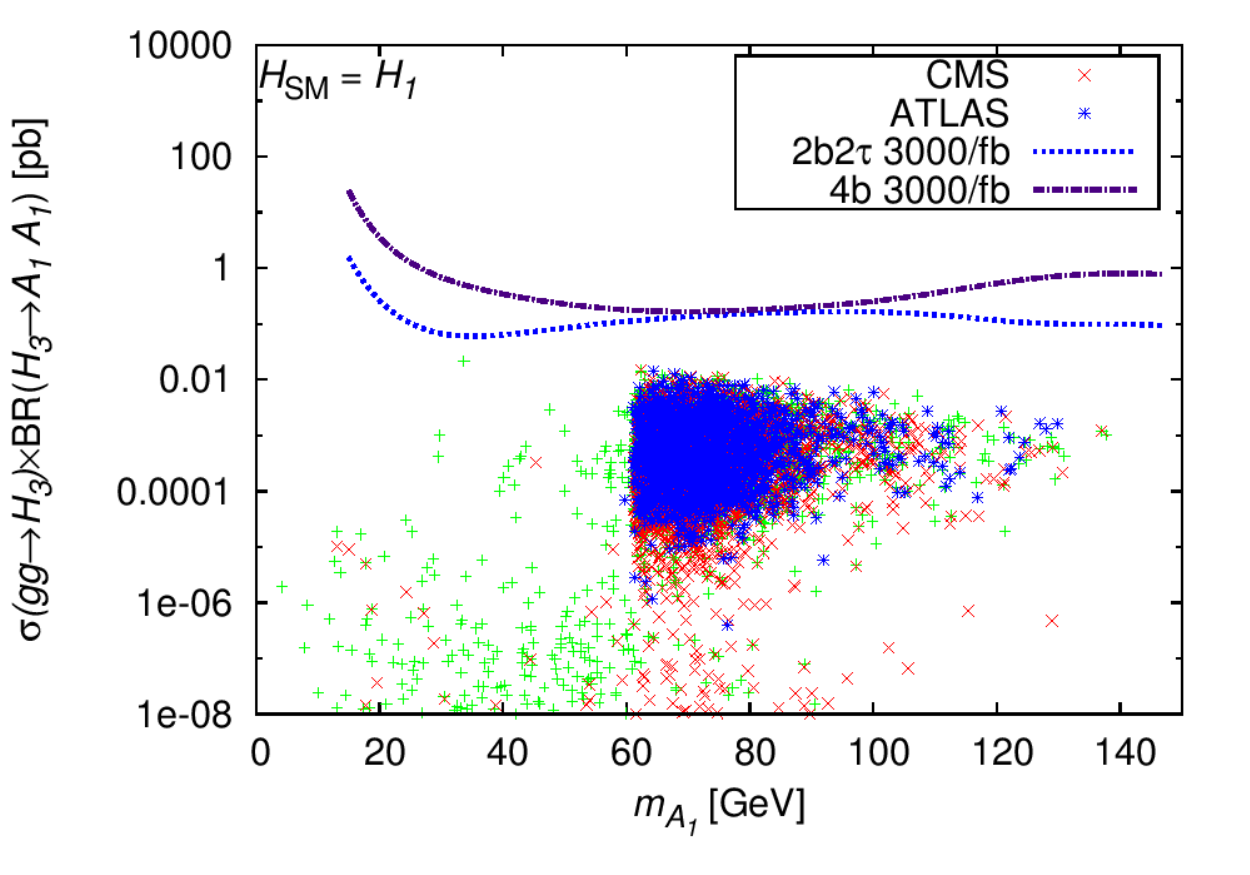}
}%
\hspace{0.5cm}%
\subfloat[]{%
\label{fig:-b}%
\includegraphics*[width=7.3cm]{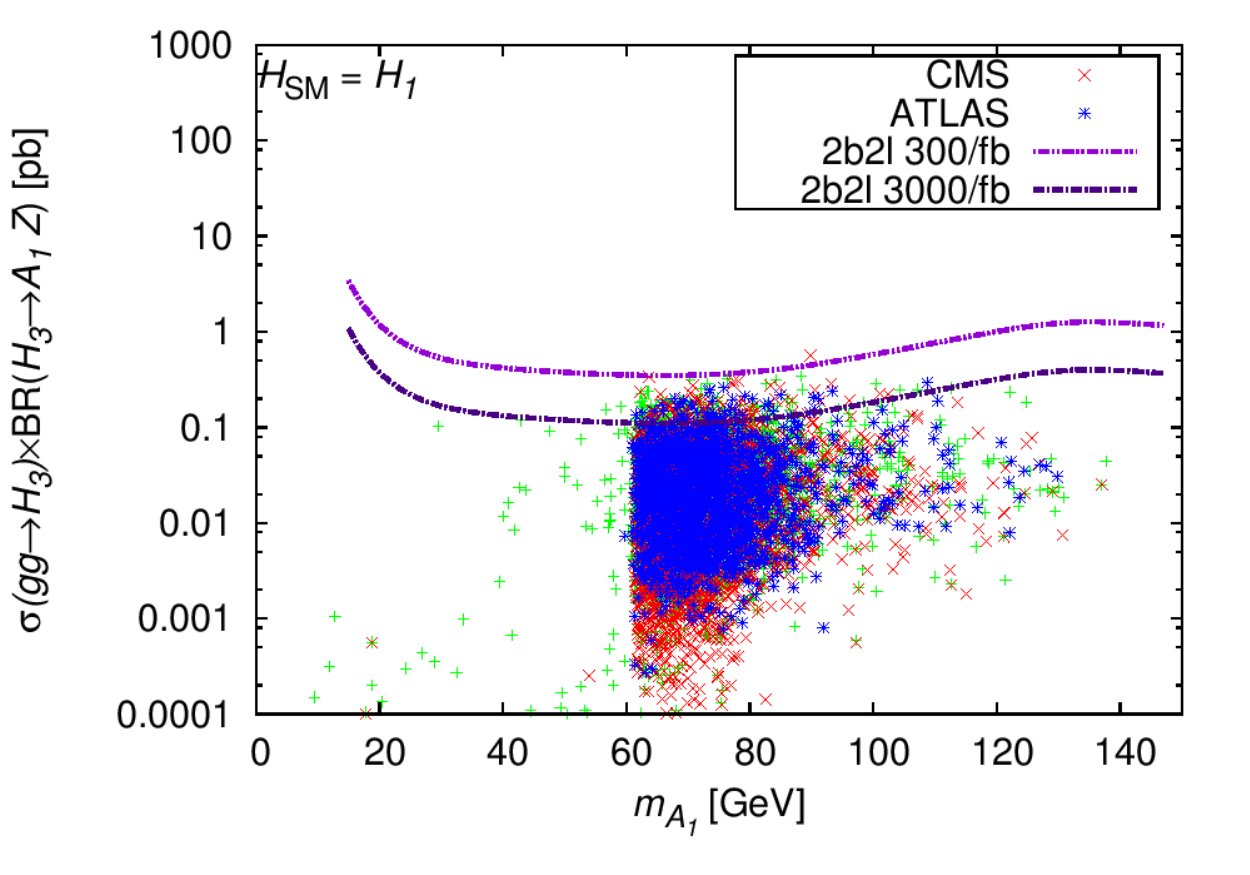}
}
\caption{Total cross sections in the case with $H_{\rm SM} =
 H_1$ for (a) the $gg \to H_3 \to  A_1A_1$ process and (b)
 the $gg \to H_3 \to A_1Z$ process. The color convention
for the points is the same as in figure
 \ref{fig:bba1h1}. See text for details about the sensitivity curves.}
\label{fig:h3a1zForH1}
\end{figure}

In the case of $H_3$ decays the overall prospects are quite different
from those noted for the lighter scalars above. In
figure~\ref{fig:h3a1zForH1}(a) one can see that the $H_3\to A_1A_1$
channel will be inaccessible even for maximum assumed luminosity.
The sensitivity curves shown
correspond to an $H_3$ mass of 350\gev\ and a combination of the fat jet and
the two single $b$-jets analyses, which always shows the highest
possible reach for a given $m_{A_1}$.
The reason for the poor prospects in the $H_3\to A_1A_1$ channel is to
a large extent a consequence of the high mass of $H_3$, partly because
such a mass reduces the production cross section and partly because
a number of other decay channels like $t\bar t$, $W^+W^-$, $ZZ$
and lighter Higgs scalars are available. These other channels are relatively
unsuppressed and hence dominate over the $A_1A_1$ channel.

In figure~\ref{fig:h3a1zForH1}(b) we see that, in contrast, the $H_3\to
A_1Z$ channel shows some promise. A fraction of the points
lies above the $\mathcal{L}=3000$/fb sensitivity curve corresponding to the $2b2\ell$ final state,
with one point lying above the 300/fb curve. Most of
these points satisfy the CMS and/or ATLAS constraints on $R_X$. The
total cross section for this process can reach
detectable levels due to the fact that the $H_3 A_1Z$ coupling is not
subject to the peculiar cancellation noted earlier for the $H_1A_1 Z$ coupling.
Since $A_1$ with mass larger than $\sim 60$\gev\ will
not be accessible in any other production mode, this channel will
be very crucial for its discovery, which may be possible even at
300/fb integrated luminosity at the LHC.

\subsection{SM-like $H_2$}

\begin{figure}[tbp]
\centering
\subfloat[]{%
\label{fig:-a}%
\includegraphics*[width=7.7cm]{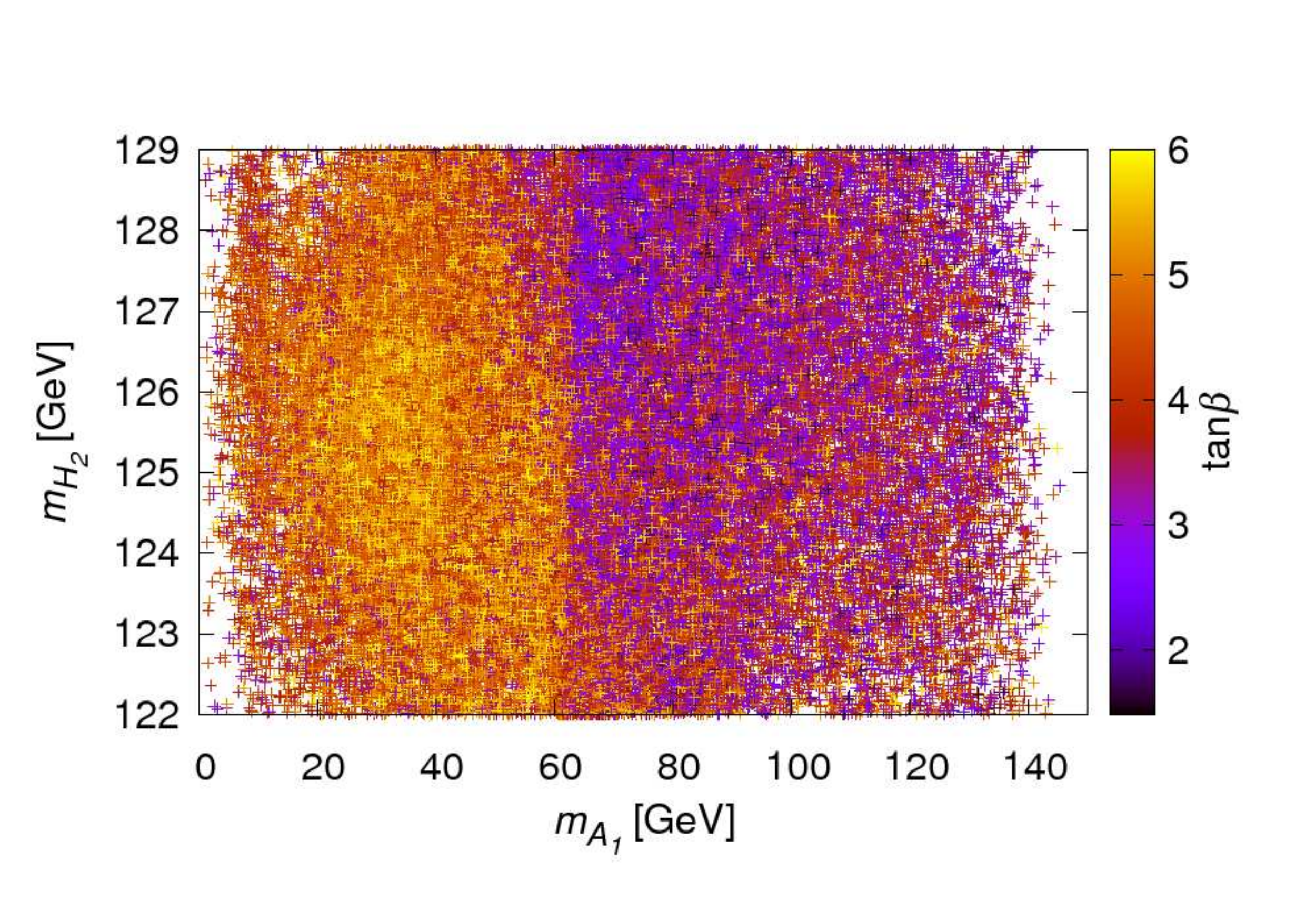}
}%
\subfloat[]{%
\label{fig:-b}%
\includegraphics*[width=7.7cm]{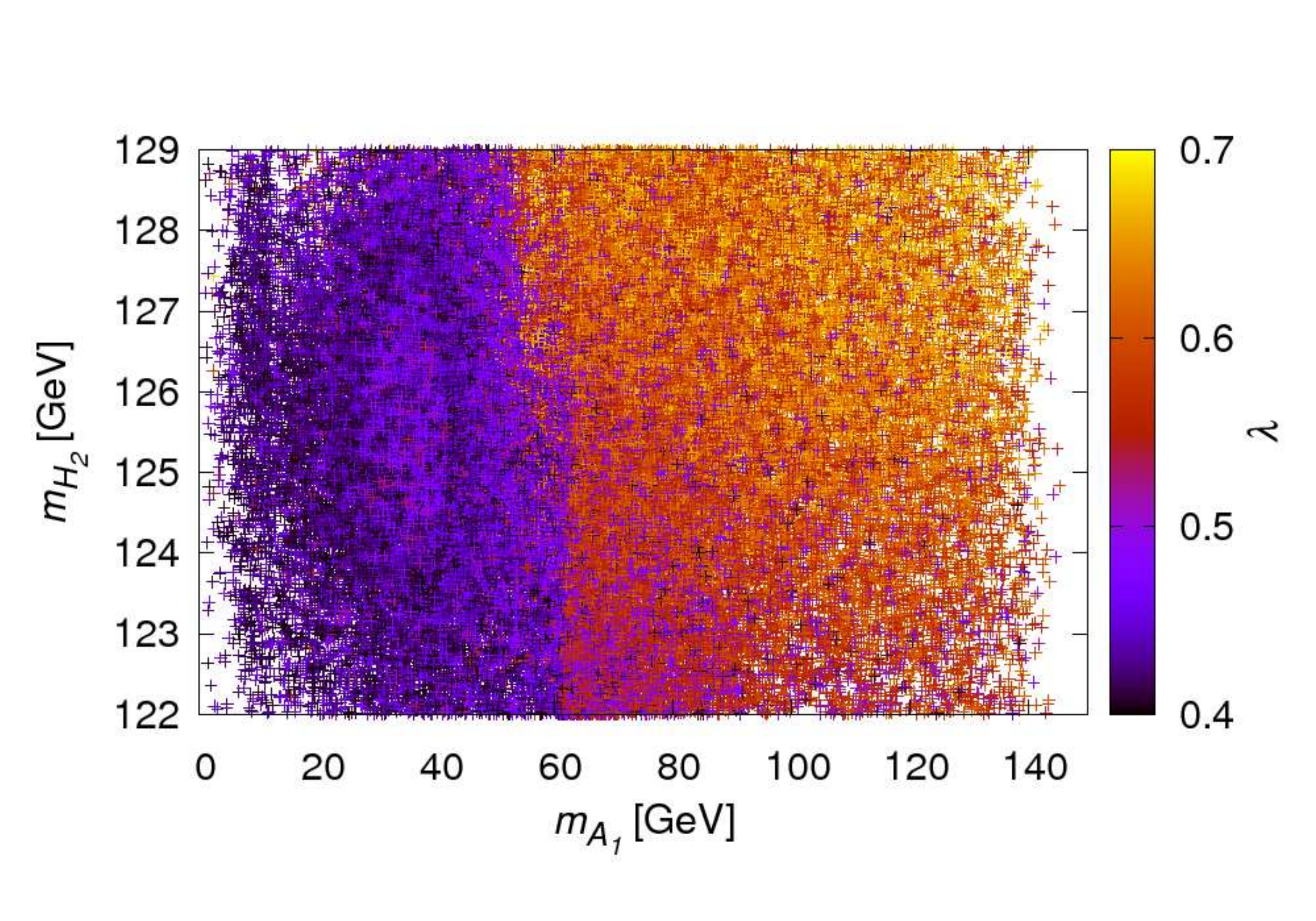}
}%
\caption[]{Mass of $H_2$ vs. that of $A_1$ for the case with $H_{\rm
    SM} = H_2$. The heat map
 shows (a) the distribution of \tanb\ and (b) the distribution of \lam.}
\label{fig:h2params1}
\end{figure}

For the case $H_{\rm SM}=H_2$, we noted in the initial wide-ranged
scans that a vast majority of the points with $m_{H_2} \sim 125$\gev\ that survived
  the constraints imposed within \nmssmtools\  corresponded to
 the naturalness limit. We therefore performed a dedicated scan
  of the reduced parameter ranges seen in table\,\ref{tab:params}, and will
  only analyse such points here. Contrary to the $H_{\rm SM}=H_1$
  case, here $m_{A_1}$ can be much smaller even in the
  naturalness limit, as seen in figure~\ref{fig:h2params1}(a), where we show the
  distribution of $m_{A_1}$ against that of $m_{H_2}$, with the heat map
  corresponding to the parameter \tanb. One sees in the figure that lower values
  of $m_{A_1}$ prefer slightly larger values of \tanb. In
  figure~\ref{fig:h2params1}(b) we note that $m_{A_1}$ falls also with decreasing
  \lam, shown by the heat map.

\begin{figure}[tbp]
\centering
\subfloat[]{%
\label{fig:-a}%
\includegraphics*[width=7.7cm]{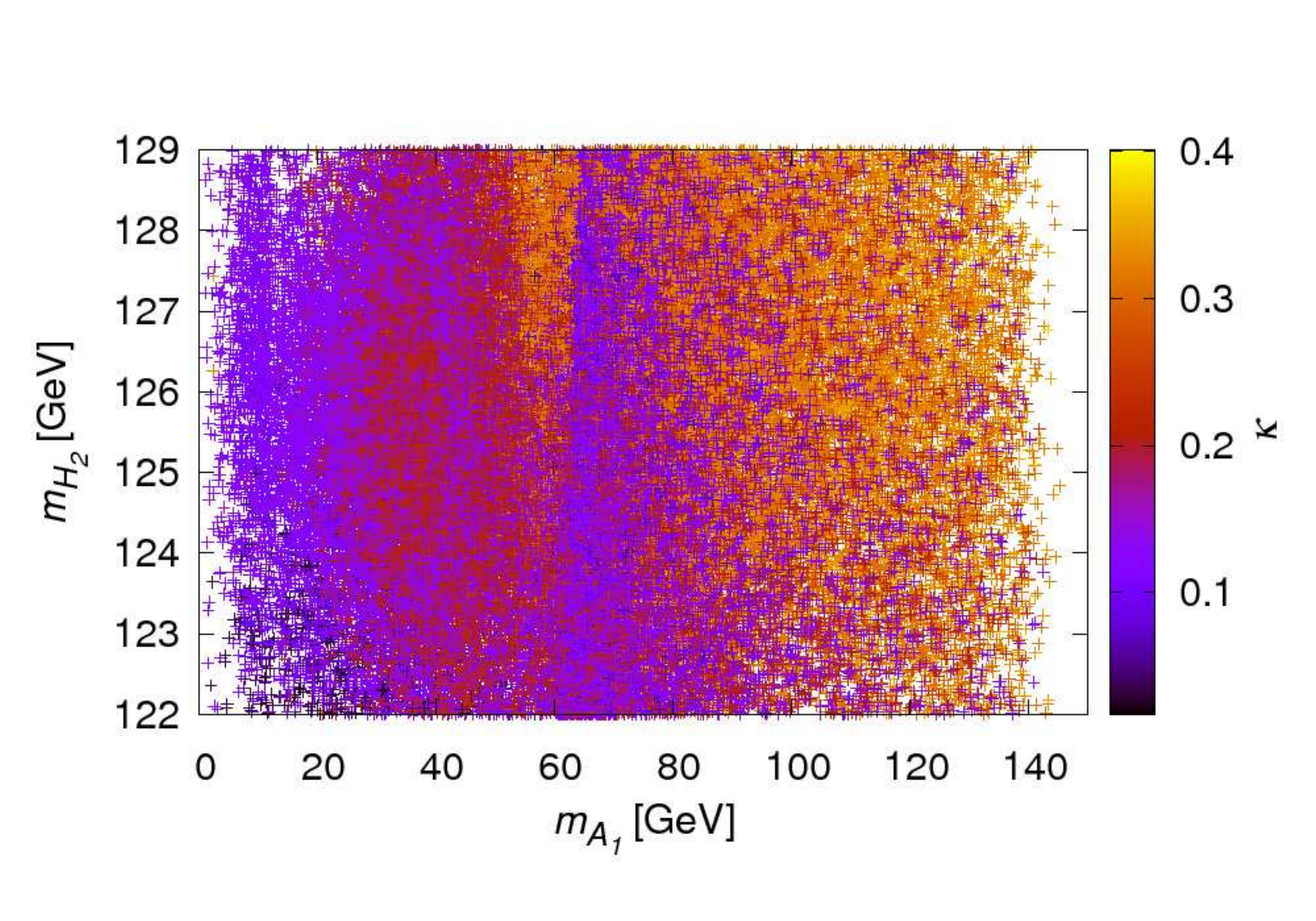}
}%
\subfloat[]{%
\label{fig:-b}%
\includegraphics*[width=7.7cm]{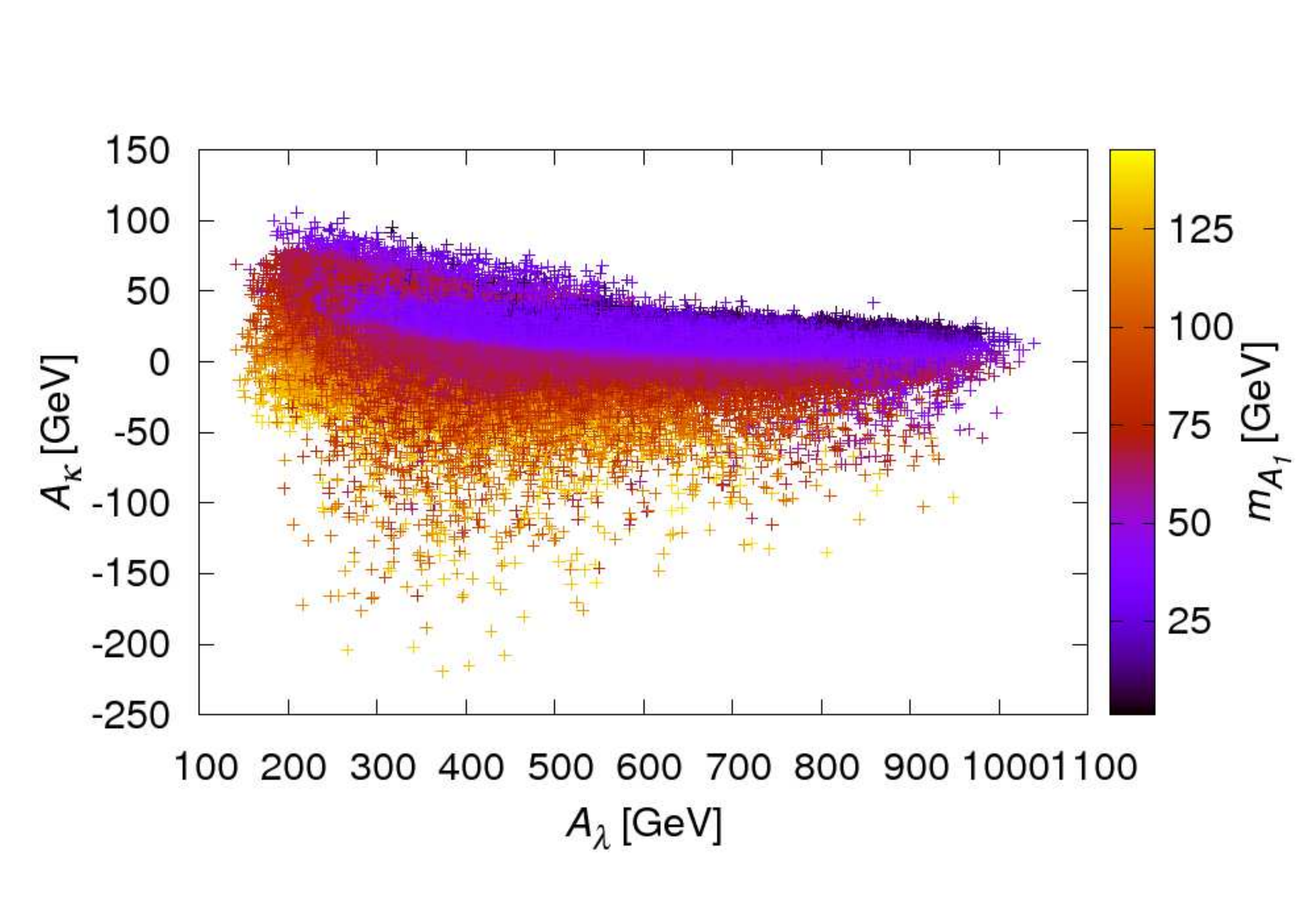}
}%
\caption[]{Case with $H_{\rm SM} = H_2$: (a) mass of $H_2$ vs.
that of $A_1$, with the heat map
  showing the distribution of \kap; (b)  the parameter \alam\
  vs. the parameter \akap\, with the heat map showing the
  distribution of the mass of $A_1$.}
\label{fig:h2params2}
\end{figure}

An important thing to note here is that, despite the comparatively
smaller \lam\ and larger \tanb, $m_{H_2}$ in this case can
 easily reach as high as 129\gev. This is because, as explained
in\cite{Badziak:2013bda},
the mixing of the SM-like $H_2$ with the lighter singlet-like $H_1$
can result in a 6--8\gev\ raise in the tree-level mass of $H_2$, even
for moderate values of \tanb\ and \lam.
  Such relatively small \lam\ and small-to-moderate \tanb,
 together with small \kap, as observed in the
  heat map of figure~\ref{fig:h2params2}(a), in turn also help
  reduce $m_{A_1}$. Importantly also, $m_{A_1}$ in this case can reach
  comparatively much smaller values without such points being
  severely constrained due to the opening of the $H_{\rm SM}\to A_1 A_1$
  channel, as was noted for the $H_{\rm SM}=H_1$ case. The relatively
  smaller values of \lam\ imply that
  the BR($H_{\rm SM}\to  A_1 A_1$) never gets overwhelming enough to
  completely diminish the other BRs and consequently the signal rates of
  $H_2$ in the $\gamma\gamma$ and $ZZ$ channels. However, the  
 onset of the $H_2\to A_1A_1$ channel does suppress points with 
  large \lam\ below $M_{H_2}/2$ and hence gives rise to sharp separations 
  between the violet and orange regions in figure~\ref{fig:h2params1}(a) and (b).

In figure~\ref{fig:h2params2}(b) one sees that $\alam $ is always positive
 although the mass of $A_1$, shown by the heat map, is almost
 independent of \alam, while \akap\ can be both negative and
 positive. Furthermore, $m_{A_1}$ clearly falls with increasing $+|\akap|$
 (and, equivalently, decreasing $-|\akap|$), which is
 again in accordance with eq.~(\ref{eq:ma2}).

\subsubsection{$bbA_1$ production}

\begin{figure}[tbp]
\begin{center}
\includegraphics[width=7.3cm]{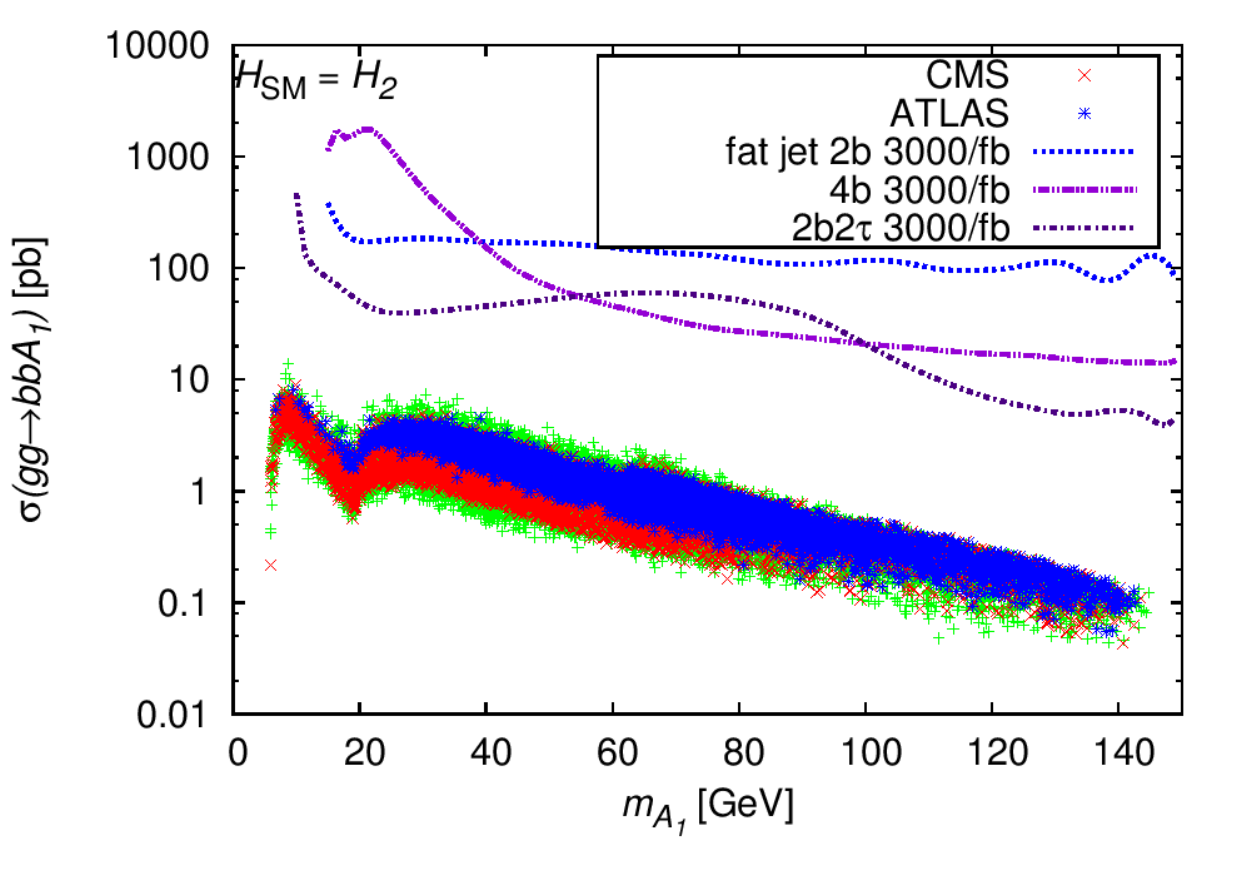}
\caption{$b\bar b A_1$ production cross section as a function of
  $m_{A_1}$ in the case with $H_{\rm SM} = H_2$. The color convention
for the points is the same as in figure
 \ref{fig:bba1h1}. See text for details about the sensitivity curves.}
\label{fig:bba1h2}
\end{center}
\end{figure}

In figure~\ref{fig:bba1h2} we show the $gg\rightarrow bbA_1$ rates for all
the good points from the scan for this case. The sensitivity
curves shown correspond to the $2b2\tau$ and $4b$ final states for an
expected integrated luminosity of 3000/fb at the LHC. As in the case with $H_{\rm
  SM}=H_1$ discussed above, none of the good points from the scan have
a cross section large enough to be discoverable in any of the final state
combinations.

\subsubsection{Production via $H_{\rm SM} \rightarrow A_1A_1/Z$}

\begin{figure}[tbp]
\centering
\subfloat[]{%
\label{fig:-a}%
\includegraphics*[width=7.3cm]{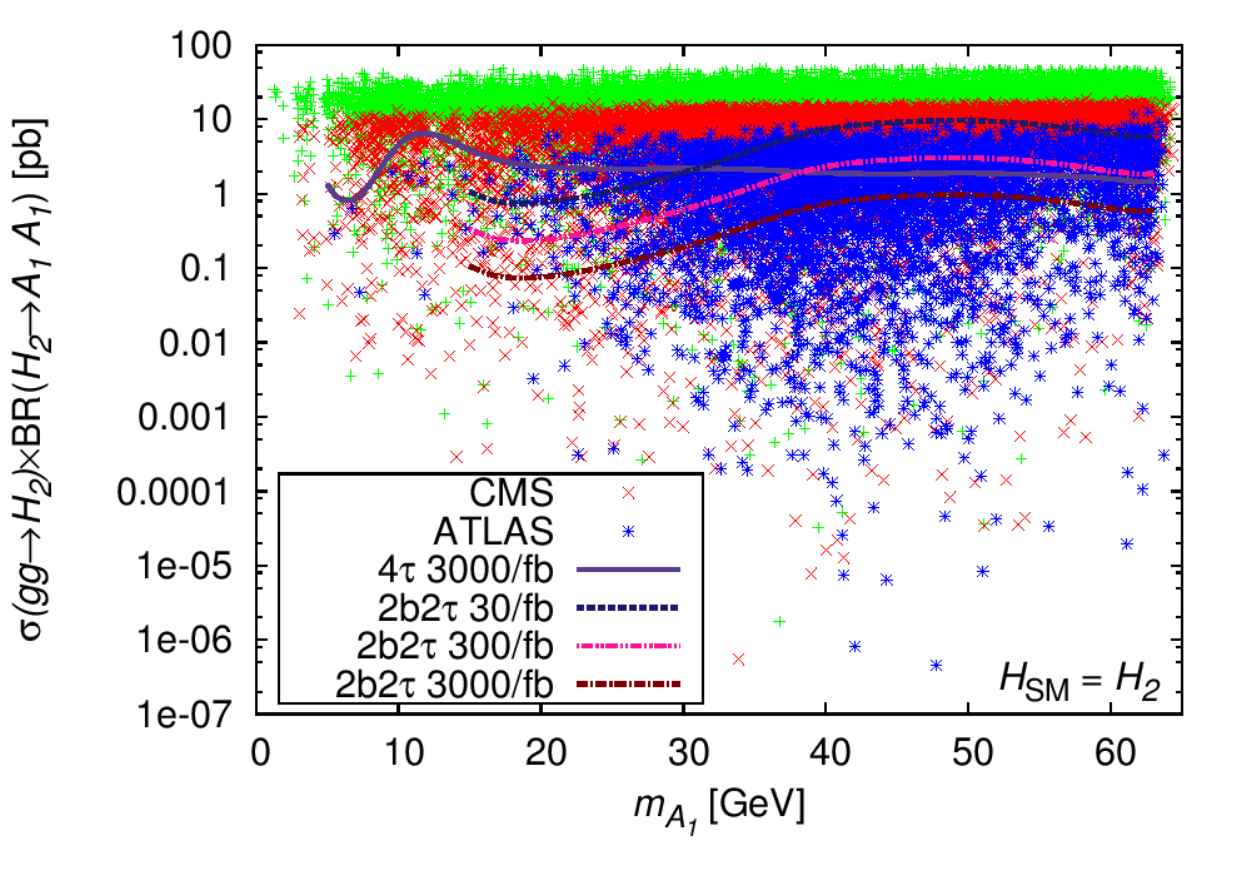}
}%
\hspace{0.5cm}%
\subfloat[]{%
\label{fig:-b}%
\includegraphics*[width=7.3cm]{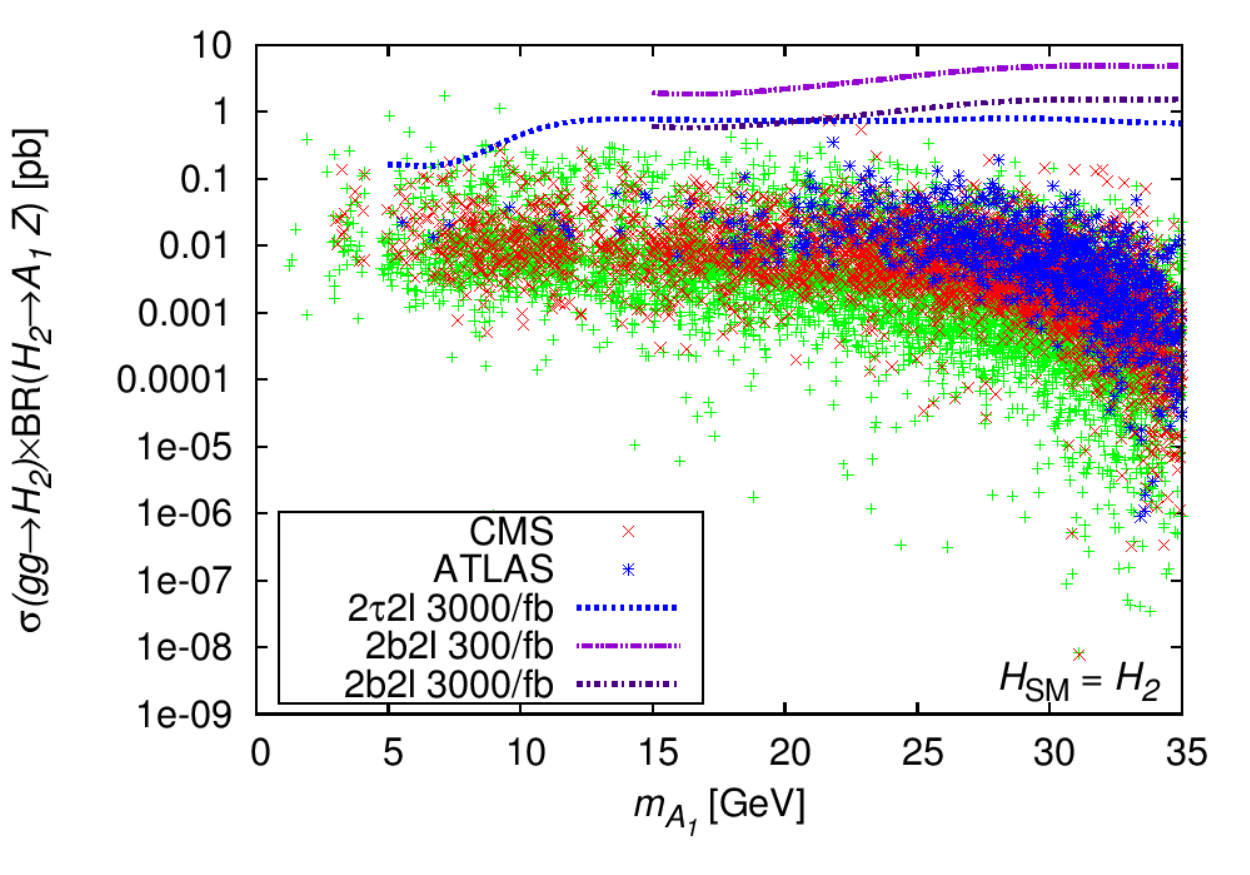}
}
\caption{Total cross sections in the case with $H_{\rm SM} =
 H_2$ for (a) the $gg \to H_{\rm SM}\to  A_1A_1$ process and (b)
 the $gg \to H_{\rm SM}\to A_1Z$ process. The color convention
for the points is the same as in figure
 \ref{fig:bba1h1}. See text for details about the sensitivity curves.}
\label{fig:H2toA1}
\end{figure}

In figure~\ref{fig:H2toA1}(a) we show the prospects for the $H_2 \to
A_1A_1$ channel when $H_2$ is SM-like. The sensitivity lines shown
correspond to the same final state combinations as in
figure~\ref{fig:h1a1zForH1}, with the addition of
a line for the $2b2\tau$ final state at $\mathcal{L}=30$/fb.
We see that, compared with the
$H_{\rm SM}=H_1$ case, a much larger part of the
parameter space can be probed in this channel at the LHC, even at as
low as 30/fb integrated luminosity. The reason is clearly that in
this case the points with $m_{A_1}<m_{H_{\rm SM}}/2$ belong to the
lower edge of the naturalness limit ($\lam \sim 0.4$), where
BR($H_2 \to A_1A_1$) is sufficiently enhanced
without causing the $H_{\rm SM}$ for these points to deviate too
much from a SM-like behaviour. This is the reason why
a large fraction of the points with large cross section is
consistent also with the CMS and ATLAS measurements of
$\mu_{\gamma\gamma / ZZ}$.

In figure~\ref{fig:H2toA1}(b) we see that the prospects in the $H_2\to
A_1Z$ channel are still poor. They are, nevertheless, much better than when
$H_{SM}=H_1$, despite the same peculiar cancellation in the $H_iA_jZ$
couplings discussed earlier. Here this cancellation is not as exact
because the relation $S_{i2}/S_{i1}\approx\tanb$ is not satisfied as strictly.

\subsubsection{Production via $H' \rightarrow A_1 A_1/Z$}

\begin{figure}[tbp]
\centering
\subfloat[]{%
\label{fig:-a}%
\includegraphics*[width=7.3cm]{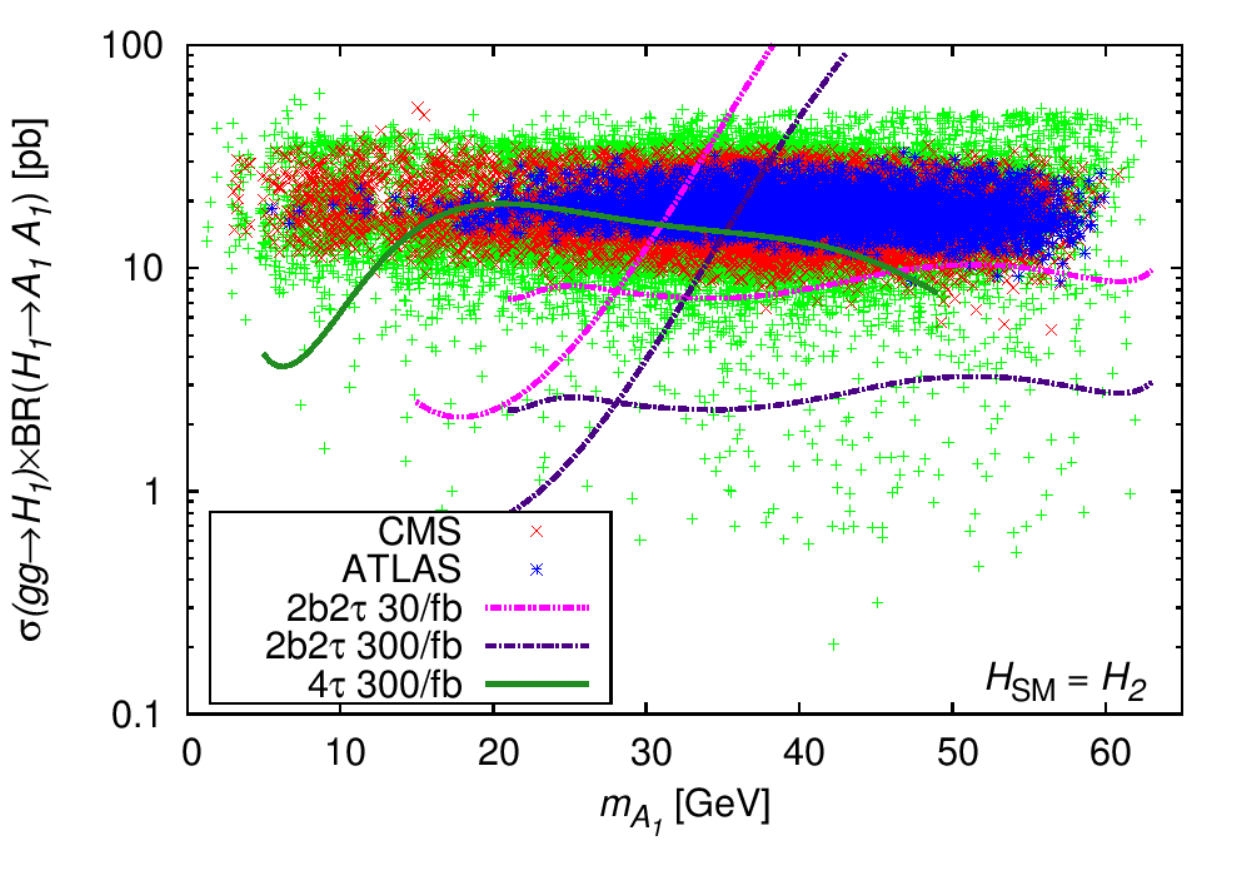}
}%
\hspace{0.5cm}%
\subfloat[]{%
\label{fig:-b}%
\includegraphics*[width=7.3cm]{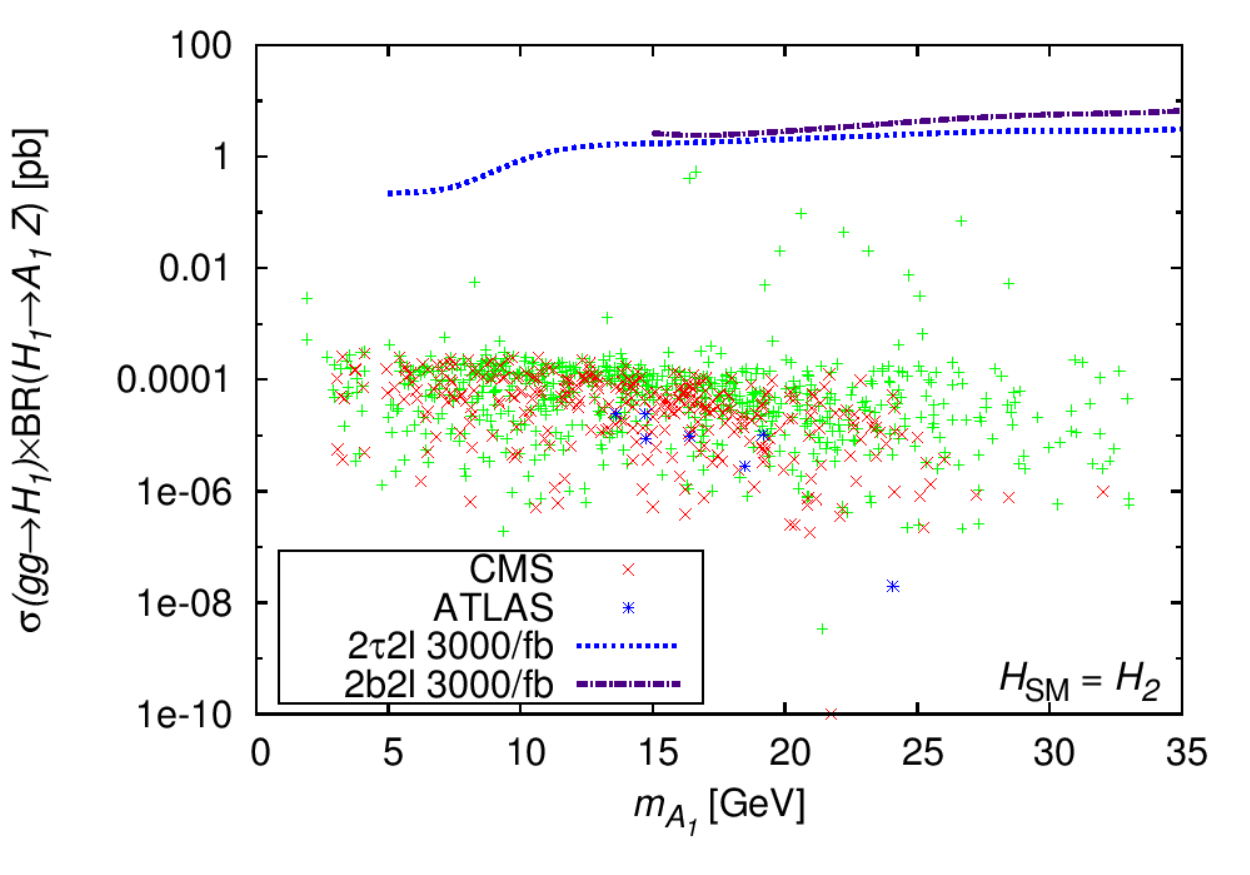}
}
\caption{Total cross sections in the case with $H_{\rm SM} =
 H_2$ for (a) the $gg \to H_1 \to  A_1A_1$ process and (b)
 the $gg \to H_1 \to A_1Z$ process. The color convention
for the points is the same as in figure
 \ref{fig:bba1h1}. See text for details about the sensitivity curves.}
\label{fig:H1toA1}
\end{figure}

The prospects for the discovery of a light pseudoscalar in the $H_1 \to
A_1A_1$ and $H_1 \to A_1Z$ decay channels, for a singlet-like $H_1$, are illustrated in
figures~\ref{fig:H1toA1}(a) and (b), respectively. In the panel (a) the
sensitivity curves correspond to the one fat jet plus two $\tau$-jets,
the two single $b$-jets plus two $\tau$-jets and the $4\tau$ final
states. These are compared with the cross sections of the good points
from the scans. For the fat jet plus two $\tau$-jets as well as the $4\tau$ curves
$m_{H_1}=100$\gev, while for the two $b$-jets plus two $\tau$-jets curve
$m_{H_1}=125$\gev, which allows the coverage of points with large
$m_{A_1}$. \footnote{Note that this should not be taken as a claim
  that for such points $H_1$ is mass-degenerate with $H_2$; the
  chosen $H_1$ mass is merely for illustration.} 
However, in neither of these cases is the mass used to
constrain the kinematics. The $4\tau$ line has been added since, for
$m_{A_1} < 2m_b$, the $A_1$ can only be accessed via this
channel. Understandably, this line is cut off at the kinematical upper
limit of $m_{H_1}/2=50$\gev. The lines corresponding to the fat jet
analysis are cut off where the efficiency for extracting the signal
becomes too bad for the result to be reliable.

One sees in figure~\ref{fig:H1toA1}(a) that almost all the points
complying with the current CMS and/or ATLAS constraints on $R_X$
are potentially discoverable, even at $\mathcal{L}=30$/fb. Thus a large part
of the scanned NMSSM parameter space can be probed via this decay
channel. In particular, since such light pseudoscalars cannot be obtained in the
naturalness limit for the case with $H_{\rm SM}=H_1$, it should
essentially be possible to exclude $m_{A_1} \lesssim 60$\gev\ in the
natural NMSSM at the LHC via this channel.
Although a light $A_1$ may
still be obtained with slightly smaller \lam\, it will be difficult
for $H_{\rm SM}$ to reach $\sim125$\gev\ without large radiative corrections.
Note also that such an exclusion will not cover the narrow
regions of the parameter space where the $A_1\to H_1H_1$ decay is kinematically allowed.
Finally, In figure~\ref{fig:H1toA1}(b), we see that the prospects for
the discovery of $A_1$ via the $H_1\to A_1 Z$ channel are non-existent
in this case too.

\begin{figure}[tbp]
\centering
\subfloat[]{%
\label{fig:-a}%
\includegraphics*[width=7.3cm]{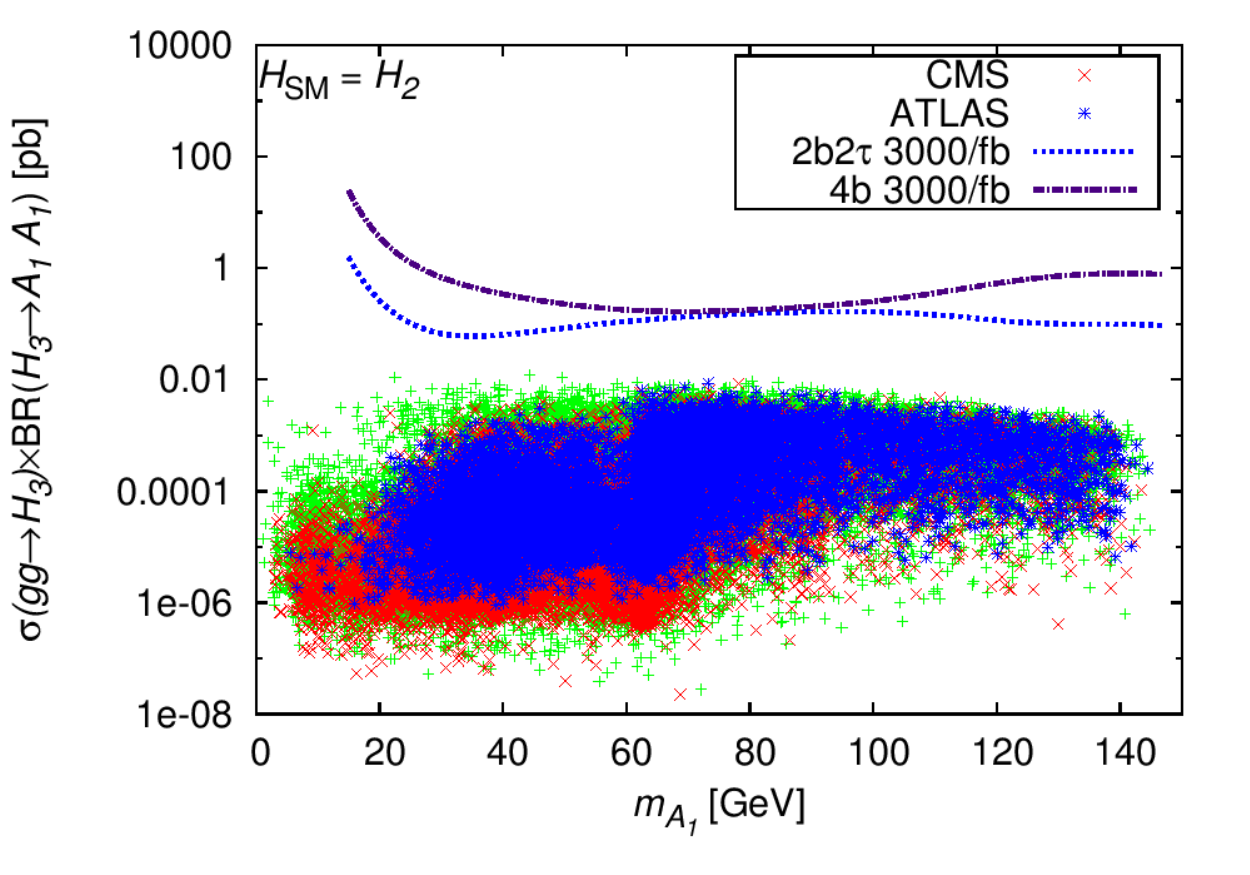}
}%
\hspace{0.5cm}%
\subfloat[]{%
\label{fig:-b}%
\includegraphics*[width=7.3cm]{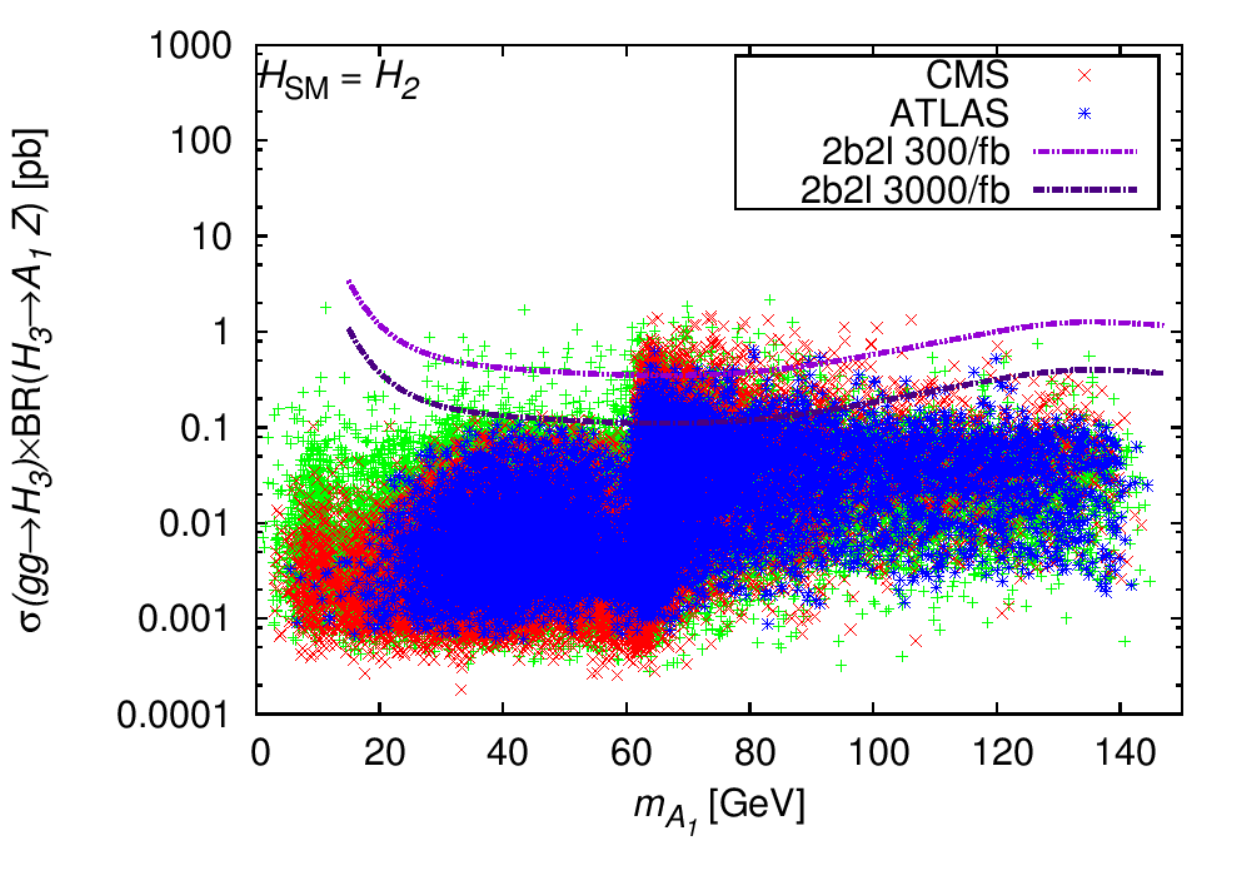}
}
\caption{Total cross sections in the case with $H_{\rm SM} =
 H_1$ for (a) the $gg \to H_3 \to  A_1A_1$ process and (b)
 the $gg \to H_3 \to A_1Z$ process. The color convention
for the points is the same as in figure
 \ref{fig:bba1h1}. See text for details about the sensitivity curves.}
\label{fig:H3toA1}
\end{figure}

For the decay chain starting from $H_3$, the situation is similar to
the one in the case with $H_{\rm SM}=H_1$. In figure~\ref{fig:H3toA1}(a) we
see that the $H_3\to A_1A_1$ channel is inaccessible here also due
to the fact that, as mentioned before, for such high masses the
production cross section gets diminished. Moreover, other decay channels of
$H_3$ dominate over this channel. The sensitivity curves in the figure
correspond to the $2b2\tau$  and $4\tau$ final states for
$\mathcal{L}=3000$/fb. In fact, the cross sections obtainable here are
even marginally smaller than those seen earlier for the $H_3\to
A_1A_1$ channel in the $H_{\rm SM}=H_1$ case, owing, again, to the
generally smaller values of \lam\ in this case.

Conversely, the $H_3\to A_1Z$ channel, shown in
figure~\ref{fig:H3toA1}, shows much more promise than before. As pointed out before, this has to do with the increased sensitivity in the fat jet analysis when the involved masses are high, as well as the relatively large $H_3A_1Z$ coupling, which is actually somewhat larger here than in the $H_{\rm SM}=H_1$ case, due to a somewhat larger doublet
component of $A_1$. Again, there are hardly any points with $m_{A_1} <
m_{H_{\rm SM}}/2$ for \lam\ large enough to yield a high total cross
section, for the same reasons as discussed earlier. We emphasise again that
this channel will be an extremely important probe for
an NMSSM $A_1$ with mass greater than $\sim 60$\gev.

\subsection{Benchmark points}

As pointed out in the previous sections, of the channels discussed in
this paper, $gg\to H_3\to A_1Z$ appears to be the only one in which an
$A_1$ lying in the interval $\sim 60\gev$--$120$\gev\ could be
discovered with 300/fb integrated luminosity at the LHC. 
Note though that the $H^\pm\to A_1W^\pm$ channel has also been found to be
of interest for such pseudoscalar masses\cite{Das:2014fha}.

While the $H_i \to A_1A_1$ decay channel has been
previously visited in the literature, this channel has not been
emphasised upon much. Therefore, in order to facilitate further investigation
of this channel, we single out a few benchmark points that
should be within the reach of the LHC at $\mathcal{L}=300$/fb, and 
provide their details in table\,\ref{tab:Bench}. Besides yielding
large cross sections for
this channel, these points also satisfy the CMS and/or ATLAS
constraints on $R_X$. Point 1 corresponds to the case with
$H_{\rm SM}=H_1$, while Points 2, 3 and 4 correspond to the $H_{\rm
 SM}=H_2$ case and are intended to cover the entire accessible
range of $m_{A_1}$.

\begin{table}[tbp]
\begin{center}
\begin{tabular}{|c|c|c|c|c|}
\hline
Case & $H_{\rm SM} = H_1$ & \multicolumn{3}{c|}{$H_{\rm SM} = H_2$} \\
 \hline
& Point 1 & Point 2 & Point 3 & Point 4\\
\hline
Input parameters & \multicolumn{4}{c|}{}\\
\hline
\mzero\,(GeV) & 991.42 & 1321.2 & 1817.8 & 1358.1\\
\mhalf\,(GeV)  & 737.59 & 752.07 & 977.07 &  947.62\\
\azero\,(GeV)  & $-729.51$ & $-88.81$ & $-1112.88$  &  $-699.12$\\
\mueff\,(GeV)  & 172.43 & 150.52 & 167.17  &  185.77\\
\tanb &  1.807 & 1.636  & 1.661  &  1.549\\
\lam & 0.66 & 0.619  & 0.6248   &  0.5978 \\
\kap & 0.1772 & 0.0692 & 0.0506  &  0.1096\\
\alam$^*$\,(TeV) &  59.136 & 426.49 & $-466.9$  & $-115.9$\\
\akap$^*$\,(TeV) & 474.85 & 444.7 & $-142.41$ &  275.31\\
\hline
Observables & \multicolumn{4}{c|}{}\\
\hline
$m_{A_1}$\,(GeV) & 89.88 & 106.22 & 73.71  &  63.95\\
$m_{H_1}$\,(GeV) & 127.1 & 85.1 & 94.7  &   116.4\\
$m_{H_2}$\,(GeV) & 130.3 & 124.8 & 126.85  &  125.41\\
$m_{H_3}$\,(GeV) & 378.47 & 327.85 & 360.58  & 388.21\\
\hline
$R_{\gamma\gamma}$ & 1.156 & 1.114 & 0.964  & 1.099\\
$R_{ZZ}$ & 0.824 & 1.063 & 0.982  &  0.972\\
$\sigma\times{\rm BR}(H_3\to A_1Z)$ (pb)  & 0.57 & 1.33 & 1.47  &  0.83\\
\hline
\end{tabular}
\caption{Some specifics of the four benchmark points for the $gg\to H_3\to A_1Z$ channel.}
\label{tab:Bench}
\end{center}
\end{table}

\section{\label{concl} Conclusions}

In this article we have revisited the discovery prospects of a
light NMSSM pseudoscalar, $A_1$, at the forthcoming 14\tev\ LHC run. We have
shown the dependence of the mass expression for $A_1$ on the
input parameters of the model and discussed its most prominent
production and decay modes. Through scans of the CNMSSM-NUHM
parameter space, we have identified the regions
where $A_1$ with mass $\lesssim 150$\gev\ can be obtained while
requiring one of the two light CP-even Higgs bosons to have a mass
around 125\gev\ and SM-like signal rates. We have then discussed in
detail the salient features of these regions, separately for the case
when the SM-like Higgs boson is $H_1$ and the case when it is $H_2$.

To connect to LHC physics we have performed detailed MC analyses of the various $A_1$ production
channels of interest. These channels can be divided into
direct and indirect ones, with the former referring to the
$b\bar{b}$-associated production mode and the latter encompassing the decays
of the heavier CP-even Higgs bosons into $A_1A_1$ or $A_1Z$
pairs. The $A_1$ thus produced is assumed to decay via the channels
with the highest BRs, i.e., $b\bar{b}$ and $\tau^+\tau^-$, while the
$Z$ boson decays leptonically. For the $A_1 \to b\bar{b}$
decays, we have adopted a jet substructure method, which improves
the experimental sensitivity for a low mass $A_1$ through boosted $b$-jet pairs and is
particularly useful for the decays of the heavy CP-even Higgs bosons.

We have found that, contrary to the findings of some earlier studies,
the direct production channel, $bbA_1$, has now been rendered ineffective by the
observed properties of the SM-like Higgs boson discovered at the
LHC. However, some of the indirect production channels still show plenty of promise.
Specifically, the decays of the scalars, including in particular
the SM-like Higgs boson, whether $H_1$ or $H_2$, carry the potential
to reveal an $A_1$ with mass $\lesssim 60$\gev\ for the integrated luminosity at the LHC
 as low as 30/fb. In case of non-discovery, this channel could
help practically exclude large portions of the NMSSM parameter
space. This is particularly true when the SM-like $H_2$ is required to
achieve a mass close to 125\gev\ in a natural way (without needing
large radiative corrections), while the $H_1$ is not light enough to allow
$A_1$ decays into its pairs.

Most notably though, when the $A_1$ is heavier than $\sim 60$\gev, while its
pair production via decays of the two lightest CP-even
Higgs bosons also becomes inaccessible, the $gg\to H_3\to A_1Z$ channel
takes over as the most promising one. This channel is, therefore, of great
importance and warrants dedicated probes in future analyses at the
LHC. We strongly advocate such studies, for which we have provided
details of some benchmark points which give significantly
large $gg\to H_3\to A_1Z$ cross sections and at the same time show
consistency with the current data from the LHC Higgs boson searches.

We summarise all our findings in table~\ref{tab:Channels}, quoting the
integrated luminosity at which a reasonable number of model points
should be discoverable at the LHC in at least one of the given final
state combinations. Also given are the mass ranges of $A_1$ over which
 the discovery in a given production channel is plausible.

\begin{table}[tbp]
\begin{center}
\begin{tabular}{|c|l|c|c|}\hline
Production mode  & Final states & Accessibility, for $\mathcal{L}$ & Mass range\,(GeV) \\\hline
$b\bar b A_1$  &  $4b$, $2b2\tau$  & \textcolor{red}{x} &\\
$H_1\to A_1A_1$ ($H_1=H_{\rm SM}$)  &  $4b$, $2b2\tau$, $4\tau$  &   \textcolor{green}{\checkmark}  300/fb &  $m_{A_1}<63$  \\
$H_1\to A_1A_1$ ($H_2=H_{\rm SM}$)  &  $4b$, $2b2\tau$, $4\tau$  &   \textcolor{green}{\checkmark}  30/fb & $m_{A_1}<60$   \\
$H_1\to A_1Z$  &   $2b2\ell$, $2\tau 2\ell$   &   \textcolor{red}{x} & \\
$H_2\to A_1A_1$ ($H_1=H_{\rm SM}$)  &    $4b$, $2b2\tau$, $4\tau$  &   \textcolor{green}{\checkmark}  300/fb  &  $60< m_{A_1} <80$\\
$H_2\to A_1A_1$ ($H_2=H_{\rm SM}$)  &    $4b$, $2b2\tau$, $4\tau$  &   \textcolor{green}{\checkmark}  30/fb  &   $m_{A_1}<63$  \\
$H_2\to A_1Z$   & $2b2\ell$, $2\tau 2\ell$  &   \textcolor{red}{x}  &  \\
$H_3\to A_1A_1$  &  $4b$, $2b2\tau$, $4\tau$  &   \textcolor{red}{x}  & \\
$H_3\to A_1Z$   & $2b2\ell$, $2\tau 2\ell$  &   \textcolor{green}{\checkmark}  300/fb  &   $60 < m_{A_1} <120$ \\\hline
\end{tabular}
\caption{List of the $A_1$ production channels included in this
  study. The second column shows the final
state combinations of interest for each channel, while the third column shows the
integrated luminosity at which the $A_1$ can be accessible at the LHC
in at least one of these combinations. In the fourth column we provide
the mass range within which a signature of $A_1$ can be established
in the given channel.}\label{tab:Channels}
\end{center}
\end{table}%

Finally, the discovery prospects of the NMSSM Higgs bosons,
including also via their decays into other Higgs states, were studied
recently in\cite{King:2014xwa}. The cross sections calculated
by us are generally consistent with those provided there, in
the cases where a comparison is meaningful. Furthermore, in that study
some points with unusually large BR$(A_1\to
\gamma\gamma)$ have been emphasised. For some of these points the
$2b2\gamma$ final state has been found to be interesting. However, due 
possibly to the fact that we analyse a different model, the CNMSSM-NUHM,
as compared to the phenomenological NMSSM studied
in\cite{King:2014xwa}, we did not see any points with substantial 
BR$(A_1\to \gamma\gamma)$ in our scans. Hence we 
did not consider the $2b2\gamma$ final state in our analyses.

\begin{center}
  \textbf{ACKNOWLEDGMENTS}
\end{center}

\noindent This work has been funded in part by the Welcome Programme
of the Foundation for Polish Science. S.~Moretti is supported in part through the NExT
Institute. S.~Munir is supported in part by the Swedish Research
Council under contracts 2007-4071 and 621-2011-5107. L.~Roszkowski is
also supported in part by an STFC consortium grant of Lancaster,
Manchester and Sheffield Universities. The use of the CIS computer cluster at NCBJ is
gratefully acknowledged.

\bibliographystyle{JHEP}	
\bibliography{NMSSM_A1_refs}

\providecommand{\href}[2]{#2}\begingroup\raggedright\begin{thebibliography}{10}

\bibitem{Fayet:1974pd}
P.~Fayet, {\it {Supergauge Invariant Extension of the Higgs Mechanism and a
  Model for the electron and Its Neutrino}},  {\em Nucl.Phys.} {\bf B90} (1975)
  104--124.

\bibitem{Ellis:1988er}
J.~R. Ellis, J.~Gunion, H.~E. Haber, L.~Roszkowski, and F.~Zwirner, {\it {Higgs
  Bosons in a Nonminimal Supersymmetric Model}},  {\em Phys.Rev.} {\bf D39}
  (1989) 844.

\bibitem{Durand:1988rg}
L.~Durand and J.~L. Lopez, {\it {Upper Bounds on Higgs and Top Quark Masses in
  the Flipped SU(5) x U(1) Superstring Model}},  {\em Phys.Lett.} {\bf B217}
  (1989) 463.

\bibitem{Drees:1988fc}
M.~Drees, {\it {Supersymmetric Models with Extended Higgs Sector}},  {\em
  Int.J.Mod.Phys.} {\bf A4} (1989) 3635.

\bibitem{Miller:2003ay}
D.~Miller, R.~Nevzorov, and P.~Zerwas, {\it {The Higgs sector of the
  next-to-minimal supersymmetric standard model}},  {\em Nucl.Phys.} {\bf B681}
  (2004) 3--30, [\href{http://arxiv.org/abs/hep-ph/0304049}{{\tt
  hep-ph/0304049}}].

\bibitem{Aad:2012tfa}
{\bf ATLAS Collaboration}, G.~Aad et~al., {\it {Observation of a new particle
  in the search for the Standard Model Higgs boson with the ATLAS detector at
  the LHC}},  {\em Phys.Lett.} {\bf B716} (2012) 1--29,
  [\href{http://arxiv.org/abs/1207.7214}{{\tt arXiv:1207.7214}}].

\bibitem{Chatrchyan:2012ufa}
{\bf CMS Collaboration}, S.~Chatrchyan et~al., {\it {Observation of a new boson
  at a mass of 125 GeV with the CMS experiment at the LHC}},  {\em Phys.Lett.}
  {\bf B716} (2012) 30--61, [\href{http://arxiv.org/abs/1207.7235}{{\tt
  arXiv:1207.7235}}].

\bibitem{Love:2008aa}
{\bf CLEO Collaboration}, W.~Love et~al., {\it {Search for Very Light CP-Odd
  Higgs Boson in Radiative Decays of Upsilon(S-1)}},  {\em Phys.Rev.Lett.} {\bf
  101} (2008) 151802, [\href{http://arxiv.org/abs/0807.1427}{{\tt
  arXiv:0807.1427}}].

\bibitem{Tung:2008gd}
{\bf E391a Collaboration}, Y.~Tung et~al., {\it {Search for a light
  pseudoscalar particle in the decay $K^0_L \rightarrow \pi^0 \pi^0 X$}},  {\em
  Phys.Rev.Lett.} {\bf 102} (2009) 051802,
  [\href{http://arxiv.org/abs/0810.4222}{{\tt arXiv:0810.4222}}].

\bibitem{Lees:2012iw}
{\bf BaBar collaboration}, J.~Lees et~al., {\it {Search for di-muon decays of a
  low-mass Higgs boson in radiative decays of the $\Upsilon(1S)$}},  {\em
  Phys.Rev.} {\bf D87} (2013), no.~3 031102,
  [\href{http://arxiv.org/abs/1210.0287}{{\tt arXiv:1210.0287}}].

\bibitem{Lees:2012te}
{\bf BaBar Collaboration}, J.~Lees et~al., {\it {Search for a low-mass scalar
  Higgs boson decaying to a tau pair in single-photon decays of
  $\Upsilon(1S)$}},  {\em Phys.Rev.} {\bf D88} (2013), no.~7 071102,
  [\href{http://arxiv.org/abs/1210.5669}{{\tt arXiv:1210.5669}}].

\bibitem{Lees:2013vuj}
{\bf BaBar Collaboration}, J.~Lees et~al., {\it {Search for a light Higgs boson
  decaying to two gluons or $s\bar{s}$ in the radiative decays of
  $\Upsilon(1S)$}},  {\em Phys.Rev.} {\bf D88} (2013), no.~3 031701,
  [\href{http://arxiv.org/abs/1307.5306}{{\tt arXiv:1307.5306}}].

\bibitem{Peruzzi:2014ksa}
{\bf BABAR Collaboration}, I.~Peruzzi, {\it {Recent BABAR results on dark
  matter and light Higgs searches and on CP and T violation}},  {\em EPJ Web
  Conf.} {\bf 71} (2014) 00108.

\bibitem{Domingo:2008rr}
F.~Domingo, U.~Ellwanger, E.~Fullana, C.~Hugonie, and M.-A. Sanchis-Lozano,
  {\it {Radiative Upsilon decays and a light pseudoscalar Higgs in the NMSSM}},
   {\em JHEP} {\bf 0901} (2009) 061,
  [\href{http://arxiv.org/abs/0810.4736}{{\tt arXiv:0810.4736}}].

\bibitem{Schael:2010aw}
{\bf ALEPH Collaboration}, S.~Schael et~al., {\it {Search for neutral Higgs
  bosons decaying into four taus at LEP2}},  {\em JHEP} {\bf 1005} (2010) 049,
  [\href{http://arxiv.org/abs/1003.0705}{{\tt arXiv:1003.0705}}].

\bibitem{Chatrchyan:2012am}
{\bf CMS Collaboration}, S.~Chatrchyan et~al., {\it {Search for a light
  pseudoscalar Higgs boson in the dimuon decay channel in $pp$ collisions at
  $\sqrt{s}=7$ TeV}},  {\em Phys.Rev.Lett.} {\bf 109} (2012) 121801,
  [\href{http://arxiv.org/abs/1206.6326}{{\tt arXiv:1206.6326}}].

\bibitem{Chatrchyan:2012cg}
{\bf CMS Collaboration}, S.~Chatrchyan et~al., {\it {Search for a
  non-standard-model Higgs boson decaying to a pair of new light bosons in
  four-muon final states}},  {\em Phys.Lett.} {\bf B726} (2013) 564--586,
  [\href{http://arxiv.org/abs/1210.7619}{{\tt arXiv:1210.7619}}].

\bibitem{Ellwanger:2003jt}
U.~Ellwanger, J.~F. Gunion, C.~Hugonie, and S.~Moretti, {\it {Towards a no lose
  theorem for NMSSM Higgs discovery at the LHC}},
  \href{http://arxiv.org/abs/hep-ph/0305109}{{\tt hep-ph/0305109}}.

\bibitem{Ellwanger:2005uu}
U.~Ellwanger, J.~F. Gunion, and C.~Hugonie, {\it {Difficult scenarios for NMSSM
  Higgs discovery at the LHC}},  {\em JHEP} {\bf 0507} (2005) 041,
  [\href{http://arxiv.org/abs/hep-ph/0503203}{{\tt hep-ph/0503203}}].

\bibitem{Moretti:2006hq}
S.~Moretti, S.~Munir, and P.~Poulose, {\it {Another step towards a no-lose
  theorem for NMSSM Higgs discovery at the LHC}},  {\em Phys.Lett.} {\bf B644}
  (2007) 241--247, [\href{http://arxiv.org/abs/hep-ph/0608233}{{\tt
  hep-ph/0608233}}].

\bibitem{Forshaw:2007ra}
J.~Forshaw, J.~Gunion, L.~Hodgkinson, A.~Papaefstathiou, and A.~Pilkington,
  {\it {Reinstating the 'no-lose' theorem for NMSSM Higgs discovery at the
  LHC}},  {\em JHEP} {\bf 0804} (2008) 090,
  [\href{http://arxiv.org/abs/0712.3510}{{\tt arXiv:0712.3510}}].

\bibitem{Belyaev:2008gj}
A.~Belyaev, S.~Hesselbach, S.~Lehti, S.~Moretti, A.~Nikitenko, et~al., {\it
  {The Scope of the 4 tau Channel in Higgs-strahlung and Vector Boson Fusion
  for the NMSSM No-Lose Theorem at the LHC}},
  \href{http://arxiv.org/abs/0805.3505}{{\tt arXiv:0805.3505}}.

\bibitem{Almarashi:2011te}
M.~Almarashi and S.~Moretti, {\it {Reinforcing the no-lose theorem for NMSSM
  Higgs discovery at the LHC}},  {\em Phys.Rev.} {\bf D84} (2011) 035009,
  [\href{http://arxiv.org/abs/1106.1599}{{\tt arXiv:1106.1599}}].

\bibitem{Ellwanger:2013ova}
U.~Ellwanger, {\it {Higgs pair production in the NMSSM at the LHC}},  {\em
  JHEP} {\bf 1308} (2013) 077, [\href{http://arxiv.org/abs/1306.5541}{{\tt
  arXiv:1306.5541}}].

\bibitem{Cheung:2008rh}
K.~Cheung and T.-J. Hou, {\it {Light Pseudoscalar Higgs boson in Neutralino
  Decays in the Next-to-Minimal Supersymmetric Standard Model}},  {\em
  Phys.Lett.} {\bf B674} (2009) 54--58,
  [\href{http://arxiv.org/abs/0809.1122}{{\tt arXiv:0809.1122}}].

\bibitem{Almarashi:2010jm}
M.~Almarashi and S.~Moretti, {\it {Low Mass Higgs signals at the LHC in the
  Next-to-Minimal Supersymmetric Standard Model}},  {\em Eur.Phys.J.} {\bf C71}
  (2011) 1618, [\href{http://arxiv.org/abs/1011.6547}{{\tt arXiv:1011.6547}}].

\bibitem{Almarashi:2011hj}
M.~M. Almarashi and S.~Moretti, {\it {Muon Signals of Very Light CP-odd Higgs
  states of the NMSSM at the LHC}},  {\em Phys.Rev.} {\bf D83} (2011) 035023,
  [\href{http://arxiv.org/abs/1101.1137}{{\tt arXiv:1101.1137}}].

\bibitem{Almarashi:2011bf}
M.~Almarashi and S.~Moretti, {\it {Very Light CP-odd Higgs bosons of the NMSSM
  at the LHC in 4b-quark final states}},  {\em Phys.Rev.} {\bf D84} (2011)
  015014, [\href{http://arxiv.org/abs/1105.4191}{{\tt arXiv:1105.4191}}].

\bibitem{Almarashi:2012ri}
M.~M. Almarashi and S.~Moretti, {\it {Scope of Higgs production in association
  with a bottom quark pair in probing the Higgs sector of the NMSSM at the
  LHC}},  \href{http://arxiv.org/abs/1205.1683}{{\tt arXiv:1205.1683}}.

\bibitem{Almarashi:2011qq}
M.~M. Almarashi and S.~Moretti, {\it {LHC Signals of a Heavy CP-even Higgs
  Boson in the NMSSM via Decays into a $Z$ and a Light CP-odd Higgs State}},
  {\em Phys.Rev.} {\bf D85} (2012) 017701,
  [\href{http://arxiv.org/abs/1109.1735}{{\tt arXiv:1109.1735}}].

\bibitem{Ellwanger:2011aa}
U.~Ellwanger, {\it {A Higgs boson near 125 GeV with enhanced di-photon signal
  in the NMSSM}},  {\em JHEP} {\bf 1203} (2012) 044,
  [\href{http://arxiv.org/abs/1112.3548}{{\tt arXiv:1112.3548}}].

\bibitem{King:2012is}
S.~King, M.~Muhlleitner, and R.~Nevzorov, {\it {NMSSM Higgs Benchmarks Near 125
  GeV}},  {\em Nucl.Phys.} {\bf B860} (2012) 207--244,
  [\href{http://arxiv.org/abs/1201.2671}{{\tt arXiv:1201.2671}}].

\bibitem{Ellwanger:2012ke}
U.~Ellwanger and C.~Hugonie, {\it {Higgs bosons near 125 GeV in the NMSSM with
  constraints at the GUT scale}},  {\em Adv.High Energy Phys.} {\bf 2012}
  (2012) 1, [\href{http://arxiv.org/abs/1203.5048}{{\tt arXiv:1203.5048}}].

\bibitem{Gherghetta:2012gb}
T.~Gherghetta, B.~von Harling, A.~D. Medina, and M.~A. Schmidt, {\it {The
  Scale-Invariant NMSSM and the 126 GeV Higgs Boson}},  {\em JHEP} {\bf 1302}
  (2013) 032, [\href{http://arxiv.org/abs/1212.5243}{{\tt arXiv:1212.5243}}].

\bibitem{Cao:2012fz}
J.~Cao et~al., {\it {A SM-like Higgs near 125 GeV in low energy SUSY: a
  comparative study for MSSM and NMSSM}},  {\em JHEP} {\bf 1203} (2012) 086,
  [\href{http://arxiv.org/abs/1202.5821}{{\tt arXiv:1202.5821}}].

\bibitem{Kim:2012az}
J.~E. Kim, H.~P. Nilles, and M.-S. Seo, {\it {Singlet Superfield Extension of
  the Minimal Supersymmetric Standard model with Peccei-Quinn symmetry and a
  Light Pseudoscalar Higgs Boson at the LHC}},  {\em Mod.Phys.Lett.} {\bf A27}
  (2012) 1250166, [\href{http://arxiv.org/abs/1201.6547}{{\tt
  arXiv:1201.6547}}].

\bibitem{Munir:2013wka}
S.~Munir, L.~Roszkowski, and S.~Trojanowski, {\it {Simultaneous enhancement in
  $\gamma\gamma$, $b\bar{b}$ and $\tau^+ \tau^-$ rates in the NMSSM with nearly
  degenerate scalar and pseudoscalar Higgs bosons}},  {\em Phys. Rev.} {\bf
  D88} (2013) 055017, [\href{http://arxiv.org/abs/1305.0591}{{\tt
  arXiv:1305.0591}}].

\bibitem{Cerdeno:2013cz}
D.~G. Cerdeno, P.~Ghosh, and C.~B. Park, {\it {Probing the two light Higgs
  scenario in the NMSSM with a low-mass pseudoscalar}},  {\em JHEP} {\bf 1306}
  (2013) 031, [\href{http://arxiv.org/abs/1301.1325}{{\tt arXiv:1301.1325}}].

\bibitem{Stal:2011cz}
O.~Stal and G.~Weiglein, {\it {Light NMSSM Higgs bosons in SUSY cascade decays
  at the LHC}},  {\em JHEP} {\bf 1201} (2012) 071,
  [\href{http://arxiv.org/abs/1108.0595}{{\tt arXiv:1108.0595}}].

\bibitem{Das:2012rr}
D.~Das, U.~Ellwanger, and A.~M. Teixeira, {\it {Modified Signals for
  Supersymmetry in the NMSSM with a Singlino-like LSP}},  {\em JHEP} {\bf 1204}
  (2012) 067, [\href{http://arxiv.org/abs/1202.5244}{{\tt arXiv:1202.5244}}].

\bibitem{Cerdeno:2013qta}
D.~G. Cerdeño, P.~Ghosh, C.~B. Park, and M.~Peiró, {\it {Collider signatures
  of a light NMSSM pseudoscalar in neutralino decays in the light of LHC
  results}},  {\em JHEP} {\bf 1402} (2014) 048,
  [\href{http://arxiv.org/abs/1307.7601}{{\tt arXiv:1307.7601}}].

\bibitem{Djouadi:2008uw}
A.~Djouadi, M.~Drees, U.~Ellwanger, R.~Godbole, C.~Hugonie, et~al., {\it
  {Benchmark scenarios for the NMSSM}},  {\em JHEP} {\bf 0807} (2008) 002,
  [\href{http://arxiv.org/abs/0801.4321}{{\tt arXiv:0801.4321}}].

\bibitem{King:2012tr}
S.~King, M.~Mühlleitner, R.~Nevzorov, and K.~Walz, {\it {Natural NMSSM Higgs
  Bosons}},  {\em Nucl.Phys.} {\bf B870} (2013) 323--352,
  [\href{http://arxiv.org/abs/1211.5074}{{\tt arXiv:1211.5074}}].

\bibitem{Curtin:2013fra}
D.~Curtin, R.~Essig, S.~Gori, P.~Jaiswal, A.~Katz, et~al., {\it {Exotic decays
  of the 125 GeV Higgs boson}},  {\em Phys.Rev.} {\bf D90} (2014), no.~7
  075004, [\href{http://arxiv.org/abs/1312.4992}{{\tt arXiv:1312.4992}}].

\bibitem{King:2014xwa}
S.~King, M.~Mühlleitner, R.~Nevzorov, and K.~Walz, {\it {Discovery Prospects
  for NMSSM Higgs Bosons at the High-Energy Large Hadron Collider}},  {\em
  Phys.Rev.} {\bf D90} (2014) 095014,
  [\href{http://arxiv.org/abs/1408.1120}{{\tt arXiv:1408.1120}}].

\bibitem{Ellwanger:2009dp}
U.~Ellwanger, C.~Hugonie, and A.~M. Teixeira, {\it {The Next-to-Minimal
  Supersymmetric Standard Model}},  {\em Phys.Rept.} {\bf 496} (2010) 1--77,
  [\href{http://arxiv.org/abs/0910.1785}{{\tt arXiv:0910.1785}}].

\bibitem{Maniatis:2009re}
M.~Maniatis, {\it {The Next-to-Minimal Supersymmetric extension of the Standard
  Model reviewed}},  {\em Int.J.Mod.Phys.} {\bf A25} (2010) 3505--3602,
  [\href{http://arxiv.org/abs/0906.0777}{{\tt arXiv:0906.0777}}].

\bibitem{Kowalska:2012gs}
K.~Kowalska, S.~Munir, L.~Roszkowski, E.~M. Sessolo, S.~Trojanowski, et~al.,
  {\it {The Constrained NMSSM with a 125 GeV Higgs boson -- A global
  analysis}},  {\em Phys. Rev.} {\bf D87} (2013) 115010,
  [\href{http://arxiv.org/abs/1211.1693}{{\tt arXiv:1211.1693}}].

\bibitem{Feroz:2008xx}
F.~Feroz, M.~Hobson, and M.~Bridges, {\it {MultiNest: an efficient and robust
  Bayesian inference tool for cosmology and particle physics}},  {\em
  Mon.Not.Roy.Astron.Soc.} {\bf 398} (2009) 1601--1614,
  [\href{http://arxiv.org/abs/0809.3437}{{\tt arXiv:0809.3437}}].

\bibitem{NMSSMTools}
{\url{http:/http://www.th.u-psud.fr/NMHDECAY/nmssmtools.html}}.

\bibitem{CMS-PAS-HIG-14-009}
{\it Precise determination of the mass of the higgs boson and studies of the
  compatibility of its couplings with the standard model},  Tech. Rep.
  CMS-PAS-HIG-14-009, CERN, Geneva, Jul, 2014.

\bibitem{ATLAS-CONF-2014-009}
{\it Updated coupling measurements of the higgs boson with the atlas detector
  using up to 25\,fb$^{-1}$ of proton-proton collision data},  Tech. Rep.
  ATLAS-CONF-2014-009, CERN, Geneva, May, 2014.

\bibitem{Bechtle:2008jh}
P.~Bechtle, O.~Brein, S.~Heinemeyer, G.~Weiglein, and K.~E. Williams, {\it
  {HiggsBounds: Confronting Arbitrary Higgs Sectors with Exclusion Bounds from
  LEP and the Tevatron}},  {\em Comput.Phys.Commun.} {\bf 181} (2010) 138--167,
  [\href{http://arxiv.org/abs/0811.4169}{{\tt arXiv:0811.4169}}].

\bibitem{Bechtle:2011sb}
P.~Bechtle, O.~Brein, S.~Heinemeyer, G.~Weiglein, and K.~E. Williams, {\it
  {HiggsBounds 2.0.0: Confronting Neutral and Charged Higgs Sector Predictions
  with Exclusion Bounds from LEP and the Tevatron}},  {\em Comput.Phys.Commun.}
  {\bf 182} (2011) 2605--2631, [\href{http://arxiv.org/abs/1102.1898}{{\tt
  arXiv:1102.1898}}].

\bibitem{Bechtle:2013gu}
P.~Bechtle, O.~Brein, S.~Heinemeyer, O.~Stal, T.~Stefaniak, et~al., {\it
  {Recent Developments in HiggsBounds and a Preview of HiggsSignals}},  {\em
  PoS} {\bf CHARGED2012} (2012) 024,
  [\href{http://arxiv.org/abs/1301.2345}{{\tt arXiv:1301.2345}}].

\bibitem{Bechtle:2013wla}
P.~Bechtle, O.~Brein, S.~Heinemeyer, O.~Stål, T.~Stefaniak, et~al., {\it
  {$\mathsf{HiggsBounds}-4$: Improved Tests of Extended Higgs Sectors against
  Exclusion Bounds from LEP, the Tevatron and the LHC}},  {\em Eur.Phys.J.}
  {\bf C74} (2014) 2693, [\href{http://arxiv.org/abs/1311.0055}{{\tt
  arXiv:1311.0055}}].

\bibitem{Djouadi:1994ge}
A.~Djouadi and P.~Gambino, {\it {Leading electroweak correction to Higgs boson
  production at proton colliders}},  {\em Phys.Rev.Lett.} {\bf 73} (1994)
  2528--2531, [\href{http://arxiv.org/abs/hep-ph/9406432}{{\tt
  hep-ph/9406432}}].

\bibitem{Degrassi:2004mx}
G.~Degrassi and F.~Maltoni, {\it {Two-loop electroweak corrections to Higgs
  production at hadron colliders}},  {\em Phys.Lett.} {\bf B600} (2004)
  255--260, [\href{http://arxiv.org/abs/hep-ph/0407249}{{\tt hep-ph/0407249}}].

\bibitem{Actis:2008ug}
S.~Actis, G.~Passarino, C.~Sturm, and S.~Uccirati, {\it {NLO Electroweak
  Corrections to Higgs Boson Production at Hadron Colliders}},  {\em
  Phys.Lett.} {\bf B670} (2008) 12--17,
  [\href{http://arxiv.org/abs/0809.1301}{{\tt arXiv:0809.1301}}].

\bibitem{Dittmaier:2011ti}
{\bf LHC Higgs Cross Section Working Group}, S.~Dittmaier et~al., {\it
  {Handbook of LHC Higgs Cross Sections: 1. Inclusive Observables}},
  \href{http://arxiv.org/abs/1101.0593}{{\tt arXiv:1101.0593}}.

\bibitem{superiso}
A.~Arbey and F.~Mahmoudi, {\it {SuperIso Relic: A program for calculating relic
  density and flavor physics observables in Supersymmetry}},  {\em
  Comput.Phys.Commun.} {\bf 176} (2007) 367--382,
  [\href{http://arxiv.org/abs/0906.0369}{{\tt arXiv:0906.0369}}].

\bibitem{Beringer:1900zz}
{\bf Particle Data Group}, J.~Beringer et~al., {\it {Review of Particle Physics
  (RPP)}},  {\em Phys.Rev.} {\bf D86} (2012) 010001.

\bibitem{Ade:2013zuv}
{\bf Planck Collaboration}, P.~Ade et~al., {\it {Planck 2013 results. XVI.
  Cosmological parameters}},  {\em Astron.Astrophys.} {\bf 571} (2014) A16,
  [\href{http://arxiv.org/abs/1303.5076}{{\tt arXiv:1303.5076}}].

\bibitem{micromegas}
G.~Belanger, F.~Boudjema, A.~Pukhov, and A.~Semenov, {\it {micrOMEGAs2.0: a
  program to calculate the relic density of dark matter in a generic model}},
  {\em Comput.Phys.Commun.} {\bf 181} (2010) 1277--1292,
  [\href{http://arxiv.org/abs/hep-ph/0607059}{{\tt hep-ph/0607059}}].

\bibitem{Alwall:2014hca}
J.~Alwall, R.~Frederix, S.~Frixione, V.~Hirschi, F.~Maltoni, et~al., {\it {The
  automated computation of tree-level and next-to-leading order differential
  cross sections, and their matching to parton shower simulations}},  {\em
  JHEP} {\bf 1407} (2014) 079, [\href{http://arxiv.org/abs/1405.0301}{{\tt
  arXiv:1405.0301}}].

\bibitem{cteq6}
J.~Pumplin, D.~Stump, J.~Huston, H.~Lai, P.~M. Nadolsky, et~al., {\it {New
  generation of parton distributions with uncertainties from global QCD
  analysis}},  {\em JHEP} {\bf 0207} (2002) 012,
  [\href{http://arxiv.org/abs/hep-ph/0201195}{{\tt hep-ph/0201195}}].

\bibitem{Harlander:2012pb}
R.~V. Harlander, S.~Liebler, and H.~Mantler, {\it {SusHi: A program for the
  calculation of Higgs production in gluon fusion and bottom-quark annihilation
  in the Standard Model and the MSSM}},  {\em Computer Physics Communications}
  {\bf 184} (2013) 1605--1617, [\href{http://arxiv.org/abs/1212.3249}{{\tt
  arXiv:1212.3249}}].

\bibitem{Djouadi:1991tka}
A.~Djouadi, M.~Spira, and P.~Zerwas, {\it {Production of Higgs bosons in proton
  colliders: QCD corrections}},  {\em Phys.Lett.} {\bf B264} (1991) 440--446.

\bibitem{Dawson:1990zj}
S.~Dawson, {\it {Radiative corrections to Higgs boson production}},  {\em
  Nucl.Phys.} {\bf B359} (1991) 283--300.

\bibitem{Spira:1995rr}
M.~Spira, A.~Djouadi, D.~Graudenz, and P.~Zerwas, {\it {Higgs boson production
  at the LHC}},  {\em Nucl.Phys.} {\bf B453} (1995) 17--82,
  [\href{http://arxiv.org/abs/hep-ph/9504378}{{\tt hep-ph/9504378}}].

\bibitem{Harlander:2002wh}
R.~V. Harlander and W.~B. Kilgore, {\it {Next-to-next-to-leading order Higgs
  production at hadron colliders}},  {\em Phys.Rev.Lett.} {\bf 88} (2002)
  201801, [\href{http://arxiv.org/abs/hep-ph/0201206}{{\tt hep-ph/0201206}}].

\bibitem{Anastasiou:2002yz}
C.~Anastasiou and K.~Melnikov, {\it {Higgs boson production at hadron colliders
  in NNLO QCD}},  {\em Nucl.Phys.} {\bf B646} (2002) 220--256,
  [\href{http://arxiv.org/abs/hep-ph/0207004}{{\tt hep-ph/0207004}}].

\bibitem{Ravindran:2003um}
V.~Ravindran, J.~Smith, and W.~L. van Neerven, {\it {NNLO corrections to the
  total cross-section for Higgs boson production in hadron hadron collisions}},
   {\em Nucl.Phys.} {\bf B665} (2003) 325--366,
  [\href{http://arxiv.org/abs/hep-ph/0302135}{{\tt hep-ph/0302135}}].

\bibitem{Marzani:2008az}
S.~Marzani, R.~D. Ball, V.~Del~Duca, S.~Forte, and A.~Vicini, {\it {Higgs
  production via gluon-gluon fusion with finite top mass beyond next-to-leading
  order}},  {\em Nucl.Phys.} {\bf B800} (2008) 127--145,
  [\href{http://arxiv.org/abs/0801.2544}{{\tt arXiv:0801.2544}}].

\bibitem{Harlander:2009mq}
R.~V. Harlander and K.~J. Ozeren, {\it {Finite top mass effects for hadronic
  Higgs production at next-to-next-to-leading order}},  {\em JHEP} {\bf 0911}
  (2009) 088, [\href{http://arxiv.org/abs/0909.3420}{{\tt arXiv:0909.3420}}].

\bibitem{Pak:2009dg}
A.~Pak, M.~Rogal, and M.~Steinhauser, {\it {Finite top quark mass effects in
  NNLO Higgs boson production at LHC}},  {\em JHEP} {\bf 1002} (2010) 025,
  [\href{http://arxiv.org/abs/0911.4662}{{\tt arXiv:0911.4662}}].

\bibitem{Maltoni:2003pn}
F.~Maltoni, Z.~Sullivan, and S.~Willenbrock, {\it {Higgs-boson production via
  bottom-quark fusion}},  {\em Phys.Rev.} {\bf D67} (2003) 093005,
  [\href{http://arxiv.org/abs/hep-ph/0301033}{{\tt hep-ph/0301033}}].

\bibitem{Harlander:2003ai}
R.~V. Harlander and W.~B. Kilgore, {\it {Higgs boson production in bottom quark
  fusion at next-to-next-to leading order}},  {\em Phys.Rev.} {\bf D68} (2003)
  013001, [\href{http://arxiv.org/abs/hep-ph/0304035}{{\tt hep-ph/0304035}}].

\bibitem{Dittmaier:2003ej}
S.~Dittmaier, M.~Kramer, and M.~Spira, {\it {Higgs radiation off bottom quarks
  at the Tevatron and the CERN LHC}},  {\em Phys.Rev.} {\bf D70} (2004) 074010,
  [\href{http://arxiv.org/abs/hep-ph/0309204}{{\tt hep-ph/0309204}}].

\bibitem{Liu:2012qu}
N.~Liu, L.~Wu, P.~W. Wu, and J.~M. Yang, {\it {Complete one-loop effects of
  SUSY QCD in $b\bar{b}h$ production at the LHC under current experimental
  constraints}},  {\em JHEP} {\bf 1301} (2013) 161,
  [\href{http://arxiv.org/abs/1208.3413}{{\tt arXiv:1208.3413}}].

\bibitem{Nhung:2013lpa}
D.~T. Nhung, M.~Muhlleitner, J.~Streicher, and K.~Walz, {\it {Higher Order
  Corrections to the Trilinear Higgs Self-Couplings in the Real NMSSM}},  {\em
  JHEP} {\bf 1311} (2013) 181, [\href{http://arxiv.org/abs/1306.3926}{{\tt
  arXiv:1306.3926}}].

\bibitem{Sjostrand:2007gs}
T.~Sjostrand, S.~Mrenna, and P.~Z. Skands, {\it {A Brief Introduction to PYTHIA
  8.1}},  {\em Comput.Phys.Commun.} {\bf 178} (2008) 852--867,
  [\href{http://arxiv.org/abs/0710.3820}{{\tt arXiv:0710.3820}}].

\bibitem{Cacciari:2011ma}
M.~Cacciari, G.~P. Salam, and G.~Soyez, {\it {FastJet User Manual}},  {\em
  Eur.Phys.J.} {\bf C72} (2012) 1896,
  [\href{http://arxiv.org/abs/1111.6097}{{\tt arXiv:1111.6097}}].

\bibitem{Elagin:2010aw}
A.~Elagin, P.~Murat, A.~Pranko, and A.~Safonov, {\it {A New Mass Reconstruction
  Technique for Resonances Decaying to di-tau}},  {\em Nucl.Instrum.Meth.} {\bf
  A654} (2011) 481--489, [\href{http://arxiv.org/abs/1012.4686}{{\tt
  arXiv:1012.4686}}].

\bibitem{Gripaios:2012th}
B.~Gripaios, K.~Nagao, M.~Nojiri, K.~Sakurai, and B.~Webber, {\it
  {Reconstruction of Higgs bosons in the di-tau channel via 3-prong decay}},
  {\em JHEP} {\bf 1303} (2013) 106, [\href{http://arxiv.org/abs/1210.1938}{{\tt
  arXiv:1210.1938}}].

\bibitem{Bianchini:2014vza}
L.~Bianchini, J.~Conway, E.~K. Friis, and C.~Veelken, {\it {Reconstruction of
  the Higgs mass in $H \to \tau\tau$ Events by Dynamical Likelihood
  techniques}},  {\em J.Phys.Conf.Ser.} {\bf 513} (2014) 022035.

\bibitem{Ellis:1987xu}
R.~K. Ellis, I.~Hinchliffe, M.~Soldate, and J.~van~der Bij, {\it {Higgs Decay
  to tau+ tau-: A Possible Signature of Intermediate Mass Higgs Bosons at the
  SSC}},  {\em Nucl.Phys.} {\bf B297} (1988) 221.

\bibitem{Butterworth:2008iy}
J.~M. Butterworth, A.~R. Davison, M.~Rubin, and G.~P. Salam, {\it {Jet
  substructure as a new Higgs search channel at the LHC}},  {\em
  Phys.Rev.Lett.} {\bf 100} (2008) 242001,
  [\href{http://arxiv.org/abs/0802.2470}{{\tt arXiv:0802.2470}}].

\bibitem{deLima:2014dta}
D.~E. Ferreira~de Lima, A.~Papaefstathiou, and M.~Spannowsky, {\it {Standard
  model Higgs boson pair production in the ($ b\overline{b} $)($ b\overline{b}
  $) final state}},  {\em JHEP} {\bf 1408} (2014) 030,
  [\href{http://arxiv.org/abs/1404.7139}{{\tt arXiv:1404.7139}}].

\bibitem{Cambridge}
Y.~L. Dokshitzer, G.~Leder, S.~Moretti, and B.~Webber, {\it {Better jet
  clustering algorithms}},  {\em JHEP} {\bf 9708} (1997) 001,
  [\href{http://arxiv.org/abs/hep-ph/9707323}{{\tt hep-ph/9707323}}].

\bibitem{Aachen}
M.~Wobisch and T.~Wengler, {\it {Hadronization corrections to jet
  cross-sections in deep inelastic scattering}},
  \href{http://arxiv.org/abs/hep-ph/9907280}{{\tt hep-ph/9907280}}.

\bibitem{antikT}
M.~Cacciari, G.~P. Salam, and G.~Soyez, {\it {The Anti-k(t) jet clustering
  algorithm}},  {\em JHEP} {\bf 0804} (2008) 063,
  [\href{http://arxiv.org/abs/0802.1189}{{\tt arXiv:0802.1189}}].

\bibitem{CMS-PAS-HIG-14-013}
{\bf CMS Collaboration}, {\it {Search for di-Higgs resonances decaying to 4
  bottom quarks}},  Tech. Rep. CMS-PAS-HIG-14-013, CERN, Geneva, 2014.

\bibitem{Gunion:2012gc}
J.~F. Gunion, Y.~Jiang, and S.~Kraml, {\it {Could two NMSSM Higgs bosons be
  present near 125 GeV?}},  {\em Phys.Rev.} {\bf D86} (2012) 071702,
  [\href{http://arxiv.org/abs/1207.1545}{{\tt arXiv:1207.1545}}].

\bibitem{Gunion:2012he}
J.~F. Gunion, Y.~Jiang, and S.~Kraml, {\it {Diagnosing Degenerate Higgs Bosons
  at 125 GeV}},  {\em Phys.Rev.Lett.} {\bf 110} (2013) 051801,
  [\href{http://arxiv.org/abs/1208.1817}{{\tt arXiv:1208.1817}}].

\bibitem{Cao:2013gba}
J.~Cao, F.~Ding, C.~Han, J.~M. Yang, and J.~Zhu, {\it {A light Higgs scalar in
  the NMSSM confronted with the latest LHC Higgs data}},  {\em JHEP} {\bf 1311}
  (2013) 018, [\href{http://arxiv.org/abs/1309.4939}{{\tt arXiv:1309.4939}}].

\bibitem{Aad:2014eha}
{\bf ATLAS Collaboration}, G.~Aad et~al., {\it {Measurement of Higgs boson
  production in the diphoton decay channel in pp collisions at center-of-mass
  energies of 7 and 8 TeV with the ATLAS detector}},  {\em Phys.Rev.} {\bf D90}
  (2014) 112015, [\href{http://arxiv.org/abs/1408.7084}{{\tt
  arXiv:1408.7084}}].

\bibitem{Badziak:2013bda}
M.~Badziak, M.~Olechowski, and S.~Pokorski, {\it {New Regions in the NMSSM with
  a 125 GeV Higgs}},  {\em JHEP} {\bf 1306} (2013) 043,
  [\href{http://arxiv.org/abs/1304.5437}{{\tt arXiv:1304.5437}}].

\bibitem{Das:2014fha}
D.~Das, L.~Mitzka, and W.~Porod, {\it {Discovery of Charged Higgs through
  {$\gamma\gamma$} final states}},  \href{http://arxiv.org/abs/1408.1704}{{\tt
  arXiv:1408.1704}}.

\end{thebibliography}\endgroup

\end{document}